
\input epsf.tex




\font\twelverm=cmr10 scaled 1200    \font\twelvei=cmmi10 scaled
1200 \font\twelvesy=cmsy10 scaled 1200   \font\twelveex=cmex10
scaled 1200 \font\twelvebf=cmbx10 scaled 1200
\font\twelvesl=cmsl10 scaled 1200 \font\twelvett=cmtt10 scaled
1200   \font\twelveit=cmti10 scaled 1200
\font\twelvesc=cmcsc10 scaled 1200  
\skewchar\twelvei='177   \skewchar\twelvesy='60


\def\twelvepoint{\normalbaselineskip=12.4pt plus 0.1pt minus 0.1pt
  \abovedisplayskip 12.4pt plus 3pt minus 9pt
  \belowdisplayskip 12.4pt plus 3pt minus 9pt
  \abovedisplayshortskip 0pt plus 3pt
  \belowdisplayshortskip 7.2pt plus 3pt minus 4pt
  \smallskipamount=3.6pt plus1.2pt minus1.2pt
  \medskipamount=7.2pt plus2.4pt minus2.4pt
  \bigskipamount=14.4pt plus4.8pt minus4.8pt
  \def\rm{\fam0\twelverm}          \def\it{\fam\itfam\twelveit}%
  \def\sl{\fam\slfam\twelvesl}     \def\bf{\fam\bffam\twelvebf}%
  \def\mit{\fam 1}                 \def\cal{\fam 2}%
  \def\sc{\twelvesc}               \def\tt{\twelvett}
  \def\sf{\twelvesf}
  \textfont0=\twelverm   \scriptfont0=\tenrm   \scriptscriptfont0=\sevenrm
  \textfont1=\twelvei    \scriptfont1=\teni    \scriptscriptfont1=\seveni
  \textfont2=\twelvesy   \scriptfont2=\tensy   \scriptscriptfont2=\sevensy
  \textfont3=\twelveex   \scriptfont3=\twelveex  \scriptscriptfont3=\twelveex
  \textfont\itfam=\twelveit
  \textfont\slfam=\twelvesl
  \textfont\bffam=\twelvebf \scriptfont\bffam=\tenbf
  \scriptscriptfont\bffam=\sevenbf
  \normalbaselines\rm}



\def\beginlinemode{\endmode
  \begingroup\parskip=0pt \obeylines\def\\{\par}\def\endmode{\par\endgroup}}
\def\beginparmode{\endmode
  \begingroup \def\endmode{\par\endgroup}}
\let\endmode=\par
{\obeylines\gdef\ {}}
\def\singlespace{\baselineskip=\normalbaselineskip}

\def\oneandahalfspace{\baselineskip=\normalbaselineskip
  \multiply\baselineskip by 3 \divide\baselineskip by 2}
\def\doublespace{\baselineskip=\normalbaselineskip \multiply\baselineskip by 2}

\newcount\firstpageno
\firstpageno=2
\footline={\ifnum\pageno<\firstpageno{\hfil}\else{\hfil\twelverm\folio\hfil}\fi}
\def\toppageno{\global\footline={\hfil}\global\headline
  ={\ifnum\pageno<\firstpageno{\hfil}\else{\hfil\twelverm\folio\hfil}\fi}}
\let\rawfootnote=\footnote              
\def\footnote#1#2{{\rm\singlespace\parindent=0pt\parskip=0pt
  \rawfootnote{#1}{#2\hfill\vrule height 0pt depth 6pt width 0pt}}}
\def\raggedcenter{\leftskip=4em plus 12em \rightskip=\leftskip
  \parindent=0pt \parfillskip=0pt \spaceskip=.3333em \xspaceskip=.5em
  \pretolerance=9999 \tolerance=9999
  \hyphenpenalty=9999 \exhyphenpenalty=9999 }
\def\dateline{\rightline{\ifcase\month\or
  January\or February\or March\or April\or May\or June\or
  July\or August\or September\or October\or November\or December\fi
  \space\number\year}}
\def\received{\vskip 3pt plus 0.2fill
 \centerline{\sl (Received\space\ifcase\month\or
  January\or February\or March\or April\or May\or June\or
  July\or August\or September\or October\or November\or December\fi
  \qquad, \number\year)}}


\hsize=6.5truein
\vsize 8.5truein
\voffset=0.0truein
\parskip=\medskipamount
\def\\{\cr}
\twelvepoint            
\doublespace            
\overfullrule=0pt       




\def\title                      
  {\null\vskip 3pt plus 0.2fill
   \beginlinemode \doublespace \raggedcenter \bf}

\def\author                     
  {\vskip 3pt plus 0.2fill \beginlinemode
   \singlespace \raggedcenter\sc}

\def\affil                      
  {\vskip 3pt plus 0.1fill \beginlinemode
   \oneandahalfspace \raggedcenter \sl}

\def\abstract                   
  {\vskip 3pt plus 0.3fill \beginparmode
   \singlespace ABSTRACT: }

\def\endtopmatter               
  {\endpage                     
   \body}

\def\body                       
  {\beginparmode}               

\def\head#1{                    
  \goodbreak\vskip 0.5truein    
  {\immediate\write16{#1}
   \raggedcenter \uppercase{#1}\par}
   \nobreak\vskip 0.25truein\nobreak}

\def\subhead#1{                 
  \vskip 0.25truein             
  {\raggedcenter {#1} \par}
   \nobreak\vskip 0.25truein\nobreak}

\def\beginitems{
\par\medskip\bgroup\def\i##1 {\item{##1}}\def\ii##1 {\itemitem{##1}}
\leftskip=36pt\parskip=0pt}
\def\enditems{\par\egroup}

\def\beneathrel#1\under#2{\mathrel{\mathop{#2}\limits_{#1}}}


\def\references                 
  {\head{{\bf References}}            
   \beginparmode
   \frenchspacing \parindent=0pt \leftskip=1truecm
   \parskip=8pt plus 3pt \everypar{\hangindent=\parindent}}



\gdef\journal#1, #2, #3, 1#4#5#6{               
    {\sl #1~}{\bf #2}, #3 (1#4#5#6)}            

\gdef\refa#1, #2, #3, #4, 1#5#6#7.{\noindent#1, #2 {\bf #3}, #4
(1#5#6#7).\rm}

\gdef\refb#1, #2, #3, #4, 1#5#6#7.{\noindent#1 (1#5#6#7), #2 {\bf
#3}, #4.\rm}

\def\pr{\journal Phys.Rev., }

\def\prl{\journal Phys.Rev.Lett., }

\def\jmp{\journal J.Math.Phys., }

\def\rmp{\journal Rev.Mod.Phys., }

\def\np{\journal Nucl.Phys., }

\def\pl{\journal Phys.Lett., }

\def\annp{\journal Ann.Phys.(N.Y.), }

\def\cqg{\journal Class.Quantum Grav., }

\def\grg{\journal Gen.Rel.Grav., }

\def\figurecaptions             
  {\endpage
   \beginparmode
   \head{Figure Captions}
}

\def\endpage                    
  {\vfill\eject}

\def\endpaper                   
  {\endmode\vfill\supereject}


\def\heading                            
  {\vskip 0.5truein plus 0.1truein      
   \beginparmode \def\\{\par} \parskip=0pt \singlespace \raggedcenter}

\def\subheading                         
  {\vskip 0.25truein plus 0.1truein     
   \beginlinemode \singlespace \parskip=0pt \def\\{\par}\raggedcenter}

\def\tag#1$${\eqno(#1)$$}

\def\align#1$${\eqalign{#1}$$}

\def\aligntag#1$${\gdef\tag##1\\{&(##1)\cr}\eqalignno{#1\\}$$
  \gdef\tag##1$${\eqno(##1)$$}}

\def\overset #1\to#2{{\mathop{#2}\limits^{#1}}}
\def\underset#1\to#2{{\let\next=#1\mathpalette\undersetpalette#2}}
\def\undersetpalette#1#2{\vtop{\baselineskip0pt
\ialign{$\mathsurround=0pt
#1\hfil##\hfil$\crcr#2\crcr\next\crcr}}}


\def\ref#1{Ref.~#1}                     
\def\[#1]{[\cite{#1}]}
\def\cite#1{{#1}}
\def\(#1){(\call{#1})}
\def\call#1{{#1}}
\def\taghead#1{}
\def\frac#1#2{{#1 \over #2}}
\def\half{{\frac 12}}

\def\12{{1\over2}}

\def\sla{\raise.15ex\hbox{$/$}\kern-.57em}
\def\leaderfill{\leaders\hbox to 1em{\hss.\hss}\hfill}
\def\twiddle{\lower.9ex\rlap{$\kern-.1em\scriptstyle\sim$}}
\def\bigtwiddle{\lower1.ex\rlap{$\sim$}}
\def\gtwid{\mathrel{\raise.3ex\hbox{$>$\kern-.75em\lower1ex\hbox{$\sim$}}}}
\def\ltwid{\mathrel{\raise.3ex\hbox{$<$\kern-.75em\lower1ex\hbox{$\sim$}}}}
\def\square{\kern1pt\vbox{\hrule height 1.2pt\hbox{\vrule width 1.2pt\hskip 3pt
   \vbox{\vskip 6pt}\hskip 3pt\vrule width 0.6pt}\hrule height 0.6pt}\kern1pt}
\def\tdot#1{\mathord{\mathop{#1}\limits^{\kern2pt\ldots}}}

\def\pmb#1{\setbox0=\hbox{#1}%
  \kern-.025em\copy0\kern-\wd0
  \kern  .05em\copy0\kern-\wd0
  \kern-.025em\raise.0433em\box0 }

\oneandahalfspace

\centerline {\bf INTRODUCTORY LECTURES ON QUANTUM
COSMOLOGY\footnote{$^{\dag}$}{\rm Published in {\it Proceedings of
the Jerusalem Winter School on Quantum Cosmology and Baby
Universes}, edited by S.Coleman, J.B.Hartle, T.Piran and
S.Weinberg (World Scientific, Singapore, 1991)}}

\vskip 0.5in
\author Jonathan J.Halliwell
\affil Center for Theoretical Physics Laboratory for Nuclear
Science Massachusetts Institute of Technology Cambridge, MA 02139,
U.S.A. \vskip 0.5in \centerline {Bitnet address: Halliwell@MITLNS}
\vskip 0.5in \centerline {March, 1990} \vskip 0.5in
\centerline{CTP \# 1845}

\vskip 1.0in \abstract We describe the modern approach to quantum
cosmology, as initiated by Hartle and Hawking, Linde, Vilenkin and
others. The primary aim is to explain how one determines the
consequences for the late universe of a given quantum theory of
cosmological initial or boundary conditions. An extensive list of
references is included, together with a guide to the literature.
\endtopmatter

\def\a{\alpha}
\def\x{{\bf x}}
\def\D{{\cal D}}
\def\R{{I\kern-0.3em R}}

\head{\bf 1. INTRODUCTION}

My intention in these lectures is to describe the practical
business of actually doing quantum cosmology. That is, I will
describe how, in the context of particular models, one determines
the consequences for the late universe of a given theory of
initial conditions.

What is the motivation for studying quantum cosmology? One
possible motivation comes from quantum gravity. Cosmological
models are simple examples to which quantum gravity ideas may be
applied. Moreover, the very early universe is perhaps the only
laboratory in which quantum gravity may be tested. A second
motivation, and the main one for the purposes of these lectures,
concerns initial conditions in cosmology. Although the hot big
bang model explains some of the features of the observed universe,
there are a number of features that it did not explain, such as
its flatness, absence of horizons, and the origin of the density
fluctuations required to produce galaxies. The inflationary
universe scenario (Guth, 1981), which involves quantized matter
fields on a classical gravitational background, provided a
possible solution to the horizon and flatness problems. Moreover,
by assuming that the matter fields start out in a particular
quantum state, the desired density fluctuation spectrum may be
obtained.\footnote{$^{\dag}$} {For a review, see, for example,
Brandenberger (1987, 1989).} However, in the inflationary universe
scenario, the question of initial conditions was largely ignored.
Whilst it is certainly true that, as a result of inflation, the
observed universe could have arisen from a much larger class of
initial conditions than in the hot big bang model, it is certainly
not true that it could have arisen from \it any \rm initial state
-- one could choose an initial quantum state for the matter which
did not lead to the correct density perturbation spectrum, and
indeed, one could choose initial conditions for which inflation
does not occur. In order to have a complete explanation of the
presently observed state of the universe, therefore, it is
necessary to face up to the question of initial conditions.

Now, as the evolution of the universe is followed backwards in
time, the curvatures and densities approach the Planck scale, at
which one would expect quantum gravitational effects to become
important. Quantum cosmology, in which both the matter and
gravitational fields are quantized, is therefore the natural
framework in which to address the question of initial conditions.

In a sentence, quantum cosmology is the application of quantum
theory to the dynamical systems describing closed cosmologies.
Historically, the earliest investigations into quantum cosmology
were primarily those of by DeWitt (1967), Misner (1969a, 1969b,
1969c, 1970, 1972, 1973) and Wheeler (1963, 1968) in the 1960's.
This body of work I shall refer to as the ``old" quantum
cosmology, and will not be discussed here. It is discussed in the
articles by MacCallum (1975), Misner (1972) and Ryan (1972).

After the initial efforts by the above authors, quantum cosmology
went \break through a bit of a lull in the 1970's. However, it was
re-vitalized in the 1980's, primarily by Hartle and Hawking
(Hartle and Hawking, 1983; Hawking, 1982, 1984a), by Vilenkin
(1984, 1986, 1988) and by Linde (1984a, 1984b, 1984c). There were
two things that these authors added to the old approach. Firstly,
Hartle and Hawking introduced Euclidean functional integrals, and
used a blend of canonical and path integral methods. Secondly, all
of the above authors faced up squarely to the issue of boundary or
initial conditions on the wave function of the universe. It is
this modern approach to quantum cosmology that will be the subject
of these lectures.

The central object of interest in quantum cosmology is the wave
function of a closed universe,
$$
\Psi[h_{ij}(\x), \Phi(\x), B] \eqno(1.1)
$$
This is the amplitude that the universe contains a three-surface
$B$ on which the three-metric is $h_{ij}(\x)$ and the matter field
configuration is $\Phi(\x)$. From such an amplitude one would hope
to extract various predictions concerning the outcome of large
scale observations. To fix the amplitude (1.1), one first needs a
theory of dynamics, such as general relativity. From this one can
derive an equation analagous to the Schrodinger equation, called
the Wheeler-DeWitt equation, which the wave function of the
universe must satisfy. The Wheeler-DeWitt equation will have many
solutions, so in order to have any predictive power, it is
necessary to propose a law of initial or boundary conditions to
single out just one solution. And fnally, one needs some kind of
scheme to interpret the wave function. So these are the three
elements that go into quantum cosmology: dynamics, initial
conditions, interpretation.

One of the most basic observational facts about the universe we
observe today is that it is described by classical laws to a very
high degree of precision. Since in quantum cosmology the universe
is taken to be fundamentally quantum mechanical in nature, one of
the most primitive predictions a quantum theory of initial
conditions should make, is that the universe is approximately
classical when it is large. Indeed, what we will typically find to
be the case is that the wave function indicates the regions in
which space-time is essentially classical, and those in which it
is not. In the regions where spacetime is essentially classical,
we will find that the wave function is peaked about a set of
solutions to the classical Einstein equations and, as a
consequence of the boundary conditions on the wave function, this
set is a subset of the general solution. The boundary conditions,
through the wave function, therefore set initial conditions on the
classical solutions. We may then begin to ask whether or not the
finer details of the universe we observe, such as the existence of
an inflationary era, are consequences of the chosen theory of
initial conditions. In addition, in the approximately classical
region, we will recover from the Wheeler-DeWitt equation the
familiar quantum field theory for the matter fields on a classical
curved spacetime background. Moreover, we will find that the
boundary conditions on the wave function of the universe single
out a particular choice of vacuum state for the matter fields. We
may then ask whether or not the chosen vacuum state is the
appropriate one for the subsequent emergence of large scale
structure.

These remarks will hopefully become clearer as we progress, but in
brief, the theme of these lectures may be summarized as follows.
The inflationary universe scenario -- and indeed most other
cosmological scenarios -- will always depend to some extent on
initial conditions. I would like to try and argue that, within the
context of quantum cosmology, there exist natural quantum theories
of initial or boundary conditions from which the appropriate
initial conditions for inflation and the emergence of large scale
structure follow.

Throughout the text I will give very few references. An extensive
guide to the literature is contained in Section 13.

\head{2. A SIMPLE EXAMPLE}

Rather than begin with the general formalism of quantum cosmology,
I am going to first consider a simple inflationary universe model.
This will help clarify some of the rather vague remarks made above
concerning the need for initial conditions. The model will be
treated rather heuristically; the details will be attended to
later.

Consider a universe described by a homogeneous isotropic
Robertson-Walker metric
$$
ds^2 = \sigma^2 \left[ -N^2(t) dt^2 + e^{2\a(t)} d\Omega_3^2(k)
\right] \eqno(2.1)
$$
where $\sigma^2 = 2/(3\pi m_p^2) $ and $d\Omega_3^2(k)$ is the
metric on the spatial sections which have constant curvature
$k=-1,0,+1$. In quantum cosmology one is generally interested in
closed ($k=+1$) universes, but for the moment we will retain all
three values of $k$. The metric is described by a single scale
factor, $e^{\alpha(t)}$. As matter source we will use a
homogeneous minimally coupled scalar field $ \sqrt{2} \pi \sigma
\phi(t) $ with potential $2 \pi^2 \sigma^2 V(\phi)$. The
Einstein-scalar action for this system is
$$
S= \half \int dt N e^{3\a} \left[ - {\dot \a^2 \over N^2} + {\dot
\phi^2 \over N^2 } - V(\phi) + k e^{-2\a} \right] \eqno(2.2)
$$
(the full form of the Einstein-scalar action is given in the next
section). By varying with respect to $ \alpha $, $\phi$ and $N$,
one may derive the field equations and constraint, which, after
some rearrangement, are conveniently written,
$$
\ddot \phi = - 3 \dot \a \dot \phi - \half V'(\phi) \eqno(2.3)
$$
$$
\ddot \a = -2 \dot \phi^2 - \dot \a^2 + V(\phi) \eqno(2.4)
$$
$$
- \dot \a^2 + \dot \phi^2 +V(\phi) = k e^{-2\a} \eqno(2.5)
$$
in the gauge $N=1$. We will not assume a precise form for
$V(\phi)$, except that it is of the inflationary type; that is,
that for some range of values of $ \phi $, $ V(\phi) $ is large
and $ |V'(\phi)/V(\phi)| << 1 $. This is satisfied, for example,
for large $ \phi $ in chaotic models, with $ V(\phi)=m^2\phi^2 $
or $ \lambda \phi^4 $, and for $ \phi $ near the origin in models
with a Coleman-Weinberg potential. It is important to note that
the general solution to the system (2.3)-(2.5) will involve {\it
three} arbitrary parameters.

For models in which the potential satisfies the above conditions,
it is easily seen that there exist solutions for which $ \dot \phi
\approx 0 $ and the potential then acts like a cosmological
constant; thus the model undergoes inflation, $ e^{\alpha} \approx
e^{V^{\half}t} $. However, whether or not such a solution arises
is clearly a question of {\it initial conditions}: one needs to
choose the initial value of $ \dot \phi $ to be small, and one
needs to choose the initial value of $ \phi $ to be in the region
for which $ |V'(\phi)/V(\phi)| << 1 $. It is therefore pertinent
to ask, to what extent is inflation generic in a model of this
type?

To address this question, one needs a complete picture of the
classical solutions. Clearly it would be very difficult to solve
the field equations exactly, even for very simple choices of $
V(\phi) $. However, one can often obtain useful information using
the qualitative theory of dynamical systems. The sort of
differential equations one encounters in cosmology can frequently
be cast in the form
$$
\dot x = f(x,y,z...), \quad \dot y = g(x,y,z...), \quad \dot z =
... \eqno(2.6)
$$
Eq.(2.6) gives the direction of the solutions at every point
$(x,y,z...)$. By drawing arrows at a selection of points one may
thus construct a complete picture of the entire family of
trajectories which solve (2.6) without integrating explicitly.

This method may be applied to the field equations (2.3), (2.4) by
writing $ x = \dot \phi $, $ y = \dot \alpha $,  $ z = \phi $ (the
constraint (2.5) is not normally used so that the three cases
$k=0,-1,+1$ may be treated simultaneously). The resulting
three-dimensional phase portrait is, however, rather difficult to
construct.\footnote{$^{\dag}$} {In the case $k=0$, one can
eliminate $\dot \alpha $ using the constraint, and the phase
portrait becomes two-dimensional. This has been constructed for
various inflationary potentials by Belinsky et al.(1985) and Piran
and Williams (1985).} Let us therefore make a simplification,
which is to go straightaway to a region where the
$\phi$-dependence of $V(\phi)$ is negligible. This is like having
a massless scalar field and a cosmological constant. One then has
a two-dimensional system,
$$
\dot x = - 3xy, \quad \dot y = -2x^2 - y^2 + V \eqno(2.7)
$$
The constraint equation
$$
x^2 - y^2 + V = k e^{-2\a} \eqno(2.8)
$$
simply indicates that the $k=0$ solutions are the two curves $
y=\pm \sqrt{x^2+V} $, the $k=+1$ solutions lie between these
curves and the $k=-1$ solutions lie outside these curves.

The phase portrait for this two-dimensional system is shown in
Fig.1. The point of particular interest is the point $ \dot \alpha
= V^{\half} $, $ \dot \phi = 0 $, on the $ k=0 $ curve, because at
this point the model undergoes inflation. This point is an
attractor for all the expanding $k=0$ and $k=-1$ solutions. The
$k=+1$ solutions, however, with which one is primarily concerned
in quantum cosmology, do not all end up on the attractor: if they
start out away from the $k=0$ curve with $|\dot \phi|$ large they
recollapse before getting anywhere near the attractor. Inflation
occurs, therefore, only for the subset of $k=+1$ solutions with
reasonably small initial $ \dot \phi $. Furthermore, when
$V(\phi)$ is allowed to vary with $\phi$, there is also the issue
of {\it sufficient} inflation. In the massive scalar field model,
for example, even if $\dot \phi \approx 0 $ initially, it is known
that the universe inflates by the required factor $e^{65}$ only
for initial values of $\phi$ greater than about $4$ (in Planck
units) (Hawking, 1984a; Page, 1986a).

So this simple model allows one to see quite clearly how the
occurence of inflation depends rather crucially on the initial
values of $ \phi $ and $ \dot \phi $. Now let us consider the
quantization of this model, still proceeding heuristically, to see
how quantum cosmology may shed some light on this issue.

We wish to quantize the dynamical system described by the action
(2.2), for the case $k=+1$. We begin by finding the Hamiltonian of
the theory. The momenta conjugate to $ \alpha $ and $ \phi $ are
defined in the usual way and are given by
$$
\pi_{\a} = - e^{3\a} {\dot\a \over N}, \quad \pi_{\phi} = e^{3\a}
{\dot\phi \over N} \eqno(2.9)
$$
The canonical Hamiltonian is defined in the usual way and is given
by
$$
H_c = \half N e^{-3\a} \left[ - \pi_{\a}^2 + \pi_{\phi}^2 +
e^{6\a}V(\phi) - e^{4\a} \right] \equiv N H \eqno(2.10)
$$
The Hamiltonian form of the action is given by
$$
S = \int dt \left[ \dot \a \pi_{\a} + \dot \phi \pi_{\phi} - N H
\right] \eqno(2.11)
$$
This form of the action exposes the fact that the lapse function
$N$ is a Lagrange multiplier which enforces the constraint
$$
H=0 \eqno(2.12)
$$
This is just the phase-space form of the constraint (2.5). The
constraint indicates the presence of a symmetry, in this case
reparametrization invariance, about which we will have more to say
later.

Proceeding naively, we quantize this system by introducing a wave
function $ \Psi (\a, \phi, t ) $ and asking that it satisfy a
time-dependent Schr\"odinger equation constructed from the
canonical Hamiltonian (2.10):
$$
i { \partial \Psi \over \partial t } = H_c \Psi \eqno(2.13)
$$
To ensure that the symmetry corresponding to the constraint (2.12)
be imposed at the quantum level, we will also ask that the wave
function is annihilated by the operator version of (2.12):
$$
H \Psi = \half e^{-3\a} \left[ {\partial^2 \over \partial \a^2} -
{\partial^2 \over \partial \phi^2} + e^{6\a} V(\phi) - e^{4\a}
\right] \Psi = 0 \eqno(2.14)
$$
where the momenta in (2.12) have been replaced by operators using
the usual substitutions. However, since $H_c=NH$, it follows from
(2.13) and (2.14) that the wave function is independent of $t$;
thus the entire dynamics of the wave function is in fact contained
in (2.14) with $\Psi=\Psi(\a,\phi)$. The fact that the wave
function does not depend on the time parameter $t$ explicitly is
actually characteristic of parametrized theories such as general
relativity. (2.14) is called the Wheeler-DeWitt equation and is
the central equation of interest in quantum cosmology.

Let us find some simple solutions to this equation. Let us go to a
region for which $ |V'(\phi)/V(\phi)|<<1 $ and look for solutions
which do not depend very much on $ \phi $, so we may ignore the $
\phi $ derivative term in (2.14). The problem is then a standard
one-dimensional WKB problem in $\alpha$ with a potential $ U =
e^{6\alpha} V(\phi) -e^{4\alpha} $. In the region $ U<<0 $, where
the scale factor is small, there are WKB solutions of the form
$$
\Psi(\a,\phi) \approx \exp\left( \pm {1 \over 3V(\phi)}
(1-e^{2\a}V(\phi))^{3/2} \right) \eqno(2.15)
$$
This region, in which the wave function is exponential, is
normally regarded as some kind of tunneling or classically
forbidden region. In the region $ U>>0 $, where the scale factor
is large, there are WKB solutions of the form
$$
\Psi(\a,\phi) \approx \exp\left( \pm {i \over 3V(\phi)}
(e^{2\a}V(\phi)-1)^{3/2} \right) \eqno(2.16)
$$
This region, in which the wave function is oscillatory, is usually
thought of as a classically allowed region. One can impose
boundary conditions in either region, and then match the solutions
in the two regions using the usual WKB matching procedure.

Consider in a little more detail the oscillatory region, including
the $ \phi $ dependence. Let us look for solutions of the form $
\Psi =e^{iS} $, where $S$ is a rapidly varying function of $
\alpha $ and $\phi $. Inserting this in the Wheeler-DeWitt
equation, one finds that, to leading order,  $ S $ must obey the
Hamilton-Jacobi equation
$$
- \left( {\partial S \over \partial \a} \right)^2 + \left(
{\partial S \over \partial \phi} \right)^2 +U(\a,\phi)=0
\eqno(2.17)
$$
We will assume that some set of boundary conditions are imposed on
$\Psi$; thus a particular solution of the Hamilton-Jacobi equation
(2.17) will be picked out. Compare (2.17) with the Hamiltonian
constraint,
$$
-\pi_{\a}^2 + \pi_{\phi}^2 + U(\a,\phi) = 0 \eqno(2.18)
$$
It invites the identification
$$
\pi_{\a} = {\partial S \over \partial \a}, \quad \pi_{\phi} =
{\partial S \over \partial \phi} \eqno(2.19)
$$
More precisely, one can in fact show that a wave function of the
form $e^{iS}$ predicts a strong correlation between coordinates
and momenta of the form (2.19). Furthermore, using the
relationship between velocities and momenta (2.9), and the fact
that $S$ obeys the Hamilton-Jacobi equation (2.17), one may show
that (2.19) defines a set of trajectories in the $\a\phi$ plane
which are solutions to the classical field equations and
constraint, (2.3)-(2.5). That is, the wave function $e^{iS}$ is
strongly peaked about a set of solutions to the classical field
equations.

For a given solution $S$ of the Hamilton-Jacobi equation the first
integral of the field equations (2.19) about which the wave
function is peaked involves just {\it two} arbitrary parameters.
Recall, however, that the general solution to the full field
equations (2.3)-(2.5) involved {\it three} arbitrary parameters.
For given $S$, therefore, {\it the wave function} $e^{iS}$ {\it is
strongly peaked about the two-parameter subset of the
three-parameter general solution}. By imposing boundary conditions
on the wave function a particular solution $\Psi $ to the
Wheeler-DeWitt equation is picked out, which in the WKB
approximation picks out a particular solution $S$ to the
Hamilton-Jacobi equation; this in turn defines a two-parameter
subset of the three-parameter general solutions. It is in this way
that boundary conditions on the wave function of the universe
effectively imply initial conditions on the classical solutions.

Let us see how this works for the particular solution (2.16). For
$ e^{2\a}V>>1 $, it is of the form $e^{iS}$ with $S\approx
-{1\over 3} e^{3\a}V^{\half}$. According to the above analysis,
this wave function is peaked about the trajectories defined by
$$
\dot \a \approx V^{\half}, \quad \dot \phi \approx 0 \eqno(2.20)
$$
(we could of course have taken the opposite sign for $S$ -- this
leads to a set of contracting solutions). Eq.(2.20) integrates to
yield
$$
e^{\a} \approx e^{V^{\half}(t-t_0)}, \quad \phi \approx
\phi_0=constant \eqno(2.21)
$$
Here $t_0$ and $\phi_0$ are the two arbitrary constants
parametrizing this set of solutions. The constant $t_0$ is in fact
irrelevant, because it is just the origin of unobservable
parameter time. From (2.20) one may see that the wave function is
peaked right on the inflationary attractor in Fig.1. So {\it this
particular wave function} picks out the inflationary solutions.

One can actually get a little more out of the wave function in
addition to (2.20). The wave function more generally is of the
form $C(\a,\phi)e^{iS}$. The $e^{iS}$ part, as we have discussed,
shows that the wave function is peaked about a set of
trajectories. These trajectories may be labeled by the value of
the arbitrary constant $\phi_0$. The prefactor effectively
provides a measure on the set of possible values of $\phi_0$, and
may therefore be used to assess the relative likelihood of
inflation. We will describe this in a lot more detail later.

From this simple model we have learned a few things that are in
fact quite general. They are as follows:

\item{1)} Classical cosmology needs initial conditions. This is
illustrated rather clearly using the phase-portrait of classical
solutions, allowing one to see what sort of features are generic,
and what sort of features are dependent on a specific choice of
initial conditions.

\item{2)} In the quantized model, there is a region in which the
wave function is exponential, indicating that this region is
classically forbidden.\footnote{$^{\dag}$} {In this particular
model, and for the particular solution to the Wheeler-DeWitt
equation we looked at, the classically forbidden region is at
small values of the scale factor. This is in accord with the
general belief that ``quantum gravity effects become important
when the universe is very small". This is, however, dependent on
boundary conditions. There are other solutions to the
Wheeler-DeWitt equation which are oscillatory for small scale
factors. We shall return to this point in Section 6.}

\item{3)} There is a region in which the wave function is
oscillatory, indicating that this region classically allowed. To
be precise, the wave function in the oscillatory region is
strongly peaked about a {\it set} of solutions to the classical
field equations.

\item{4)} The set of solutions about which a given WKB wave
function is peaked is a {\it subset} of the general solution to
the field equations. That is, a particular solution to the
Wheeler-DeWitt equation is peaked about a particular subset of the
full set of solutions to the field equations. Moreover, the wave
function provides a measure on the classical trajectories within
this set. A general solution to the Wheeler-DeWitt equation would
be peaked about a general solution to the field equations, so by
simply quantizing the model one does not necessarily learn
anything about initial conditions. One merely transfers the
question of initial conditions on the classical solutions to the
question of boundary conditions on the wave function of the
universe. To have complete predictive power, therefore, one needs
a quantum theory of boundary conditions.

In connection with the fourth point above, one might ask, have we
really improved the situation with regard to initial conditions by
going to the quantum theory? The answer is, I believe, yes, for at
least two reasons. Firstly, as the simple model above indicates, a
classical description of cosmology is not always valid. In
attempting to impose classical initial conditions at small
three-geometries, therefore, one might be imposing them in a
region in which, from the point of view of the quantum theory, a
classical description is not really appropriate. Secondly, a
somewhat more aesthetic point. Classically, there is no obvious
reason for choosing one set of initial conditions over another. No
one choice stands out as being more natural or elegant than any
other. In quantum cosmology, however, one can argue that certain
quantum states for the universe have considerably more appeal than
others on the grounds of simplicity or naturalness. I will leave
this to the reader to judge for themselves when we come to discuss
particular proposals for quantum theories of initial conditions.

This ends what has really been an introductory tour of quantum
cosmology. In the following sections, we will go over essentially
the same points but in greater generality and detail.

\head{\bf 3. THE HAMILTONIAN FORMULATION OF GENERAL RELATIVITY}

We now procede to the general formalism of quantum cosmology. This
begins with the Hamiltonian formulation of general relativity
(Hanson et al., 1976; Misner et al.,1970; Teitelboim, 1990). One
considers a three-surface on which the three-metric is $h_{ij}$,
with some matter field configuration. We will take the
three-surface to be compact, since we are considering only closed
universes. The three-surface is embedded in a four-manifold on
which the four-metric is $g_{\mu\nu}$. This embedding is described
by the standard $(3+1)$ form of the four-metric,
$$
ds^2= g_{\mu\nu} dx^{\mu} dx^{\nu} =  -\left(N^2-N_iN^i\right)
dt^2+2N_i dx^i dt+h_{ij} dx^i dx^j \eqno(3.1)
$$
where $ N $ and $ N_i $ are the lapse and shift functions. (Our
conventions are $\mu,\nu = 0,1,2,3 $ and $ i,j = 1,2,3$). They
describe the way in which the choice of coordinates on one
three-surface is related to the choice on an adjacent
three-surface, and are therefore arbitrary.

The action will be taken to be the standard Einstein-Hilbert
action coupled to matter,
$$
S = {m_p^2 \over 16 \pi } \left[ \int _M d^4 x (-g)^{\half}
(R-2\Lambda) + 2 \int _{\partial M} d^3 x \ h^{\half} K \right] +
S_{matter} \eqno(3.2)
$$
where $ K $ is the trace of the extrinsic curvature $K_{ij}$ at
the boundary $ \partial M $ of the four-manifold $ M $, and is
given by
$$
K_{ij} = {1 \over 2N} \left[ - {\partial h_{ij} \over \partial t}
+ 2 D_{(i} N_{j)} \right] \eqno(3.3)
$$
Here, $D_i$ is the covariant derivative in the three-surface. For
a scalar field $\Phi$, the matter action is
$$
S_{matter} = - \half \int d^4 x (-g)^{\half} \left[ g^{\mu\nu}
\partial_{\mu} \Phi \partial_{\nu} \Phi +V(\Phi) \right]
\eqno(3.4)
$$
In terms of the (3+1) variables, the action takes the form
$$
S= {m_p^2 \over 16 \pi} \int d^3x \ dt \ N h^{\half} \left[ K_{ij}
K^{ij} -K^2 + {^3 R} - 2 \Lambda \right] + S_{matter} \eqno(3.5)
$$

In a perfectly standard way, one may derive the Hamiltonian form
of the action,
$$
S=\int d^3x \ dt\left[ \dot h_{ij} \pi^{ij} + \dot \Phi \pi_{\Phi}
-N{\cal H}-N^i{\cal H}_i\right] \eqno(3.6)
$$
where $\pi^{ij}$ and $ \pi_{\Phi} $ are the momenta conjugate to
$h_{ij}$ and $ \Phi $ respectively. The Hamiltonian is a sum of
constraints, with the lapse $ N $ and shift $ N^i$ playing the
role of lagrange multipliers. There is the momentum constraint,
$$
{\cal H}_i=-2D_j \pi^j_i +{\cal H}^{matter}_i=0 \eqno(3.7)
$$
and the Hamiltonian constraint
$$
{\cal H}=  {16 \pi \over m_p^2} G_{ijkl}\pi^{ij} \pi^{kl}- {m_p^2
\over 16 \pi }   h^{\half} ({^3}R-2 \Lambda) +{\cal H}^{matter}=0
\eqno(3.8)
$$
where $ G_{ijkl} $ is the DeWitt metric and is given by
$$
G_{ijkl}={1\over 2} h^{-\half} \left(h_{ik} h_{jl}+h_{il}
h_{jk}-h_{ij} h_{kl}\right) \eqno(3.9)
$$
These constraints are equivalent, respectively, to the time-space
and time-time components of the classical Einstein equations. The
constraints play a central role in the canonical quantization
procedure, as we shall see.

The arena in which the classical dynamics takes place is called
superspace, the space of all three-metrics and matter field
configurations $ (h_{ij}(\x), \Phi(\x)) $ on a
three-surface\footnote{$^{\dag}$} {This superspace has nothing to
do with the superspace of supersymmetry. Also, earlier authors in
quantum cosmology used a different definition of superspace: they
defined it to be the space of all three-metrics, but factored out
by the three-dimensional diffeomorphisms.}. Superspace is infinite
dimensional, with a finite number of coordinates $(h_{ij}(\x),
\Phi(\x))$ at every point $\x$ of the three-surface. The DeWitt
metric (plus some suitable metric on the matter fields) provides a
metric on superspace. It has the important property that its
signature is hyperbolic at every point $\x$ in the three-surface.
The signature of the DeWitt metric is independent of the signature
of spacetime.

\head{\bf 4. QUANTIZATION}

In the canonical quantization procedure, the quantum state of the
system is represented by a wave functional $ \Psi [ h_{ij} , \Phi
] $, a functional on superspace. An important feature of this wave
function is that is does not depend explicitly on the coordinate
time label $ t $.  This is because the three-surfaces are compact,
and thus their intrinsic geometry, specified by the three-metric,
fixes more-or-less uniquely their relative location in the
four-manifold. Another way of saying essentially the same thing,
is to say that general relativity is an example of a \it
parametrized \rm theory, which means that ``time" is already
contained amongst the dynamical variables describing it, $ h_{ij},
\Phi $.

According to the Dirac quantization procedure, the wave function
is annihilated by the operator versions of the classical
constraints. That is, if one makes the usual substitutions for
momenta
$$
\pi^{ij}\to -i{\delta\over \delta h_{ij}}\qquad
\pi_\Phi\to-i{\delta \over \delta\Phi} \eqno(4.1)
$$
one obtains the following equations for $ \Psi $. There is the
momentum constraint
$$
{\cal H}_i\Psi= 2i D_j {\delta \Psi \over \delta h_{ij} } + {\cal
H}^{matter}_i \Psi = 0 \eqno(4.2)
$$
and the Wheeler-DeWitt equation
$$
{\cal H}\Psi = \left[ - G_{ijkl} {\delta \over \delta h_{ij} }
{\delta \over \delta h_{kl} } -h^{\half} ({^3}R-2 \Lambda) +{\cal
H}^{matter} \right] \Psi = 0 \eqno(4.3)
$$
where we have ignored operator ordering problems.

The momentum constraint implies that the wave function is the same
for configurations $(h_{ij}(\x),\Phi(\x))$ that are related by
coordinate transformations in the three-surface. To see this, let
us restrict attention to the case of no matter, and consider the
effect of shifting the argument of the wave function by a
diffeomorphism in the three-surface, $x^i \rightarrow x^i - \xi^i
$. One has
$$
\Psi[h_{ij}+D_{(i}\xi_{j)}] = \Psi[h_{ij}] + \int d^3 {\bf x} \
D_{(i}\xi_{j)} {\delta \Psi \over \delta h_{ij} } \eqno(4.4)
$$
Integrating by parts in the last term, and dropping the boundary
term (since the three-manifold is compact), one finds that the
change in $ \Psi $ is given by
$$
\delta \Psi = - \int d^3 {\bf x} \ \xi_j D_i \left( {\delta \Psi
\over \delta h_{ij}} \right) = {1\over 2i} \int d^3 {\bf x} \
\xi_i {\cal H}^i \Psi \eqno(4.5)
$$
showing that wave functions satisfying (4.2) are unchanged. The
momentum constraint (4.2) is therefore the quantum mechanical
expression of the invariance of the theory under three-dimensional
diffeomorphisms.\footnote{$^*$} {This was first shown by Higgs
(1958).} Similarly, the Wheeler-DeWitt equation (4.3) is connected
with the reparametrization invariance of the theory. This is a lot
harder to show and we will not go into it here\footnote{$^{\dag}$}
{The difficulty is essentially due to the fact that although wave
functions $\Psi[h_{ij}]$ carry a representation of the
three-dimensional diffeomorphism group, they do not carry a
representation of the four-dimensional diffeomorphisms. A closely
related fact is that the Poisson bracket algebra of the
constraints is not that of the four-dimensional diffeomorphsims.
For a discussion of these issues and their resolution, see Isham
and Kucha{$\rm \check r$} (1985a, 1985b), Kucha{$\rm \check r$}
(1986).}.

The Wheeler-DeWitt equation is a second order hyperbolic
functional differential equation describing the dynamical
evolution of the wave function in superspace. The part of the
three-metric corresponding to the minus sign in the hyperbolic
signature, and so to the ``time" part, is the volume of the
three-metric, $ h^{\half} $. The Wheeler-DeWitt equation will in
general have a vast number of solutions, so in order to have any
predictive power we need boundary conditions to pick out just one
solution. This might involve, for example, giving the value of the
wave function at the boundary of superspace.

As an alternative to the canonical quantization procedure, one can
construct the wave function using a path integral. In the path
integral method, the wave function (or more precisely, some kind
of propagator) is represented by a Euclidean functional integral
over a certain class of four-metrics and matter fields, weighted
by $e^{-I}$, where $ I $ is the Euclidean action of the gravity
plus matter system. Formally, one writes
$$
\Psi [\tilde h_{ij},\tilde\Phi,B]= \sum_M \int {\cal D} g_{\mu\nu}
{\cal D} \Phi e^{-I}. \eqno(4.6)
$$
The sum is taken over some class of manifolds $M$ for which $B$ is
part of their boundary, and over some class of four-metrics $
g_{\mu\nu} $ and matter fields $\Phi$ which induce the
three-metric $\tilde h_{ij}$ and matter field configuration $
\tilde \Phi$ on the three-surface $B$ (see Fig.2.). The sum over
four-manifolds is actually very difficult to define in practice,
so one normally considers each admissable four-manifold
separately. The path integral permits one to construct far more
complicated amplitudes than the wave function for a single
three-surface (Hartle, 1990), but this is the simplest and most
frequently used amplitude, and it is the only one that will be
discussed here.

When the four-manifold has topology $\R \times B$, the path
integral has the explicit form
$$
\Psi[\tilde h_{ij}, \tilde \Phi,B]  = \int \D N^{\mu} \int \D
h_{ij} \D \Phi \ \delta [ \dot N^{\mu} - \chi^{\mu} ] \
\Delta_{\chi} \ \exp(-I[g_{\mu\nu}, \Phi] ) \eqno(4.7)
$$
Here, the delta-functional enforces the gauge-fixing condition
$\dot N^{\mu} = \chi^{\mu}$ and $\Delta_{\chi}$ is the associated
Faddeev-Popov determinant. The lapse and shift $N^{\mu}$ are
unrestricted at the end-points. The three-metric and matter field
are integrated over a class of paths $(h_{ij}(\x,\tau),
\Phi(\x,\tau))$ with the restriction that they match the argument
of the wave function on the three-surface $B$, which may be taken
to be the surface $\tau=1$. That is,
$$
h_{ij}(\x,1)= \tilde h_{ij}(\x), \quad \Phi(\x,1)=\tilde  \Phi(\x)
\eqno(4.8)
$$
To complete the specification of the class of paths one also needs
to specify the conditions satisfied at the initial point, $\tau =
0$ say.

The expression, ``Euclidean path integral" should be taken with a
very large grain of salt for the case of gravitational systems.
One needs to work rather hard to give the expression (4.6) a
sensible meaning. In particular, in addition to the usual issues
associated with defining a functional integral over fields, one
has to deal with the fact that the gravitational action is not
bounded from below. This means that the path integral will not
converge if one integrates over real Euclidean metrics.
Convergence is achieved only by integrating along a complex
contour in the space of complex four-metrics. The sum is therefore
over complex metrics and is not even equivalent to a sum over
Euclidean metrics in any sense. Furthermore, there is generally no
unique contour and the outcome of evaluating the path integral
could depend rather crucially on which complex contour one
chooses. We will have more to say about this later on.

As we have already noted, the Wheeler-DeWitt equation and momentum
constraints, (4.2), (4.3) are normally thought of as a quantum
expression of invariance under four-dimensional diffeomorphisms.
One ought to be able to see the analagous thing in the path
integral, and in fact one can. The wave functions generated by the
path integral (4.7) may formally be shown to satisfy the
Wheeler-DeWitt equation and momentum constraints, providing that
the path integral is constructed in an {\it invariant} manner.
This means that the action, measure, and class of paths summed
over should be invariant under diffeomorphisms (Halliwell and
Hartle, 1990).

Which solution to the Wheeler-DeWitt equation is generated by the
path integral will depend on how the initial conditions on the
paths summed over are chosen, and how the contour of integration
is chosen; thus the question of boundary conditions on the wave
function in canonical quantization appears in the path integral as
the question of choosing a contour and choosing a class of paths.
No precise relationship is known, however.

\subhead {\bf Interpretation}

To complete this discussion of the general formalism of quantum
cosmology, a few words on interpretation are in order. Hartle has
covered the basic ideas involved in interpreting the wave
function. Here, I am just going to tell you how I am going to
interpret the wave function without trying to justify it. The
basic idea is that we are going to regard a strong peak in the
wave function, or in a distribution constructed from the wave
function, as a prediction. If no such peaks may be found, then we
make no prediction. This will be sufficient for our purposes.
References to the vast literature on this subject are given in
Section 13.

\def\a{\alpha}
\def\b{\beta}
\def\pp{{\prime\prime}}
\def\R{{I\kern-0.3em R}}

\head{\bf 5. MINISUPERSPACE -- GENERAL THEORY}

Since superspace, the configuration space one deals with in
quantum cosmology, is infinite dimensional, the full formalism of
quantum cosmology is very difficult to deal with in practice. In
classical cosmology, because the universe appears to be
homogeneous and isotropic on very large scales, one's
considerations are largely restricted to the region of superspace
in the immediate vicinity of homogeneity and isotropy. That is,
one begins by studying homogeneous isotropic (or sometimes
anisotropic) metrics and then goes on to consider small
inhomogeneous perturbations about them. In quantum cosmology one
does the same. To be precise, one generally begins by considering
a class of models in which all but a finite number of degrees of
freedom of the metric and matter fields are ``frozen" or
``suspended". This is most commonly achieved by restricting the
fields to be homogeneous. Such models are known as
``minisuperspace" models and are characterized by the fact that
their configuration space, minisuperspace, is finite dimensional.
One is thus dealing with a problem of quantum mechanics, not of
field theory. A very large proportion of the work done in quantum
cosmology has concentrated on models of this type.

Clearly in the quantum theory there are considerable difficulties
associated with the restriction to minisuperspace. Setting most of
the field modes and their momenta to zero identically violates the
uncertainty principle. Moreover, the restriction to minisuperspace
is not known to be part of a systematic approximation to the full
theory. At the humblest level, one can think of minisuperspace
models not as some kind of approximation, but rather, as toy
models which retain certain aspects of the full theory, whilst
avoiding others, thereby allowing one to study certain features of
the full theory in isolation from the rest. However, in these
lectures we are interested in cosmological predictions. I am
therefore going to take the stronger point of view that these
models {\it do} have something to do with the full theory. In what
follows I will therefore try to emphasize what aspects of
minisuperspace models may be argued to transcend the restrictions
to minisuperspace. We will return to the question of the validity
of the minisuperspace ``approximation" later on.

The simple model of the previous section was of course a
minisuperspace model, in that we restricted the metric and matter
field to be homogeneous and isotropic. More generally,
minisuperspace usually involves the following: in the four-metric
(3.1), the lapse is taken to be homogeneous, $N=N(t)$, and the
shift is set to zero, $N^i=0$, so that one has
$$
ds^2 = -N^2(t) dt^2 + h_{ij}({\bf x},t) dx^i dx^j \eqno(5.1)
$$
Most importantly, the three-metric $h_{ij} $ is restricted to be
homogeneous, so that it is described by a finite number of
functions of $t$, $q^{\a}(t)$ say, where $\a=0,1,2 \cdots (n-1)$.
Some examples of possible ways in which the three-metric may be
restricted are given below.

One could take a Roberston-Walker metric as we did in Section 2,
$$
h_{ij}({\bf x},t) dx^i dx^j = a^2(t) d\Omega_3^2 \eqno(5.2)
$$
Here, $d\Omega_3^2$ is the metric on the three-sphere, and $q^{\a}
= a $. One could take an anisotropic metric with spatial sections
of topology $S^1 \times S^2 $,
$$
h_{ij}({\bf x},t) dx^i dx^j = a^2(t) dr^2 + b^2(t) d \Omega_2^2
\eqno(5.3)
$$
Here, $d\Omega_2^2$ is the metric on the two-sphere, $r$ is
periodically identified, and $q^{\a} = (a,b) $. More generally,
one could consider Bianchi-type metrics,
$$
h_{ij}({\bf x},t) dx^i dx^j = a^2(t) ( e^{\beta} )_{ij} \sigma^i
\sigma^j \eqno(5.4)
$$
Here, the $\sigma^i$ are a basis of one-forms and the $q^{\a}$
consist of the scale factor $a$ and the various components of the
matrix $\beta$, which describe the degree of anisotropy. Many more
models are cited in Section 13.

In terms of the variables describing the $(3+1)$ decomposition of
the four-metric, (3.1), the Einstein action with cosmological
constant (3.2) is
$$
S[h_{ij},N,N^i] = {m_p^2 \over 16 \pi} \int dt \ d^3x \ N
h^{\half} \left[ K_{ij} K^{ij} -K^2 + {^3 R} - 2 \Lambda \right]
\eqno(5.5)
$$
On inserting the restricted form of the metric described above one
generally obtains a result of the form
$$
S[q^{\a}(t), N(t)] = \int_0^1 dt N \left[ {1 \over 2N^2}
f_{\a\b}(q) \dot q^{\a} \dot q^{\b} - U(q) \right] \equiv \int L
dt \eqno(5.6)
$$
Here, $ f_{\a\b}(q) $ is the reduced version of the DeWitt metric,
(3.6), and has indefinite signature, $ (-,+,+,+...) $. The range
of the $t$ integration may be taken to be from $0$ to $1$ by
shifting $t$ and by scaling the lapse function. The inclusion of
matter variables, restricted in some way, also leads to an action
of this form, so that the $q^{\a}$ may include matter variables as
well as three-metric components. The $(-)$ part of the signature
in the metric always corresponds to a gravitational variable,
however.

Restricting to a metric of the form (5.1) is not the only way of
obtaining a minisuperspace model. Sometimes it will be convenient
to scale the lapse by functions of the three-metric.
Alternatively, one may wish to consider not homogeneous metrics,
but inhomogeneous metrics of a restricted type, such as
spherically symmetric metrics. Or, one may wish to use a
higher-derivative action in place of (5.5). In that case, the
action can always be reduced to first order form by the
introduction of extra variables ({\it e.g.} $Q= \ddot a$, etc.).
One way or another, one always obtains an action of the form
(5.6). We will therefore take this action to be the defining
feature of minisuperspace models. So from here onwards, our task
is to consider the quantization of systems described by an action
of the form (5.6).

The action (5.6) has the form of that for a relativistic point
particle moving in a curved space-time of $n$ dimensions with a
potential. Varying with respect to $q^{\a}$ one obtains the field
equations
$$
{1 \over N} {d \over dt} \left( {\dot q^{\a} \over N} \right) + {1
\over N^2} \Gamma^{\a}_{\b\gamma} \dot q^{\b} \dot q^{\gamma} +
f^{\a\b} {\partial U \over \partial q^{\b} } = 0 \eqno(5.7)
$$
where $\Gamma^{\a}_{\b\gamma} $ is the usual Christoffel
connection constructed from the metric $f_{\a\b}$. Varying with
respect to $N$ one obtains the constraint
$$
{1 \over 2 N^2} f_{\a\b} \dot q^{\a} \dot q^{\b} + U(q) = 0
\eqno(5.8)
$$
These equations describe geodesic motion in minisuperspace with a
forcing term.

It is important to note that the general solution to (5.7), (5.8)
will involve $(2n-1)$ arbitrary parameters.\footnote{$^{\dag}$}
{As in the model of Section 2, one of the parameters will be
$t_0$, the origin of unobservable parameter time, so effectively
one has $(2n-2)$ physically relevant parameters.}

For consistency, (5.7) and (5.8) ought to be equivalent,
respectively, to the $00$ and $ij$ components of the full Einstein
equations,
$$
R_{\mu\nu} - \half R g_{\mu\nu} + \Lambda g_{\mu\nu} = {8 \pi
\over m_p^2} T_{\mu\nu} \eqno(5.9)
$$
This is not, however, guaranteed. Inserting an ansatz for the
metric into the action and then taking variations to derive the
minisuperspace field equations does not necessarily yield the same
field equations as are obtained by inserting the minisuperspace
ansatz directly into (5.9). Our analysis is therefore restricted
to minisuperspace models for which these two paths to the field
equation give the same result. This excludes, for example, metrics
in Bianchi Class B, but does include a sufficent number of
interesting examples. When studying minisuperspace models, one
should always check that the acts of taking variations and
inserting an ansatz commute (and also that the $0j$ components of
(5.9) are trivially satisfied).

The Hamiltonian is found in the usual way. One first defines
canonical momenta
$$
p_{\a} = {\partial L \over \partial \dot q^{\a} } = f_{\a\b} {\dot
q^{\b} \over N} \eqno(5.10)
$$
and the canonical Hamiltonian is
$$
H_c = p_{\a} \dot q^{\a} - L = N \left[ \half f^{\a\b} p_{\a}
p_{\b} + U(q) \right] \equiv N H \eqno(5.11)
$$
where $f^{\a\b}(q)$ is the inverse metric on minisuperspace. The
Hamiltonian form of the action is
$$
S = \int_0^1 dt \left[ p_{\a} \dot q^{\a} - NH \right] \eqno(5.12)
$$
This indicates that the lapse function $N$ is a Lagrange
multiplier enforcing the Hamiltonian constraint
$$
H(q^{\a},p_{\a}) = \half f^{\a\b} p_{\a} p_{\b} + U(q) = 0
\eqno(5.13)
$$
This is equivalent to the Hamiltonian constraint of the full
theory (3.8), integrated over the spatial hypersurfaces. The
momentum constraint, (3.7), is usually satisfied identically by
the minisuperspace ansatz (modulo the above reservations).

\subhead {\bf Canonical Quantization}

Canonical quantization involves the introduction of a
time-independent wave function $ \Psi(q^{\a}) $ and demanding that
it is annihilated by the operator corresponding to the classical
constraint (5.13). This yields the Wheeler-DeWitt equation,
$$
\hat H(q^{\a}, -i {\partial \over \partial q^{\a} } ) \Psi(q^{\a})
= 0 \eqno(5.14)
$$
Because the metric $f^{\a\b}$ depends on $q$ there is a
non-trivial operator ordering issue in (5.14). This may be
partially resolved by demanding that the quantization procedure is
covariant in minisuperspace; {\it i.e.} that is is unaffected by
field redefinitions of the three-metric and matter fields, $q^{\a}
\to \tilde q^{\a}(q^{\a})$. This narrows down the possible
operator orderings to
$$
\hat H = - \half \nabla^2 + \xi \R + U(q) \eqno(5.15)
$$
where $ \nabla^2 $ and $ \R $ are the Laplacian and curvature of
the minisuperspace metric $f_{\a\b}$ and $\xi$ is an arbitrary
constant.

The constant $\xi$ may be fixed once one recognises that the
minisuperspace metric (and indeed, the full superspace metric
(3.9)) is not uniquely defined by the form of the action or the
Hamiltonian, but is fixed only up to a conformal factor.
Classically the constraint (5.13) may be multiplied by an
arbitrary function of $q$, $\Omega^{-2}(q)$ say, and the
constraint is identical in form but has metric $\tilde f_{\a\b} =
\Omega^2 f_{\a\b} $ and potential $\tilde U = \Omega^{-2} U $. The
same is true in the action (5.6) or (5.12) if, in addition to the
above rescalings, one also rescales the laspe function, $ N \to
\tilde N = \Omega^{-2} N $. Clearly the quantum theory should also
be insensitive to such rescalings. This is achieved if the metric
dependent part of the operator (5.15) is {\it conformally
covariant}; {\it i.e.} if the coefficient $\xi$ is taken to be the
conformal coupling
$$
\xi = - {(n-2) \over 8 (n-1) } \eqno(5.16)
$$
for $n \ge 2 $ (Halliwell, 1988a; Moss, 1988; Misner, 1970). In
what follows, we will be working almost exclusively in the lowest
order semi-classical approximation, for which these issues of
operator ordering are in fact irrelevant. However, I have
mentioned this partially for completeness, but also because one
often studies models in which considerable simplifications arise
by suitable lapse function rescalings and field redefinitions, and
one might wonder whether or not these changes of variables affect
the final results.

\subhead {\bf Path Integral Quantization}

The wave function may also be obtained using a path integral. To
discuss the path integral, we first need to discuss the symmetry
of the action. The Hamiltonian constraint, (5.13), indicates the
presence of a symmetry, namely reparametrization invariance. This
is the left-over of the general covariance of the full theory,
after the restriction to minisuperspace. More precisely, under the
transformations
$$
\delta_{\epsilon} q^{\a} = \epsilon(t) \{ q^{\a}, H\}, \quad
\delta_{\epsilon} p_{\a} = \epsilon(t) \{ p_{\a}, H\}, \quad
\delta_{\epsilon} N = \dot \epsilon(t) \eqno(5.17)
$$
it is not difficult to show that the action changes by an amount
$$
\delta S = \left[ \epsilon(t) \left( p_{\a} {\partial H \over
\partial p_{\a} }-H \right) \right]_0^1 \eqno(5.18)
$$
The action is therefore unchanged if the parameter $\epsilon(t)$
satisfies the boundary conditions $\epsilon (0)=0=\epsilon(1)$.
This symmetry may be completey broken by imposing a gauge-fixing
condition of the form
$$
G \equiv \dot N - \chi (p_{\a}, q^{\a}, N) = 0 \eqno(5.19)
$$
where $ \chi $ is an arbitrary function of $p_{\a},q^{\a},N$.

We may now write down the path integral. It has the form
$$
\Psi ({q^{\a}}^{\pp}) = \int {\cal D} p_{\a} {\cal D} q^{\a} {\cal
D} N \ \ \delta [ G] \ \Delta_G \ e^{iS[p,q,N]} \eqno(5.20)
$$
where $S[p,q,N]$ is the Hamiltonian form of the action (5.12) and
$\Delta_G$ is the Faddeev-Popov measure associated with the
gauge-fixing condition (5.19), and guarantees that the path
integral is independent of the choice of gauge-fixing function
$G$. The integral is taken over a set of paths $(q^{\a}(t),
p_{\a}(t), N(t))$ satisfying the boundary condition $ q^{\a}(1) =
{q^{\a}}^{\pp} $ at $t=1$ with $p_{\a}$ and $N$ free, and some yet
to be specified conditions at $t=0$.

The only really practical gauge to work in is the gauge $ \dot N =
0 $. Then it may be shown that $\Delta_G =
constant$.\footnote{$^{\dag}$} {This is easily seen: $\Delta_G$ is
basically the determinant of the operator $\delta_{\epsilon} G
/\delta \epsilon$. In the gauge $\dot N=0$, this is the operator
$d^2/dt^2$, which has constant determinant.} The functional
integral over $N$ then reduces to a single ordinary integration
over the constant $N$. One thus has
$$
\Psi ({q^{\a}}^{\pp}) = \int dN \int {\cal D} p_{\a} {\cal D}
q^{\a} \ e^{iS[p,q,N]} \eqno(5.21)
$$
Eq.(5.21) has a familiar form: it is the integral over all times
$N$ of an ordinary quantum mechanical propagator, or wave
function,
$$
\Psi ({q^{\a}}^{\pp}) = \int dN \psi ({q^{\a}}^{\pp}, N)
\eqno(5.22)
$$
where $ \psi ({q^{\a}}^{\pp}, N) $ satisfies the time-dependent
Schr\"odinger equation with time coordinate $N$. From Eq.(5.22),
it is readily shown that the wave function generated by the path
integral satisfies the Wheeler-DeWitt equation. Suppose we operate
on (5.22) with the Wheeler-DeWitt operator at ${q^{\a}}^{\pp}$.
Then, using the fact that the integrand satisfies the
Schr\"odinger equation, one has
$$
\hat H^{\pp} \Psi({q^{\a}}^{\pp}) = \int dN \ i {\partial \psi
\over \partial N} = i \left[ \psi ({q^{\a}}^{\pp}, N)
\right]_{N_1}^{N_2} \eqno(5.23)
$$
where $N_1,N_2$ are the end-points of the $N$ integral, about
which we have so far said nothing. Clearly for the wave function
to satisfy the Wheeler-DeWitt equation we have to {\it choose} the
end-points so that the right-hand side of (5.23) vanishes. $N$ is
generally integrated along a contour in the complex plane. This
contour is usually taken to be infinite, with $ \psi
({q^{\a}}^{\pp},N) $ going to zero at the ends, or closed {\it
i.e.} $N_1=N_2$. In both of these cases, the right-hand side of
(5.23) vanishes and the wave function so generated satisfies the
Wheeler-DeWitt equation. (In the closed contour case, attention to
branch cuts may be needed.) Note that these ranges are invariant
under reparametrizations of $N$. They would not be if the contour
had finite end-points and the right-hand side would then not be
zero. This is an illustration of the remarks in Section 4
concering the relationship between the Wheeler-DeWitt equation and
the invariance properties of the path integral.

The representation (5.21) of the wave function is of considerable
practical value in that it can actually be used to evaluate the
wave function directly. But first, one normally rotates to
Euclidean time, $\tau = it$. After integrating out the momenta,
the resulting Euclidean functional integral has the form
$$
\Psi( {q^{\a}}^{\pp}) = \int dN \int {\cal D} q^{\a} \exp \left(
-I[q^{\a}(\tau),N] \right) \eqno(5.24)
$$
Here, $I$ is the minisuperspace Euclidean action
$$
I[q^{\a}(\tau), N] = \int_0^1 d\tau N \left[ {1 \over 2N^2}
f_{\a\b}(q) \dot q^{\a} \dot q^{\b} + U(q) \right] \eqno(5.25)
$$
Although the part of this action which corresponds to the matter
modes is always positive definite, the gravitational part is not.
Recall that the minisuperspace metric has indefinite signature,
the $(-)$ part corresponding to the conformal part of the
three-metric, so the kinetic term is indefinite. Also, the
potential, which is the integral of $ 2\Lambda - {^3 R} $, is not
positive definite. So complex integration contours are necessary
to give meaning to (5.24).

Here, however, we will work largely in the lowest order
semi-classical approximation, which involves taking the wave
function to be (a sum of terms) of the form $e^{-I_{cl}}$, where
$I_{cl}$ is the action of the classical solution
$(q^{\a}(\tau),N)$ satisfying the prescribed boundary conditions.
This solution may in fact be complex, and indeed will need to be
if the wave function is to be oscillatory. Similarly, when working
with the Wheeler-DeWitt equation, we will work largely in the WKB
approximation, in which solutions of the above type are sought.

We are now in a position to comment on the validity of the
minisuperspace ``approximation". Providing we are sufficiently
careful in making our minisuperspace ansatz, the classical
solutions $(q^{\a}(\tau),N)$ will be solutions to the full field
equations, and thus $I_{cl}$ will be the action of a solution to
the full Einstein equations. The lowest order semi-classical
approximation to the minisuperspace wave function therefore
coincides with the lowest order semi-classical approximation to
the wave function of the full theory. This means that
minisuperspace does give some indication as to what is going on in
the full theory as long as we remain close to the lowest order
semi-classical approximation.

\subhead {\bf The Probability Measure}

Given a wave function $\Psi(q^{\a})$ for a minisuperspace model we
need to construct from it a probability measure with which to make
predictions. The question is, which probability measure do we use?
The Wheeler-DeWitt equation is a Klein-Gordon type equation. It
therefore has associated with it a conserved current
$$
J = {i \over 2 } \left( \Psi^* \nabla \Psi - \Psi \nabla \Psi^*
\right) \eqno(5.26)
$$
It satisfies
$$
\nabla \cdot J = 0 \eqno(5.27)
$$
by virtue of the Wheeler-DeWitt equation. Like the Klein-Gordon
equation, however, the probability measure constructed from the
conserved current can suffer from difficulties with negative
probabilities. For this reason, some authors have suggested that
the correct measure to use is
$$
dP = |\Psi(q^{\a})|^2 dV \eqno(5.28)
$$
where $dV$ is a volume element of minisuperspace. However, this is
also problematic, in that one of the coordinates $q^{\a}$ is, in
some crude sense, ``time", so that (5.28) is the analogue of
interpreting $|\Psi(x,t)|^2$ in ordinary quantum mechanics as the
probability of finding the particle in the space-{\it time}
interval $dxdt$. One could conceivably make sense out of (5.28),
but not before a careful discussion of the nature of time in
ordinary quantum mechanics.\footnote{$^{\dag}$} {This line of
thought has been pursued by numerous authors, including Caves
(1986, 1987), Hartle (1988b) and Page(1989b).}

For the moment we will not commit ourselves to either of these
possibilities, but will keep each one in mind. We will just look
for peaks in the wave function itself when asking for predictions.
If the peak is sufficiently strong, one would expect any sensible
measure constructed from the wave function to have the same peak.

\def\a{{\alpha}}
\def\b{{\beta}}

\head{\bf 6. CLASSICAL SPACETIME}

We have described in the previous section two ways of calculating
the wave function for minisuperspace models: the Wheeler-DeWitt
equation and the path integral. Before going on to the evaluation
of the wave function, it is appropriate to ask what sort of wave
functions we are hoping to find. If the wave function is to
correctly describe the late universe, then it must predict that
spacetime is classical when the universe is large. The first
question to ask, therefore, is ``What, in the context of quantum
cosmology, constitutes a prediction of classical spacetime?".

There are at least two requirements that must be satisfied before
a quantum system may be regarded as classical:

\item {1.} The wave function must predict that the canonical
variables are strongly correlated according to classical laws;
{\it i.e.} the wave function (or some distribution constructed
from it) must be strongly peaked about one or more classical
configurations

\item {2.} The quantum mechanical interference between distinct
such configurations should be negligible; {\it i.e.} they should
{\it decohere}.

\noindent To exemplify both of these requirements, let us first
consider a simple example from ordinary quantum mechanics. There,
the most familiar wave functions for which the first requirement
is satisfied are {\it coherent states}. These are single wave
packets strongly peaked about a single classical trajectory, $
\bar x(t) $, say. For example, for the simple harmonic oscillator,
the coherent states are of the form
$$
\psi(x,t) = e^{ipx} \exp \left( - {(x-\bar x(t))^2 \over \sigma^2
}\right) \eqno(6.1)
$$
On being presented with a solution to the Schrodinger equation of
this type, one might be tempted to say that it predicts classical
behaviour, in that on measuring the position of the particle at a
sequence of times, one would find it to be following the
trajectory $\bar x(t)$. Suppose, however, one is presented with a
solution to the Schr\"odinger equation which is a superposition of
many such states:
$$
\psi(x,t) = \sum_n c_n e^{ip_n x} \exp \left( - {(x-\bar x_n(t))^2
\over \sigma^2 }\right) \eqno(6.2)
$$
where the $\bar x_n(t) $ are a set of distinct classical
solutions. One might be tempted to say that this wave function
corresponds to classical behaviour, and that one would find the
particle to be following the classical trajectory $\bar x_n(t)$
with probability $|c_n|^2$. The problem, however, is that these
wave packets may meet up at some stage in the future and
interfere. One could not then say that the particle was following
a definite classical trajectory. To ascribe a definite classical
history to the particle, the interference between distinct states
has to be destroyed. The way in which this may be achieved is a
fascinating subject in itself, but we will say little about it
here. We will concentrate mainly on the first requirement for
classical behaviour.

Turn now to quantum cosmology. One might at first think that, in
the search for the emergence of classical behaviour, the natural
thing to do there is to try and construct the analogue of coherent
states. This is rather hard to do, but has been achieved for
certain simple models. Because the wave function does not depend
on time explicitly, the analogue of coherent states are wave
functions of the form
$$
\Psi(q^{\a}) = e^{i \phi(q^{\a}) } \exp\left( - f^2 (q^{\a})
\right) \eqno(6.3)
$$
where $f(q^{\a})=0$ is the equation of a single classical
trajectory in minisuperspace. So in a two-dimensional model, for
example, the wave function will consist of a sharply peaked ridge
in minisuperspace along a single classical trajectory.

Wave functions of this type do not arise very naturally in quantum
cosmology because they need very special boundary conditions.
However, they do highlight a particular feature typical of wave
functions in quantum cosmology that predict classical spacetime:
they are peaked about an entire history. Moreover, although the
original wave function does not carry a particular label playing
the role of time, a notion of time may emerge for certain types of
wave functions, such as (6.3): it is basically the affine
parameter along the histories about which the wave function is
peaked ({\it i.e.} the distance along the ridge in the case of
(6.3)). So time, and indeed spacetime, are only derived concepts
appropriate to certain regions of configuration spacetime and
contingent upon initial conditions. The decoherence requirement
may also be achieved in quantum cosmology, but this is rather
complicated and will not be covered here.

As we have seen in the simple model of Section 2, the sort of wave
functions most commonly arising in quantum cosmology are not of
wavepacket form, but are of WKB form, and may be broadly
classified as oscillatory, of the form $e^{iS}$, or exponential,
of the form $e^{-I}$. It is the oscillatory wave functions that
correspond to classical spacetime, whilst the exponential ones do
not. Let us discuss why this is so.

Recall that the way we are interpreting the wave function is to
regard a strong peak in the wave function, or in a distribution
constructed from it, as a prediction. Classical spacetime,
therefore, is predicted when the wave function, or some
distribution constructed from it, becomes strongly peaked about
one or more classical configurations. How do we identify such
peaks? In the most general case, the wave function will be peaked
not about some region of configuration space -- $ e^{iS} $ is most
certainly not -- but about some correlation between coordinates
and momenta. Perhaps the most transparent way of identifying such
correlations is to introduce a quantum mechanical distribution
function which depends on both coordinates and momenta, $F(p,q)$.
The Wigner function is such a distribution function, and turns out
to be very useful in quantum cosmology for identifying the
correlations present in a given wave function. However, this would
take rather a long time to explain. Here I will just report the
result that the Wigner function shows that (i) a wave function of
the form $e^{-I}$ predicts no correlation between coordinates and
momenta, and so cannot correspond to classical behaviour; and (ii)
a wave function of the form $e^{iS}$ predicts a strong correlation
between $p$ and $q$ of the form
$$
p_{\a}={\partial S \over \partial q^{\a}} \eqno(6.4)
$$
$S$ is generally a solution to the Hamilton-Jacobi equation and,
as we will demonstrate in detail below, (6.4) is then a first
integral of the equations of motion. It thus defines a set of
solutions to the field equations. A wave function of the form
$e^{iS}$, therefore, is normally thought of a being peaked about
not a single classical solution, but about a {\it set} of
solutions to the field equations. It is in this sense that it
corresponds to classical spacetime.

Given the peak about the correlation (6.4) for wave functions of
the form $e^{iS}$, it may now be explicitly verified using a
canonical transformation. For simplicity consider the
one-dimensional case. A canonical transformation from $(p,q)$ to
$(\tilde p, \tilde q)$ may be generated by a generating function
$G_0(q,\tilde p)$:
$$
p = {\partial G_0 \over \partial q}, \quad \tilde q = {\partial
G_0 \over \partial \tilde p} \eqno(6.5)
$$
In quantum mechanics, the transformation from the wave function
$\Psi(q)$ to a new wave function $\tilde \Psi (\tilde p) $ is
given by
$$
\tilde \Psi (\tilde p) = \int dq e^{-iG(q,\tilde p)} \Psi(q)
\eqno(6.6)
$$
Here, the generating function $G(q,\tilde p)$ is not actually
quite the same as $G_0(q,\tilde p)$ above, but agrees with it to
leading order in Planck's constant. Suppose $\Psi(q) = e^{iS(q)}
$. Then a transformation to new variables
$$
\tilde p = p - {\partial S \over \partial q}, \quad \tilde q = q
\eqno(6.7)
$$
may be achieved using the generating function $G_0(q, \tilde p) =
q \tilde p + S(q) $. Inserting this in (6.6), it is easily seen
that the wave function as a function of $\tilde p $ is of the form
$$
\tilde \Psi(\tilde p ) = \delta ( \tilde p ) \eqno(6.8)
$$
to leading order. As advertised, it is therefore strongly peaked
about the configuration (6.4).

It is sometimes stated that wave functions of the form $e^{-I}$
are not classical because they correpond to a Euclidean spacetime.
It is certainly true that they are not classical, and it is
certainly true that, if the wave function is a WKB solution, then
$I$ is the action of a classical Euclidean solution. However, this
does not mean that they correspond to a Euclidean spacetime. In
contrast to a wave function of the form $e^{iS}$, which is {\it
peaked} about a set of classical Lorentzian solutions, a wave
function $e^{-I}$ is {\it not peaked} about a set of Euclidean
solutions. It is not classical quite simply because it fails to
predict classical correlations between the Lorentzian momentum $p$
and its conjugate $q$.

A much better way of discussing peaks in the wave function, or
more generally, of discussing predictions arising from a given
theory of initial conditions, is to use the path integral methods
described by Hartle in his lectures (Hartle, 1990). Although
conceptually much more satisfactory, they are somewhat cumbersome
to use in practice. Moreover, they have not as yet been applied to
any simple examples in quantum cosmology. For the moment it is
therefore not inappropriate to employ the rather heuristic but
quicker methods outlined above.

\subhead{\bf The General Behaviour of the Solutions}

Having argued that classical spacetime is predicted, loosely
speaking, when the wave function is oscillatory, our next task is
to determine the regions of configuration space for which the wave
function is oscillatory, and those for which the wave function is
exponential. This will depend to some extent on boundary
conditions, which we have not yet discussed, but one can get broad
indications about the behaviour of the wave function by looking at
the potential in the Wheeler-DeWitt equation So we are considering
the Wheeler-DeWitt equation
$$
\left[ - \half \nabla^2 + U(q) \right]\Psi (q) = 0 \eqno(6.9)
$$
Here, we have assumed that the curvature term has been absorbed
into the potential. Compare (6.9) with the one-dimensional quantum
mechanical problem
$$
\left[ {d^2 \over dx^2} + U(x) \right] \Psi(x) = 0 \eqno(6.10)
$$
In this case, one immediately sees that the wave function is
exponential in the region $U<<0$ and oscillatory in the region
$U>>0$. The case of (6.9) is more complicated, however, in that
there are $n$ independent variables, and the metric has indefinite
signature.

To investigate this in a little more detail, let us divide the
minisuperspace coordinates $q^{\a}$ into a single ``timelike"
coordinate $q^0$ and $n-1$ ``spacelike" coordinates ${\bf q}$.
Then locally, the Wheeler-DeWitt equation will have the form
$$
\left[ {\partial^2 \over \partial {q^0}^2 } - { \partial^2 \over
\partial {\bf q}^2 } + U(q^0, {\bf q}) \right] \Psi(q) = 0.
\eqno(6.11)
$$
The point now, is that the broad behaviour of the solution will
depend not only on the sign of $U$, but also, loosely speaking, on
whether it is the $q^0$-dependence of $U$ or the ${\bf
q}$-dependence of $U$ that is most significant. More precisely,
one has the following. Consider the surfaces of constant $U$ in
minisuperspace. They may be timelike or spacelike in a given
region. First of all suppose that they are spacelike. Then in that
local region, one can always perform a ``Lorentz" rotation to new
coordinates such that $U$ depends only on the timelike coordinate
in that region, $U \approx U(q^0)$. One can then solve
approximately by separation of variables and, assuming one can go
sufficiently far into the regions $U>0$, $U<0$ for the potential
to dominate the separation constant, the solution will be
oscillatory for $U>>0$, exponential for $U<<0$. Similarly, in
regions where the constant $U$ surfaces are timelike, one may
Lorentz-rotate to coordinates for which the potential depends only
on the spacelike coordinates. The wave function is then
oscillatory in the region $U<<0$ and exponential in the region
$U>>0$.

The above is only a rather crude way of getting an idea of the
behaviour of the solutions. In particular, the assumptions about
the separation constant need to be checked in particular cases,
given the boundary conditions.

One may also determine the broad behaviour of the wave function by
studying the path integral. In the Euclidean path integral
representation of the wave function (5.24), one considers the
propagation amplitude to a final configuration determined by the
argument of the wave function, from an initial configuration
determined by the boundary conditions. In the saddle-point
approximation, the wave function is of the form $e^{-I_{cl}}$,
where $I_{cl}$ is the Euclidean action of the classical solution
satisfying the above boundary conditions. Finding $I_{cl}$
therefore involves the mathematical question of solving the
Einstein equations as a boundary value problem. If the solution is
real, it will have real action, and the wave function will be
exponential. However, it appears to be most commonly the case for
generic boundary data that no real Euclidean solution exists, and
the only solutions are complex, with complex action. The wave
function will then be oscillatory. The boundary value problem for
the Einstein equations is actually a rather difficult mathematical
problem about which very little appears to be known, in the
general case.

In the minisuperspace case, qualitative information about the
nature of the solution to the boundary value problem is readily
obtained by inspecting the Euclidean version of the constraint
equation (5.8). So for example, when looking for a solution
between fixed values of $q^{\a}$ that are reasonably close
together, one can see that the nature of the solution depends not
only on the sign of the potential, but also on whether the
connecting trajectory is timelike or spacelike in minisuperspace.

The saddle-point appoximation to the path integral perhaps gives a
more reliable indication than the Wheeler-DeWitt equation as to
the broad behaviour of the wave function, in that the dependence
on boundary conditions is more apparent.

At this stage it is appropriate to emphasize an important
distinction between the above discussion and tunneling processes
in ordinary quantum mechanics or field theory. In ordinary quantum
mechanics or field theory, when considering tunneling at fixed
energy, one has a constraint equation similar to (5.8), but with
the important difference that its metric is {\it positive
definite}. This has the consequence that at fixed energy, the
configuration space is divided up into classically allowed and
classically forbidden regions, and one can see immediately where
they are by inspection of the potential in the constraint.

By contrast, for gravitational systems, the constraint (5.8) (or
more generally, the Hamiltonian constraint (3.8)) has a metric of
{\it indefinite} signature. This has the consequence that
configuration space {\it is not} divided up into classically
allowed and classically forbidden regions -- the constraint alone
does not rule out the existence of real Euclidean or real
Lorentzian solutions in a given region of configuration space. One
can only determine the nature of the solution ({\it i.e.} real
Euclidean, real Lorentzian or complex) by solving the boundary
value problem.

Further discussion of complex solutions and related issues may be
found in Gibbons and Hartle (1989), Halliwell and Hartle (1989)
and Halliwell and Louko (1989a, 1989b, 1990).

\head{\bf 7. THE WKB APPROXIMATION}

Having considered the general behaviour of the solutions to the
Wheeler-DeWitt equation, we now go on to find the solutions more
explicitly in the oscillatory region, using the WKB approximation.
This will allow us to be more explicit in showing that, as we have
already hinted a few times, the correlation (6.4) about which the
wave function is peaked in the oscillatory region defines a set of
solutions to the classical field equations.

We are interested in solving the Wheeler-DeWitt equation,
$$
\left[ - {1 \over 2m_p^2} \nabla^2 + m_p^2 U(q) \right] \Psi(q) =
0 \eqno(7.1)
$$
For convenience, the Planck mass $m_p$ has been reinstated,
because we are going to use it as a large parameter in terms of
which to do the WKB expansion. (If there is a cosmological
constant in the problem one can sometimes use $\Lambda m_p^{-4}$
as a small parameter to control the WKB expansion, which has the
advantage of being dimensionless.) Normally in the WKB
approximation one looks for solutions that are strictly
exponential or oscillatory, of the form $e^{-I}$ or $e^{iS}$.
However, in quantum cosmology one often uses the Wheeler-DeWitt
equation hand-in-hand with the path integral. As noted above, in
the saddle-point approximation to the path integral, one generally
finds that the dominating saddle-points are four-metrics that are
not real Euclidean, or real Lorentzian, but complex, with complex
action. It is therefore most appropriate to look for WKB solutions
to (7.1) of the form
$$
\Psi(q) = C(q) e^{-m_p^2 I(q)} + O(m_p^{-2}) \eqno(7.2)
$$
where $I$ and $C$ are complex. Inserting (7.2) into (7.1) and
equating powers of $m_p$, one obtains
$$
- \half (\nabla I)^2 + U(q) =0 \eqno(7.3)
$$
$$
2 \nabla I \cdot \nabla C + C \nabla^2 I = 0 \eqno(7.4)
$$
Here, $\nabla$ denotes the covariant derivative with respect to
$q^{\a}$ in the metric $f_{\a\b}$, and the dot product is with
respect to this metric. Let us split $I$ into real and imaginary
parts, $I(q)=I_R(q)-iS(q)$. Then the real and imaginary parts of
(7.3) are
$$
- \half (\nabla I_R)^2 + \half (\nabla S)^2 + U(q) =0 \eqno(7.5)
$$
$$
\nabla I_R \cdot \nabla S = 0 \eqno(7.6)
$$

Consider (7.5). We will return later to (7.4) and (7.6). We are
interested in wave functions which correspond to classical
spacetime. As we have discussed, to correspond to classical
spacetime, the wave function should be of the form $e^{iS}$ where
$S$ is a solution to the Lorentzian Hamilton-Jacobi equation,
$$
\half (\nabla S)^2 + U(q) = 0 \eqno(7.7)
$$
for then $S$ defines an ensemble of classical trajectories, as
will be shown in detail below. Evidently from (7.5) this is
generally not the case for the $S$ appearing in the wave function
(7.2). However, if the imaginary part of $I$ varies with $q$ much
more rapidly than the real part, {\it i.e.} if
$$
|\nabla S|>>|\nabla I_R| \eqno(7.8)
$$
then it follows from (7.5) that $S$ will be an {\it approximate}
solution to the Lorentzian Hamilton-Jacobi equation, (7.7).
Furthermore, the wave function (7.2) will then be predominantly of
the form $e^{iS}$ and, as we have already argued, it therefore
indicates a strong correlation between coordinates and momenta of
the form
$$
p_{\a} = m_p^2 {\partial S \over \partial q^{\a} } \eqno(7.9)
$$

Now we are in a position to show explicitly that (7.9) defines a
first integral to the field equations. Clearly the momenta
$p_{\a}$ defined by (7.9) satisfy the constraint (5.13), by virtue
of the Hamilton-Jacobi equation, (7.7). To obtain the second order
field equation, differentiate (7.7) with respect to $q^{\gamma}$.
One obtains
$$
\half f^{\a\b}_{,\gamma} {\partial S \over \partial q^{\a} }
{\partial S \over \partial q^{\b} } + f^{\a\b} {\partial S \over
\partial q^{\a} } {\partial^2 S \over \partial q^{\b} \partial
q^{\gamma} } + {\partial U \over \partial q^{\gamma} } =0
\eqno(7.10)
$$
The form of the second term in (7.10) invites the introduction of
a vector
$$
{d \over ds} = f^{\a\b} {\partial S \over \partial q^{\a} }
{\partial  \over \partial q^{\b} } \eqno(7.11)
$$
When operated on $q^{\gamma}$ it implies, via (7.9), the usual
relationship between velocities and momenta, (5.10), provided that
$s$ is identified with the proper time, $ds=Ndt$. Using (7.11) and
(7.9), (7.10) may now be written
$$
{d p_{\gamma} \over ds} + {1 \over 2m_p^2} f^{\a\b}_{,\gamma}
p_{\a} p_{\b} + m_p^2 {\partial U \over \partial q^{\gamma}} = 0
\eqno(7.12)
$$
The field equation (5.7) is obtained after use of (5.10) and after
raising the indices using the minisuperspace metric. We have
therefore shown that the wave function (7.12), if it satisfies the
condition (7.8), is strongly peaked about a {\it set} of solutions
to the field equations, namely the set defined by the first
integral (7.9).

Now we come to the most important point. For a given
Hamilton-Jacobi function $S$, the solution to the first integral
(7.9) will involve $n$ arbitrary parameters. Recall, however, that
the general solution to the full field equations (5.7), (5.8) will
involve $(2n-1)$ arbitrary parameters. The wave function is
therefore strongly peaked about an $n$-parameter subset of the
$(2n-1)$-parameter general solution. By imposing boundary
conditions on the Wheeler-DeWitt equation a particular wave
function is singled out. In the oscillatory region, this picks out
a particular Hamilton-Jacobi function $S$. This in turn defines
defines an $n$-parameter subset of the $(2n-1)$ parameter general
solution. It is in this way that boundary conditions on the wave
function of the universe effectively imply initial conditions on
the classical solutions.

\subhead {\bf The Measure on the Set of Classical Trajectories}

Suppose one now chooses an $(n-1)$-dimensional surface in
minisuperspace as the beginning of classical evolution. Through
(7.9), the wave function then effectively fixes the initial
velocities on that surface. However, the wave function contains
yet more information than just the initial velocities: it provides
a probability measure on the set of classical trajectories about
which the wave function is peaked. To see how this comes about,
consider the remaining parts of the wave function, $C$ and $I_R$.
From the assumption, (7.8), (7.4) may be written
$$
\nabla \cdot \left( |C|^2 \nabla S \right) = 0 \eqno(7.13)
$$
Moreover, we can combine this with (7.6) and write
$$
\nabla \cdot \left( \exp(-2m_p^2 I_R) |C|^2 \nabla S \right) = 0
\eqno(7.14)
$$
This is a current conservation law,
$$
\nabla \cdot J = 0 \eqno(7.15)
$$
where
$$
J \equiv \exp(-2m_p^2 I_R) |C|^2 \nabla S \eqno(7.16)
$$
Loosely speaking, (7.15) implies that that the coefficient of
$\nabla S$ in (7.16) provides a conserved measure on the set of
classical trajectories about which the WKB wave function is
peaked.

Eq.(7.16) is of course a special case of the Wheeler-DeWitt
current
$$
J = {i \over 2} \left( \Psi^* \nabla \Psi - \Psi \nabla \Psi^*
\right) \eqno(7.17)
$$
which is conserved by virtue of the Wheeler-DeWitt equation (7.1),
independently of any approximation. In suggesting that part of
(7.16) provides a conserved probability measure on the set of
classical trajectories, we are therefore aiming at using the
conserved current as our probability measure. So now it is time to
be precise about how the conserved current may successfully used,
avoiding the difficulties with negative probabilities. The point
is that it may be made to work only in the WKB approximation,
although this is probably sufficient for all practical purposes.
The following is based primarily on Vilenkin (1989) (see also
Hawking and Page (1986) and Misner (1970, 1972)).

First we show how to construct a conserved measure. Consider a
pencil $B$ of the congruence of classical trajectories with
tangent (co)vector $ \nabla S$, about which the wave function is
peaked. Suppose it intersects an $(n-1)$-dimensional surface
$\Sigma_1$ in $B \cap \Sigma_1 $, and subsequently intersects a
second surface $\Sigma_2$, in $B \cap \Sigma_2$. Now consider the
volume $V$ of minisuperspace swept out by the pencil of classical
trajectories between the surfaces $\Sigma_1$ and $\Sigma_2$.
Because $\nabla \cdot J = 0$, one may write
$$
0 = \int_V d V \nabla \cdot J  = \int_{\partial V} J \cdot dA
\eqno(7.18)
$$
where $d A$ is the element of area normal to the boundary of $V$.
Since $J \cdot d A $ is non-zero only on the parts of the boundary
of $V$ consisting of the ``ends", where the pencil $B$ intersects
$\Sigma_1$ and $\Sigma_2$, it follows that
$$
\int_{B \cap \Sigma_1} J \cdot dA = \int_{B \cap \Sigma_2} J \cdot
dA \eqno(7.19)
$$
This means that the flux of the pencil of trajectories across a
hypersurface is in fact independent of the hypersurface. It
suggests that we may use the quantity
$$
dP = J \cdot  d \Sigma \eqno(7.20)
$$
as a conserved probability measure on the set of classical
trajectories with tangent vector $\nabla S$, where $\Sigma$ is
some hypersurface that cuts across the flow.

Now we need to consider whether or not this definition of the
probability measure gives positive probabilities. An intimately
related problem is the choice of the surface $\Sigma$ in (7.20).

In the case of the Klein-Gordon equation, one takes the surfaces
$\Sigma$ to be surfaces of constant physical time, $X^0 =
constant$, and thus attempts to use $J_0$, the time-like component
of the current as a probability density. As is well-known,
however, this may be negative. This is very significant in
relativistic quantum mechanics, because it opens the way to the
notion of antiparticles and second quantization.

The analagous thing to do in quantum cosmology would be to take
the surfaces $\Sigma$ to be surfaces of constant $q^0$, the
timelike coordinate on minisuperspace. These would be surfaces for
which the conformal part of the three-metric is constant. Once
again one would find that the timelike component of $J$ may be
negative. However, this does not have the same significance as the
Klein-Gordon case. It corresponds to the fact that in the
classical theory one may have both expanding and collapsing
universes. It is merely due to a bad choice of surfaces $\Sigma$,
and does not oblige one to go to third quantization (the analogue
of second quantization). For classical solutions which expand and
recollapse, the flow will intersect a surface of constant
conformal factor twice, in a different direction each time. What
one really needs is a set of surfaces $\Sigma$ which the flow
intersects once and only once. For then the flux will pass through
these surfaces always in the same direction and the probability
(7.20) will be positive.

For a typical system in minisuperspace, the trajectories will
generally go backwards and forwards in all coordinates $q^{\a}$.
However, merely by inspection of a typical congruence of classical
paths, it is easy to see that one can always find a set of
surfaces which the trajectories cross only once and in the same
direction. The probability measure (7.20) will then remain of the
same sign all along the congruence of classical trajectories (see
Fig.3.).

One particularly simple choice that one might at first think of
are the surfaces of constant $S$, which are clearly orthogonal to
the congruence of classical trajectories. These do indeed provide
a good set of surfaces for substantial regions of minisuperspace.
However, this choice breaks down when the trajectories approach
the surface $U(q)=0$. Since both $J$ and $d\Sigma$ are
proportional to $\nabla S$, $ dP $ is proportional to $(\nabla
S)^2$, which vanishes at $U=0$ by virtue of the Hamilton-Jacobi
equation. But apart from this restriction there is still
considerable freedom in the choice of surfaces $\Sigma$.

So we have seen that a sensible probability measure {\it can} be
constructed from the conserved current, by suitable choice of the
surfaces $\Sigma$. The important point to note, however, is that
{\it it works only in the semi-classical regime}, when the wave
function is of WKB form (7.2), subject to the condition (7.8).

Some words are in order on how the probability density (7.20) is
to be used. It should {\it not} be thought of as an {\it absolute}
probability on the entire surface $\Sigma$. That is, it cannot be
used to find the absolute probability that the universe will start
out at some part of the surface $\Sigma$. Indeed, it is not clear
that it is normalizable over a surface $\Sigma$ stretching right
out to the boundary of superspace; {\it i.e.} one would not expect
to be able to write
$$
\int_{\Sigma} J \cdot d A = 1 \eqno(7.21)
$$
unless very special boundary conditions were imposed on the wave
function. Rather, (7.20) should be used to compute {\it
conditional} probabilities. Such probabilities are used when
answering questions of the type, ``Given that the universe starts
out in some finite subset $s_1$ of $\Sigma$, what is the
probabilility that it will start out in the subset $s_0$ of
$s_1$?". This conditional probability would be given by an
expression of the form
$$
P(s_0|s_1) = {\int_{s_0} J \cdot d A \over \int_{s_1} J \cdot d A}
\eqno(7.22)
$$
Each integral is finite because the domains of integration $s_0$,
$s_1$ are finite, and the integrand will typically be bounded on
these domains. The theory makes a prediction when conditional
probabilities of this type are close to zero or one.

Finally, it should be noted that there is a certain element of
circularity in our use of the conserved current as the probability
measure. We have shown that the conserved current can provide a
sensible probability measure in the semi-classical approximation.
Beyond that it seems unlikely that it can be made to work. The
problem, however, is that strictly speaking one really needs a
probability measure in the first place to say what one means by
``semi-classical", and to say that a given wave function is peaked
about a given configuration. The resolution to this apparent
dilemma is to use the measure $|\Psi|^2$ dV from the very
beginning, without any kind of approximations, and it is in terms
of this that one dicusses the notion of semiclassical, and the
peaking about classical trajectories. One may then apply this
measure to non-zero volume regions consisting of slightly
``thickened" $(n-1)$-dimensional hypersurfaces intersecting the
classical flow. With care, it is then in fact possible to recover
the probability measure $J \cdot d \Sigma $ discussed above, but
only in the semi-classical approximation.

Let me now summarize this rather lengthy discussion of classical
spacetime and the WKB approximation. In certain regions of
minisuperspace, and for certain boundary conditions, the
Wheeler-DeWitt equation will have solutions of the WKB form (7.2),
for which (7.8) holds. These solutions correspond to classical
spacetime in that they are peaked about the set of solutions to
the classical field equations satisfying the first integral (7.9).
These classical solutions consist of a congruence of trajectories
in minisuperspace with tangent vector $\nabla S$. One may think of
the wave function as imposing initial conditions on the velocities
on some hypersurface $\Sigma $ cutting across the flow of $S$. In
addition, the quantity $J\cdot d \Sigma$ may be used as a
probability measure on this surface; that is, it may be used to
compute conditional probabilities that the universe will start out
in some region of the surface $\Sigma$.

We will see how this works in detail in an example in the
following sections.

\def\D{{\cal D}}
\def\x{{\bf x}}
\def\R{{I\kern-0.3em R}}

\head{\bf 8. BOUNDARY CONDITION PROPOSALS}

Throughout the course of these lectures I have tried to emphasize
the importance of boundary or initial conditions in quantum
cosmology, although nothing we have done so far depends on a
particular choice of boundary conditions. Now we come to discuss
particular proposals and investigate their consequences.

A quantum theory of initial conditions involves selecting just one
wave function of the universe from amongst the many that the
dynamics allows; {\it i.e.} choosing a particular solution to the
Wheeler-DeWitt equation. Numerous proposals have been made over
the years. As long ago as 1967, DeWitt expressed a hope that
mathematical consistency alone would lead to a unique solution to
the Wheeler-DeWitt equation (DeWitt, 1967). Regretfully such a
hope does not appear to have been realized. More recently, workers
in the field have contented themselves with offering proposals
motivated by considerations of simplicity, naturalness, analogies
with simple quantum mechanical sytems etc. Here, we will
concentrate on just two recent proposals that are the most
comprehensive and the most studied. These are the ``no-boundary"
proposal of Hartle and Hawking (Hawking 1982, 1984a; Hartle and
Hawking, 1983) and the ``tunneling" boundary condition due
primarily to Vilenkin and to Linde (Vilenkin, 1982, 1983, 1984,
1985a, 1985b, 1986, 1988; Linde, 1984a, 1984b, 1984c).

It should be stated at the outset that all known proposals for
boundary conditions in quantum cosmology may be criticised on the
grounds of lack of generality of lack of precision, and these two
are no exception. The issue of proposing a sensible theory of
initial conditions which completely specifies a unique wave
function of the universe for all conceivable situations, is to my
mind still an open one.

\subhead{\bf The No-Boundary Proposal}

The no-boundary proposal of Hartle and Hawking is expressed in
terms of a Euclidean path integral. Before stating it, recall that
a wave function $\Psi[\tilde h_{ij}, \tilde \Phi, B]$ satisfying
the Wheeler-DeWitt equation and the momentum constraint may be
generated by a path integral of the form
$$
\Psi[\tilde h_{ij}, \tilde \Phi,B] = \sum_M \int \D g_{\mu\nu} \D
\Phi \ \exp(-I[g_{\mu\nu}, \Phi] ) \eqno(8.1)
$$
The sum is over manifolds $M$ which have $B$ as part of their
boundary, and over metrics and matter fields $(g_{\mu\nu},\Phi)$
on $M$ matching the arguments of the wave function on the
three-surface $B$. When $M$ has topology $\R \times B$, this path
integral has the form
$$
\Psi[\tilde h_{ij}, \tilde \Phi,S]  = \int \D N^{\mu} \int \D
h_{ij} \D \Phi \ \delta [ \dot N^{\mu} - \chi^{\mu} ] \
\Delta_{\chi} \exp(-I[g_{\mu\nu}, \Phi] ) \eqno(8.2)
$$
The lapse and shift $N^{\mu}$ are unrestricted at the end-points.
The three-metric and matter field are integrated over a class of
paths $(h_{ij}(\x,\tau), \Phi(\x,\tau))$ with the restriction that
they match the argument of the wave function on the three-surface
$B$, which may be taken to be the surface $\tau=1$. That is,
$$
h_{ij}(\x,1)= \tilde h_{ij}(\x), \quad \Phi(\x,1)=\tilde  \Phi(\x)
\eqno(8.3)
$$
To complete the specification of the class of paths one also needs
to specify the conditions satisfied at the initial point, $\tau =
0$ say.

The no-boundary proposal of Hartle and Hawking is an essentially
toplogical statement about the class of histories summed over. To
calculate the no-boundary wave function, $\Psi_{NB}[\tilde h_{ij},
\tilde \Phi,B] $, we are instructed to regard the three-surface
$B$ as the {\it only} boundary of a compact four-manifold $M$, on
which the four-metric is $g_{\mu\nu}$ and induces $\tilde h_{ij}$
on $S$, and the matter field configuration is $\Phi$ and matches
the value $\tilde \Phi$ on $S$. We are then instructed to peform a
path integral of the form (8.1) over all such $g_{\mu\nu}$ and
$\Phi$ and over all such $M$ (see Fig.4.).

For manifolds of the form $\R \times B$, the no-boundary proposal
in principle tells us what conditions to impose on the histories
$(h_{ij}(\x,\tau), \Phi(\x,\tau))$ at the initial point $\tau =0$
in the path integral (8.2). Loosely speaking, one is to choose
initial condition ensuring the closure of the four-geometry.
However, although the four-dimensional geometric picture of what
is going on here is intuitively very clear, the initial conditions
one needs to impose on the histories in the (3+1) picture are
rather subtle. They basically involve setting the initial
three-surface volume, $h^{\half}$, to zero, but also involve
conditions on the derivatives of the remaining components of the
three-metric and the matter fields, which have only been given in
certain special cases.\footnote{$^{\dag}$} {Some earlier
statements of the Hartle-Hawking proposal also used the word
``regular", {\it i.e.} demanded that the sum be over regular
geometries and matter fields. This is surely inappropriate because
in a functional integral over fields, most of the configurations
included in the sum are not even continuous, let alone
differentiable. They may, however, be regular at the
saddle-points, and we will exploit this fact below.}

There is a further issue concerning the contour of integration. As
discussed earlier a complex contour of integration is necessary if
the path integral is to converge. Although convergent contours are
readily found, convergence alone does not lead one automatically
to a unique contour, and the value of the wave function may
depend, possibly quite crucially, on which contour one chooses.
The no-boundary proposal does not obviously offer any guidelines
as to which contour one should take.

Because of these difficulties of precision in defining the
no-boundary wave function, I am going to allow myself considerable
license in my interpretation of what this proposal actually
implies for practical calculations.

As far as the closure conditions goes, the following is, I think,
a reasonable approach to take for practical purposes. The point to
note is that one rarely goes beyond the lowest order
semi-classical approximation in quantum cosmology. That is, for
all practical purposes, one works with a wave function of the form
$\Psi=e^{-I_{cl}}$, where $I_{cl}$ is the action of a (possibly
complex) solution to the Euclidean field equations. The reason one
does this is partly because of the difficulty of computing higher
order corrections; but primarily, it is because our present
understanding of quantum gravity is rather poor and if these
models have any range of validity at all, they are unlikely to be
valid beyond the lowest order semi-classical approximation. What
this means is that in attempting to apply the no-boundary
proposal, one need only concern oneself with the question of
finding initial conditions that correspond to the no-boundary
proposal at the {\it classical} level. In particular, we are
allowed to impose regularity conditions on the metric and matter
fields. To be precise, we will impose initial conditions on the
histories which ensure that (i) the four-geometry closes, and (ii)
the {\it saddle-points} of the functional integral correspond to
metrics and matter fields which are {\it regular} solutions to the
classical field equations matching the prescribed data on the
bounding three-surface $B$. There is a lot more one could say
about this, but these conditions will be sufficient for our
purposes. For a more detailed discussion of these issues see
Halliwell and Louko (1990) and Louko (1988b).

Consider next the contour of integration. Because we will only be
working in the semiclassical approximation, we do not have to
worry about finding convergent contours. Nevertheless, the contour
becomes an issue for us if the solution to the Einstein equations
satisfying the above boundary conditions is not unique. For then
the path integral will have a number of saddle-points, each of
which may contribute to the integral an amount of order
$e^{-I_{cl}^k}$, where ${I_{cl}^k}$ is the action of the solution
corresponding to saddle-point $k$. Without choosing a contour and
performing a detailed contour analysis it is unfortunately not
possible to say which saddle-points will generally provide the
dominant contributions. We therefore have no general guidelines to
offer here.

We will see how exactly these issues arise in the simple example
discussed below.

We now calculate the no-boundary wave function explicitly for the
scalar field model described in Section 2. In the gauge $\dot
N=0$, the minisuperspace path integral for the no-boundary wave
function is
$$
\Psi_{NB}(\tilde a, \tilde \phi) = \int dN \int {\cal D} a {\cal
D} \phi \exp\left( -I[a(\tau), \phi(\tau),N] \right)
$$
where $I$ is the Euclidean action for the scalar field model,
$$
I = \half \int_0^1 d \tau N \left[ - {a \over  N^2}\left( {da
\over d\tau}\right)^2 + {a^3 \over N^2} \left( {d \phi \over d
\tau} \right)^2 - a + a^3 V(\phi) \right] \eqno(8.4)
$$
The Euclidean field equations may be written,\footnote{$^{\dag}$}
{Because the path integral representation of the wave function
involves an ordinary integral over $N$, not a functional integral,
the constraint (8.7) does not immediately follow from extremizing
the action (8.4) with respect to the variables integrated over.
Rather, the saddle-point condition is ${\partial I / \partial N} =
0 $, and one actually obtains the {\it integral over time} of
(8.7). The form of (8.7) as written is obtained once one realizes
that the integrand is in fact constant, by virtue of the other two
field equations, hence the integral sign may be dropped. However,
writing the constraint {\it with} the integral over time
highlights the fact that the field equations and constraint
contain two functions and one constants worth of information. This
is precisely the right amount of information to determine the two
functions $(a(\tau), \phi(\tau))$ and the constant $N$ in terms of
the boundary data.}
$$
{1 \over  N^2 a } {d^2a \over d \tau^2} = - {2 \over N^2} \left(
{d \phi \over d \tau} \right)^2 - V(\phi) \eqno(8.5)
$$
$$
{1 \over N^2} {d^2 \phi \over d \tau^2} + {3 \over Na} { da \over
d \tau} {d \phi \over d \tau} - \half V'(\phi) = 0 \eqno(8.6)
$$
$$
{1 \over N^2} \left( {da \over d \tau} \right)^2 - {a^2 \over N^2
} \left( {d \phi \over d \tau} \right)^2 - 1 + a^2 V(\phi)  = 0
\eqno(8.7)
$$

The integral (8.4) is taken over a class of paths $(a(\tau),
\phi(\tau),N)$ satisfying the final condition
$$
a(1) = \tilde a, \quad \phi(1) = \tilde \phi \eqno(8.8)
$$
and a set of initial conditions determined by the no-boundary
proposal, discussed below. The constant $N$ is integrated along a
closed or infinite contour in the complex plane and is not
restricted by the boundary conditions. We are interested only in
the semi-classical approximation to the above path integral, in
which the wave function is taken to be of the form
$$
\Psi( \tilde a, \tilde \phi) = \exp( - I_{cl}(\tilde a, \tilde
\phi) ) \eqno(8.9)
$$
(or possibly a sum of wave functions of this form). Here
$I_{cl}(\tilde a, \tilde \phi) $ is the action of the solution to
the Euclidean field equations $(a(\tau),\phi(\tau),N)$, which
satisfies the final condition (8.5) and, in accordance with the
above interpretation of the no-boundary proposal, is regular and
respects the closure condition.

Consider, then, the important issue of determining the initial
conditions on the paths that correspond to the closure condition
and ensure that the solution is regular. Consider first $a(\tau)$.
The Euclidean four-metric is
$$
ds^2 = N^2 d \tau^2 + a^2(\tau) d \Omega_3^2 \eqno(8.10)
$$
We want the four-geometry to close off in a regular way. Imagine
making the three-sphere boundary smaller and smaller. Then
eventually we will be able to smoothly close it off with flat
space. Compare, therefore, (8.10) with the metric on flat space in
spherical coordinates
$$
ds^2 = dr^2 + r^2 d \Omega_3^2 \eqno(8.11)
$$
From this, one may see that for (8.10) to close off in a regular
way as $a \rightarrow 0 $, we must have
$$
a(\tau) \sim N \tau, \quad as \quad \tau \rightarrow 0 \eqno(8.12)
$$
This suggests that the conditions that must be satisfied at
$\tau=0$ are
$$
a(0) = 0, \quad {1\over N} {da \over d \tau}(0) = 1 \eqno(8.13)
$$

(8.13) are the conditions that are often stated in the literature.
However, this is in general too many conditions. In general, we
would not expect to be able to find a classical solution
satisfying the boundary data of fixed $a$ on the final surface,
fixed $a$ on the initial surface {\it and} and fixed $da/d\tau$ on
the initial surface. We might of course be able to do this at the
classical level, for certain special choices of boundary data, but
such conditions could not be elevated to quantum boundary
conditions on the full path integral. One of these condition must
be dropped. Since the main requirement is that the geometry
closes, let us drop the condition on the derivative and keep the
condition that $a(0)=0$. On the face of it, this seems to allow
the possibility that the four-geometry may not close off in a
regular fashion. Consider, however, the constraint equation (8.7).
It implies that if the solution is to be regular, then $da/d\tau
\rightarrow \pm 1$ as $a \rightarrow 0$. The regularity condition
is therefore recovered when the constraint equation holds. This
guarantees that the saddle-points will indeed be regular
four-geometries, if we only impose $a(0)=0$.

Now consider the scalar field $\phi(\tau)$. Consider the equation
it satisfies, (8.6). It is not difficult to see that if the
solution is to be regular as $a\rightarrow 0$, then $\phi(\tau)$
must satisfy the initial condition
$$
{d\phi \over d \tau}(0) = 0 \eqno(8.14)
$$
So the sole content of the no-boundary proposal, for this model,
is the initial condition (8.14) and the condition $a(0)=0$.

Our task is now to solve the field equations (8.5)-(8.7) for the
solution ($a(\tau)$, $\phi(\tau)$, $N$), subject to the boundary
conditions (8.9), (8.14) and $a(0)=0$, and then calculate the
action\footnote{$^{\dag}$} {The action (8.4) is the appropriate
one when $a$ and $\phi$ are fixed on both boundaries. If one wants
to fix instead derivatives of the fields on the boundary, as
(8.14) requires, then (8.4) must have the appropriate boundary
terms added. The correct boundary term does in fact vanish in the
case under consideration here, although this is a point that
generally needs to be treated quite carefully.} of the solution.

For definiteness, let us assume that the potential $V(\phi)$ is of
the chaotic type ({\it i.e.} U-shaped) and let us go to the large
$\phi$ region at which $|V'/V|<<1$. It is not difficult to see
that the approximate solution to the scalar field equation (8.6),
subject to the boundary conditions (8.8), (8.14), is
$$
\phi(\tau) \approx \tilde \phi \eqno(8.15)
$$
Similarly, the approximate solution to the second order equation
for $a(\tau)$, (8.5), satisfying the boundary conditions $a(0)=0$,
$a(1)= \tilde a$, is
$$
a(\tau) \approx {\tilde a  \sin( V^{\half} N \tau) \over \sin(
V^{\half} N)} \eqno(8.16)
$$
Finally, we insert (8.15), (8.16) into the constraint (8.7) to
obtain a purely {\it algebraic} equation for the lapse, $N$. It is
$$
\sin^2(V^{\half}N) = {\tilde a}^2 V \eqno(8.17)
$$
There are an infinite number of solutions to this equation. If
${\tilde a}^2 V<1$, they are real, and are conveniently written
$$
N = N_n^{\pm} \equiv {1 \over V^{\half} } \left[ (n+\half) \pi \pm
\cos^{-1}(\tilde a V^{\half} ) \right] \eqno(8.18)
$$
where $n = 0, \pm 1, \pm 2,...$ and $ \cos^{-1}(\tilde a V^{\half}
) $ lies in its principal range, $(0,\pi/2)$. For the moment, we
set $n=0$. We will return later to the significance of the other
values of $n$.

With $n=0$, the solution for the lapse inserted into the solution
for $a(\tau)$, (8.16), now reads
$$
a(\tau) \approx {1 \over V^{\half} } \sin\left[\left({\pi \over 2}
\pm \cos^{-1}(\tilde a V^{\half}) \right) \tau \right] \eqno(8.19)
$$
We now have the complete solution to the field equations subject
to the above boundary conditions. It is (8.15), (8.19), together
with the solution for the lapse (8.18). The action of the solution
is readily calculated. It is
$$
I_{\pm} = - {1 \over 3 V(\tilde \phi) } \left[ 1 \pm \left( 1-
\tilde a^2 V(\tilde \phi) \right)^{3/2} \right] \eqno(8.20)
$$
It is not difficult to see that these two solutions represent the
three-sphere boundary being closed off with sections of
four-sphere. As expected, the action is negative. The $(-)/(+)$
sign corresponds to the three-sphere being closed of by less
than/more than half of a four-sphere. The classical solution is
therefore not unique.

Because the classical solution is not unique, we are faced with
the problem of which solution to take in the semi-classical
approximation to the wave function. Naively, one might note that
the (+) saddle-point has most negative action, and will therefore
provide the dominant contribution. However, as briefly mentioned
earlier, this depends on the contour of integration. One can only
say that the (+) saddle-point provides the dominant contribution
if the chosen integration contour in the path integral may be
distorted into a steepest-descent contour along which the (+)
saddle-point is the global maximum. In their original paper,
Hartle and Hawking (1983) gave heuristic arguments, based on the
conformal rotation, which suggest that the contour was such that
it could not be distorted to pass through the (+) saddle-point and
was in fact dominated by the (-) saddle-point. For the moment let
us accept these arguments. They thus obtained the following
semi-classical expression for the no-boundary wave function:
$$
\Psi_{NB} (a, \phi) \approx \exp \left({1 \over 3 V(\phi)} \left[
1 - \left( 1- a^2  V(\phi) \right)^{3/2} \right] \right)
\eqno(8.21)
$$
(where we have dropped the tildes, to avoid the notation becoming
too cumbersome). (8.21) is indeed an approximate solution to the
Wheeler-DeWitt equation for the model, (2.14), in the region $ a^2
V(\phi) < 1$. Using the WKB matching procedure, it is readily
shown that the corresponding solution in the region $ a^2 V(\phi)
> 1 $ is
$$
\Psi_{NB} (a,\phi) \approx \exp \left( {1 \over 3 V(\phi)} \right)
\cos\left[ {1 \over 3V(\phi)} \left( a^2 V(\phi) -1 \right)^{3/2}
- {\pi \over 4} \right] \eqno(8.22)
$$
This completes the calculation of the no-boundary wave
function.\footnote{$^{\dag}$} {The reader familiar with the
literature will note that this is not the derivation given by
Hartle and Hawking (1983). However, I have presented it in this
way to emphasize certain points which will be discussed in what
follows.}

Some further remarks are in order. First, the contour of
integration. The path integral for the no-boundary wave function
as discussed above has two saddle-points, and Hartle and Hawking
argued that it is the saddle-point corresponding to less than half
a four-sphere that provides the dominant contribution. However,
their heuristic argument is not, in my opinion, totally
convincing.

A more detailed analysis of this situation by myself and Jorma
Louko exposed the assumptions that Hartle and Hawking implicitly
made to arrive at the above answer (Halliwell and Louko, 1989a).
By a suitable choice of variables, and by working with a
cosmological constant instead of a scalar field, we were able to
evaluate the minisuperspace path integral for this model exactly.
In particular, we were able to determine convergent contours
explicitly for the model, and thus see whether or not certain
saddle-points did or did not yield the dominant contribution to
the path integral. What we found is that there are a number of
inequivalent contours along which the path integral converges,
each dominated by different saddle-points, and thus leading to
different forms for the wave function. No one contour was
obviously preferred. In particular, the no-boundary proposal did
not indicate which contour one was supposed to take. A contour
yielding the above form for the wave function could be found, but
it was not obvious why one should take that particular one. So the
essential conclusion here is that the no-boundary proposal as it
stands does not fix the wave function uniquely. There are, so to
speak, many no-boundary wave functions, each corresponding to a
different choice of contour. The wave function is therefore only
fixed uniquely after one has put in some {\it extra} information
fixing the contour.

As an example, in the simple model above one could {\it define}
the no-boundary wave function to be as defined by Hartle and
Hawking, with the {\it additional} piece of information that one
is to take the contour dominated by the less-than-half
saddle-point. A more general statement is however not currently
available. A possible approach to this problem is that of
Halliwell and Hartle (1989), which involved restricting the
possible contours on the grounds of mathematical consistency and
physical predictions.

The second issue that deserves further comment is the equation for
the lapse, (8.17), and there are a number of points to be made
here. Firstly, we considered only $\tilde a^2 V(\tilde \phi) <1 $,
so that the solution was real. One may allow $\tilde a^2 V(\tilde
\phi) >1$, in which case $N$, the scale factor (8.16) and the
action become complex -- the action is essentially (8.20) with $
\tilde a^2 V(\tilde \phi) $ continued into the range $\tilde a^2
V(\tilde \phi) >1$. Complex saddle-points are generally expected
in this sort of problem. Indeed, they are essential if the wave
function is to be oscillatory, and thus predict classical
spacetime. Secondly, we restricted to the solutions with $n=0$.
What is the significance of the other solutions? Consider first
the case of $n$ positive. It is not difficult to see that for
values of $n>0$, the solution (8.16) undergoes many oscillations.
More precisely, $a^2$, which appears in the metric, expands to a
maximum size and then ``bounces" each time it reaches zero. The
geometric picture of these saddle-points is therefore of linear
chains of contiguous spheres (Halliwell and Myers, 1989; Klebanov
et al., 1989).

What about the saddle-points with $n<0$? These saddle-points have
negative lapse. Because the action changes sign under $N
\rightarrow -N$, the action of these saddle-points has the
``wrong" sign. However, these saddle-points are otherwise
identical to the ones with positive lapse -- their four-metrics
are the same. Moreover, they have a perfectly legitimate place as
saddle-points of the path integral. They are not artefacts of this
model. They arise because the action, by virtue of the presence of
the $\sqrt{g}$ factor, is double-valued in the space of complex
four-metrics. Carrying the metric once around the branch point
returns one to a physically identical solution to the Einstein
equations, but with action of the opposite sign. So to every
physically significant solution there corresponds {\it two}
saddle-points. Because one has to integrate over complex metrics
for convergence, both saddle-points are candidate contributants to
the path integral.

So finally it seems sensible to ask, why did we not include any of
these extra saddle-points, {\it i.e.} $n=\pm 1, \pm 2,...$, in the
calculation of the no-boundary wave function? The answer is that
one can, by a suitable choice of contour. However, the
saddle-points with $N$ negative (or more generally, with
$Re(\sqrt{g})$ negative), lead to difficulties with the recovery
of quantum field theory in curved spacetime if they dominate the
path integral, because a normally positive matter action will
become negative definite on the gravitational background
corresponding to such a saddle-point. For this reason, the contour
should not be chosen in such a way that it is dominated by a
negative $N$ saddle-point (Halliwell and Hartle, 1989). This
leaves the saddle-points with $n=0,1,2,3..$. The saddle-points
corresponding to the linear chains of spheres, $n=1,2,..$ may
contribute with a suitable choice of contour, but one could
exclude them by taking the definition of the no-boundary wave
function offered above ({\it i.e} demand that the contour be
dominated by the saddle-point corresponding to less than half of a
four-sphere).

These issue involving the contour are still very much up in the
air, and I would regard the question of choosing a sensible
contour for the no-boundary wave function as at this moment an
open one.

\subhead{\bf The Tunneling Boundary Condition}

The other proposal we will consider here is the so-called
``tunneling" boundary condition advocated primarily by Vilenkin
(1982, 1983, 1984, 1985a, 1985b, 1986, 1988) and Linde (1984a,
1984b, 1984c). I will concentrate on Vilenkin's formulation, which
is the most comprehensive. The tunneling boundary condition
attempts to draw most strongly the analogy between the quantum
creation of the universe and tunneling in ordinary quantum
mechanics. Vilenkin has offered various formulations of this
boundary condition, not all of which are obviously equivalent. The
most detailed is the ``outgoing modes" formulation, which we now
discuss (Vilenkin, 1988).

The outgoing modes statement of the tunneling boundary condition
proposal is a statement about the behaviour of the solutions to
the Wheeler-DeWitt equation at the boundary of superspace. In
brief, the idea is as follows. In a manner analagous to that in
which solutions to the Klein-Gordon equation are classified as
positive or negative frequency, Vilenkin attempts to classify the
solutions to the Wheeler-DeWitt equation as ``ingoing" or
``outgoing" at the boundary. The proposal is then that the wave
function should consist solely of outgoing modes at the parts of
the boundary of superspace which correspond to singular
four-geometries. A regularity condition, that $\Psi$ be everywhere
bounded, is also imposed.

This is perhaps a little vague, so let us discuss it more
carefully. First, consider the nature of the boundary of
superspace. The boundary of superspace will generally consist of
configurations which are in some sense singular, {\it e.g.}
$h^{\half}$ will be zero or infinite, or quantities such as
$\Phi$, or $(\partial_i \Phi)^2$ may be infinite. However, this
does not necessarily mean that a four-geometry which has that
three-geometry as a slice is singular. For example, $h^{\half}$
vanishes at the north and south pole of a four-sphere, but the
four-geometry is perfectly regular. Let us therefore divide the
boundary of superspace into two regions. The first region consists
of three-geometries having singularities attributable to the
slicing of a regular four-geometry. That is, there exists a
regular Euclidean\footnote{$^{\dag}$} {The description
``Euclidean" was not given explicitly in Vilenkin (1988), but it
appears to have been tacitly assumed there and elsewhere}
four-geometry of which the singular three-geometry is a slice.
Call these parts of the boundary non-singular. The second part of
the boundary is what remains, and will be referred to as the
singular boundary. This part of the boundary {\it does} correspond
to singularities of the four-geometry. A more detailed
mathematical discussion of this point can be given, using Morse
theory, but the above is sufficient for our purposes.

Now let us discuss the notion of ingoing and outgoing modes.
Solutions to the Klein-Gordon equation of relativistic quantum
mechanics may be expanded in terms of mode functions $e^{ip\cdot
x}$, and these modes may be classified as positive or negative
frequency, with respect to the timelike Killing vector $ -i
\partial / \partial t $. More precisely, the mode solutions are
eigenfunctions of this Killing vector and the classification
depends on the sign of the eigenvalue. The positive and negative
frequency modes may also be characterized by the sign of $J_0$,
the timelike component of the conserved current
$$
J = {i\over 2} \left( \Psi^* \nabla \Psi - \Psi \nabla \Psi^*
\right), \quad \nabla \cdot J = 0 \eqno(8.23)
$$

One might hope to do an analagous thing for the Wheeler-DeWitt
equation. In the general case, one meets with an immediate
difficulty. This is that to say what one means by positive and
negative frequency on the whole of superspace, one needs a
timelike Killing vector. However, it is a mathematical property of
superspace that it has has no Killing vectors at all, so positive
and negative frequency modes cannot in general be defined
(Kucha{$\rm \check r$}, 1981).

Despite this obstruction, one can still make considerable progress
by restricting attention to certain approximate forms for the wave
function, or by restricting attention to certain regions of
superspace, such as close to the boundary. One is, for example,
primarily interested in the solution in the oscillatory region.
There, one expects solutions to the Wheeler-DeWitt equation of the
form
$$
\Psi = \sum_n C_n e^{iS_n} \eqno(8.24)
$$
where the $S_n$ are solutions to the Hamilton-Jacobi equation. The
current for the mode $ C_n e^{iS_n} $ is
$$
J_n =  - |C_n|^2 \nabla S_n \eqno(8.26)
$$
This mode is thus defined to be outgoing at the boundary if
$-\nabla S_n$ points outward there. If the wave function is not
oscillatory in the neighbourhood of the boundary, then the
definition of outgoing modes is more problematic, if not
impossible.

Now let us give a more precise statement of Vilenkin's outgoing
modes proposal for the tunneling wave function, $\Psi_T$:

\indent {\it $\Psi_T$ is the solution to the Wheeler-DeWitt
equation that is everywhere bounded and consists solely of
outgoing modes at singular boundaries of superspace.}

Despite the apparent vagueness in its definition, and the
obstruction of principle to making it more general, the Vilenkin
outgoing modes form of the tunneling boundary condition appears in
simple minisuperspace models to be intuitively reasonably clear,
and it has been quite successful in defining a unique solution to
the Wheeler-DeWitt equation.

Now let us calculate the tunneling wave function, using the above
proposal, for the scalar field model. The Wheeler-DeWitt equation
may be written
$$
\left[ {\partial^2 \over \partial a^2} - {1 \over a^2} {\partial^2
\over
\partial \phi^2} + a^4 V(\phi) - a^2 \right] \Psi(a,\phi) = 0
\eqno(8.27)
$$
The minisuperspace for this model is the two-dimensional space
with coordinates $(a,\phi)$, with $0<a<\infty$, $-\infty < \phi <
\infty $. The only non-singular part of the boundary is $a=0$ with
$\phi$ finite. The rest is singular, and consists of
configurations with one or both of $a$ and $\phi$ infinite.
Writing $a=e^{\alpha}$, minisuperspace, which is now just flat
two-dimensional Minkowski space, is conveniently represented on
the usual conformal diagram. The non-singular boundary is mapped
to the single point $i^-$, past timelike infinity. The remaining
singular part of the boundary is mapped to ${\cal I}^{\pm} $,
future and past null infinity, and $ i^0, i^+ $, spacelike and
future timelike infinity (see Fig.5.). The basic idea of the
outgoing modes prescription is that probability flux is injected
into superspace at $i^-$ with finite $\phi$ and $a=0$, and flows
out of superspace across the singular boundaries.

We will again work in the region for which the scalar field
potential depends only very slowly on $\phi$. So provisionally we
impose the restriction $|V'/V|<<1 $, although this condition will
be revised below. Next, note that as $a$ goes to zero, the
coefficient of the second derivative with respect to $\phi$ in
(8.27) blows up. If, as the boundary conditions demand, we are to
get a regular solution, it seems reasonable to insist that
$\Psi(a,\phi)$ becomes independent of $\phi$ for small $a$. We
will therefore neglect the second derivative with respect to
$\phi$ in (8.27).

Consider first the solution in the oscillatory region, $a^2
V(\phi)>1$. The WKB solutions are proportional to $e^{iS} $, or
$e^{-iS} $, where $S = (a^2 V(\phi)-1)^{3/2} / 3V(\phi) $. The
first has probability flux $ J \sim -\nabla S$, pointing back
towards $i^-$, the second has $ J \sim \nabla S$, pointing
outwards away from $i^-$. The latter, if evolved in the forward
direction would eventually reach the singular boundary at which it
would be outgoing. The former, however, corresponds to the time
reverse of this so is ingoing at the boundary. This means that the
outgoing modes prescription implies that only an outgoing wave
should be present in the classically allowed region, so the wave
function should be proportional to $e^{-iS}$. This implies that
the tunneling wave function in the oscillatory region is of the
form
$$
\Psi_T(a, \phi) \approx A(\phi) \exp\left(- {i \over 3V(\phi)}
\left( a^2 V(\phi) -1 \right)^{3/2} \right) \eqno(8.30)
$$
Here we have included, as we may, the slowly varying
$\phi$-dependent factor $ A(\phi) $. It turns out that this needs
to be included to ensure that the solution is regular.

By the WKB matching procedure, one may determine the solution
corresponding to (8.30) in the exponential region, $a^2 V(\phi) <
1 $. It is
$$
\Psi_T (a, \phi) \approx A(\phi) \exp \left({1 \over 3 V(\phi)}
\left( 1- a^2  V(\phi) \right)^{3/2} \right)
$$
$$
-i A(\phi) \exp \left(- {1 \over 3 V(\phi)} \left( 1- a^2  V(\phi)
\right)^{3/2} \right) \eqno(8.31)
$$
The second term is exponentially smaller than the first, so may be
neglected. Now consider what happens to the solution as $a$ goes
to zero. For regularity, we need $ {\partial \Psi / \partial \phi}
\rightarrow 0 $ as $ a \rightarrow 0 $. This can only be achieved
by choosing the function $A(\phi)$ to be
$$
A(\phi) = \exp\left( - {1 \over 3 V(\phi) } \right) \eqno(8.32)
$$
With this choice, $ \Psi_T \sim e^{-\half a^2} $ for small $ a$,
which is regular for all values of $\phi$.

We should now check that all this is consistent with the
approximation of neglecting the second derivative with respect to
$\phi$ in the Wheeler-DeWitt equation. Inserting the approximate
solution with $A(\phi)$ given by (8.32) into (8.27), it may be
shown that the solution is valid in the region for which $
|V'(\phi)|<< a^{-2} $. If $a^2 V(\phi)<1 $, this is actually an
improvement on the original condition, $|V'/V|<<1$. In particular,
it means that the solution is valid for arbitrarily rapid
dependence of the potential on $\phi$ as $a$ goes to zero. This
would not have been true had we not multiplied the wave function
by (8.32). So the revised restriction under which our
approximations are valid is
$$
|V'(\phi)| << {\rm max} \left[ |V(\phi)|, a^{-2} \right]
\eqno(8.33)
$$

The final expression for the tunneling wave function is given by
$$
\Psi_T (a, \phi) \approx \exp \left(-{1 \over 3 V(\phi)} \left[ 1
- \left( 1- a^2  V(\phi) \right)^{3/2} \right] \right) \quad for
\quad a^2 V(\phi) < 1 \eqno(8.34)
$$
$$
\Psi_T (a,\phi) \approx \exp \left( -{1 \over 3 V(\phi)} \right)
\exp\left(- {i \over 3V(\phi)} \left( a^2 V(\phi) -1 \right)^{3/2}
\right) \quad for \quad a^2 V(\phi) > 1 \eqno(8.35)
$$
This completes the calculation of the tunneling wave function.

Mention should also be made of an alternative, not so well-known
version of the tunneling boundary condition, also due to Vilenkin.
This is that the wave function is given by a {\it Lorentzian} path
integral over geometries which close off in the past,
$$
\Psi_T = \int {\cal D} g_{\mu\nu} e^{iS} \eqno(8.36)
$$
where $S$ is the Lorentzian action. The phrase ``close off in the
past" is taken to mean that the histories summed over have
vanishing initial three-volume, and also that the lapse function
in the path integral (4.7) (or (5.21)) is integrated not over an
infinite range, but over a half-infinite range, from $0$ to
$\infty$. The wave function thus calculated is then not quite a
solution to the Wheeler-DeWitt equation, but is a Green function
of the Wheeler-DeWitt operator; {\it i.e.} one obtains a
delta-function on the right-hand side of Eq.(5.23), although this
delta-function is pushed to the boundary of superspace where
$h^{\half}=0$. This is in keeping with the idea that the tunneling
wave function involves probability flux being injected into
superspace at the non-singular boundary. It seems reasonable to
interpret this proposal as being essentially the same as the
no-boundary proposal, in which a particular choice for the contour
is made. Namely, that the contour is chosen to be the complex
contour which may be distorted to lie along the real Lorentzian
axis. It is not obviously equivalent to the outgoing modes version
of the tunneling proposal, however, and actually fails to coincide
precisely in some models (Halliwell and Louko, 1990).

Linde's version of the tunneling proposal (Linde, 1984a, 1984b,
1984c) also appears to involve a Lorentzian path integral as a
starting point. Because the usual Wick rotation to Euclidean time
leads to a minus sign in front of the kinetic term for the scale
factor in the action, Linde proposed that the Wick rotation should
be performed in the ``wrong" direction. It may be argued that this
involves choosing the lapse contour to be the distortion into the
region $Re(N)<0$ of the contour running up the positive imaginary
axis (Halliwell and Louko, 1989a, 1990). This proposal is
therefore identical to Vilenkin's path integral version of the
tunneling proposal.

Finally, an interesting point. Vilenkin observed that the full
Wheeler-DeWitt equation is invariant under the transformation
$$
h_{ij} \rightarrow e^{i\pi} h_{ij}, \quad V(\Phi) \rightarrow
e^{-i\pi} V(\Phi) \eqno(8.37)
$$
That is, given a solution $\Psi[h_{ij}, \Phi]$, a second solution
may be generated from it using the above transformation. In
particular, Vilenkin noticed that the no-boundary and tunneling
wave functions for the scalar field minisuperspace model are
related by this transformation:
$$
\Psi_{NB} = \Psi_T ( V \rightarrow e^{-i \pi} V, a \rightarrow
e^{i \pi/2} a) \eqno(8.38)
$$
(to see this explicitly one has to use the Airy functions of which
(8.34) and (8.35) are asymptotic forms). The possible significance
of this observation is the following: as I have tried to
emphasize, there are considerable difficulties of precision and
generality in the definitions of the no-boundary and tunneling
wave functions. If, however, one succeeded in defining {\it one}
of these wave functions in a much more precise, more general way,
then the other could be {\it defined} by the transformation
(8.37).

\head{\bf  9. NO-BOUNDARY VS. TUNNELING}

Let us now compare the no-boundary and tunneling wave functions.
For convenience we record their explicit forms in the oscillatory
region, in a range of $\phi$ for which $ V(\phi) $ is slowly
varying. To be definite, let us take the potential $V(\phi)$ to be
of the chaotic inflationary type. Let us introduce
$$
S = {1 \over 3 V(\phi) } \left( a^2 V(\phi) - 1 \right)^{3/2} -
{\pi \over 4} \eqno(9.1)
$$
The tunneling wave function is
$$
\Psi_T \approx \exp \left( - {1 \over 3 V( \phi ) } \right)
e^{-iS} \eqno(9.2)
$$
The no-boundary wave function is
$$
\Psi_{NB} \approx \exp \left( + {1 \over 3 V( \phi ) } \right)
\left[ e^{-iS} + e^{iS} \right] \eqno(9.3)
$$
There are two differences. The first is that the no-boundary wave
function is real, being a sum of a WKB component and its complex
conjugate, whilst the Vilenkin wave function consists of just one
WKB component.\footnote{$^{\dag}$} {The fact that the no-boundary
wave function is real corresponds to the fact that it is in a
sense CPT invariant, and has implications for the arrow of time in
cosmology (Hawking, 1985; Page, 1985).} If one component
corresponds to expanding solutions, then the other corresponds to
collapsing solutions, although it is not possible to say which one
is which. It may be argued that these components have negligible
interference, so each component may be considered separately
(Halliwell, 1989b). One may thus compare a single component of the
no-boundary wave function with the Vilenkin wave function.

The second, and more important difference, is the sign difference
in the prefactor. Both wave functions are peaked about the same
set of solutions to the field equations, namely those satisfying
the first integral $p=\nabla S $, with $S$ given by (9.1). As we
have shown, these solutions are initially inflationary, with $a(t)
\approx e^{V^{\half} t} $, $\phi(t) \approx \phi_0 = constant$.
These solutions may be labeled by their initial values of $ \phi$,
$\phi_0$. Although all the solutions undergo {\it some} inflation,
the {\it amount} by which they inflate depends on $\phi_0$. For
example, if the potential is $ V(\phi) = m^2 \phi^2 $, then
sufficient inflation is obtained only for values of $\phi_0$ in
excess of about 4 (in Planck units) (Hawking, 1984a; Page, 1986a).

To see which initial values of $\phi$ are most favoured by each
wave function, we need to study the measure on the set of paths,
$J\cdot d \Sigma$. Because the trajectories have $\dot \phi
\approx 0 $, $J$ points largely in the $a$ direction. A suitable
choice of surfaces $\Sigma$ is therefore surfaces of constant $a$,
at least locally. The probability measure is thus given by
$$
dP = J \cdot d \Sigma \approx \exp\left( \pm {2 \over 3V(\phi)
}\right) d \phi \eqno(9.4)
$$
(with $(+)$ for the no-boundary wave function, $(-)$ for the
tunneling wave function). With this measure, we now have to ask
the right questions. As discussed previously, we cannot take this
to be an absolute measure on the initial values of $\phi$. Rather,
it should be thought of as a conditional probability measure. So
we must first decide what conditions to impose; that is, in what
range of values of $\phi$ are we to ask for predictions?

First of all consider what happens if $\phi$ is very small
initially, close to zero (for convenience, we restrict attention
to positive $\phi$ in what follows). Universes starting out with a
very small initial value of $\phi$ will very rapidly reach a small
maximum size and then recollapse in a short period of time. One
would not expect large scale structure and indeed, observers, to
exist in such universes. It therefore seems reasonable to impose
the condition that the universe expands out to a ``reasonable"
size. This is somewhat vague, but what it means is that we
restrict attention to initial values of $\phi$ greater than some
exceedingly small value $ \phi_{min} $, say. This restriction has
the consequence that the no-boundary $(+)$ measure (9.4) is now
bounded (it was previously unbounded at $\phi=0$) and it is peaked
about $\phi_{min}$.

Now consider very large values of $\phi$. For a chaotic potential
at least, as $\phi$ becomes very large the scalar field energy
density $ V(\phi) $ will approach the Planck energy density,
$V(\phi) \sim 1 $. If minisuperspace models are to have any
validity at all, it seems unlikely that they can be trusted in the
range of $\phi$ for which $V(\phi)>1$. So our second condition is
to ask for predictions only in the region $\phi< \phi_p $, where $
V(\phi_p)=1$. For the potential $V(\phi) =m^2 \phi^2$, $m$ is
normally taken to be about $10^{-4}$, so $ \phi_p \sim 10^{4} $.

Our task is now to ask for predictions with the condition that the
initial value of $\phi$ lies in the range $ \phi_{min} < \phi <
\phi_p $. For a chaotic potential, there will be a value of $\phi
$ in this range, larger than $\phi_{min}$, call it $ \phi_{suf} $,
for which sufficient inflation is achieved if $\phi_0 > \phi_{suf}
$, and it is not achieved if $\phi_0 < \phi_{suf} $. For the
massive scalar field, $\phi_{suf} \sim 4 $. A pertinent question
to ask, therefore is this: ``What is the probability that $\phi_0
> \phi_{suf}$, given that $ \phi_{min} < \phi_0 < \phi_p $?" It is
given, using (9.4), by the following expression.
$$
P(\phi_0>\phi_{suf} | \phi_{min} < \phi_0 < \phi_p ) =
{\int_{\phi_{suf}}^{\phi_p} d\phi \exp\left( \pm {2 \over 3
V(\phi)} \right) \over \int_{\phi_{min}}^{\phi_p} d\phi \exp\left(
\pm {2 \over 3 V(\phi)} \right)} \eqno(9.5)
$$
This is effectively the probability of sufficient inflation.

It is reasonably easy to see the result of evaluating (9.5) by
merely looking at the plot of the two probability distributions,
$\exp\left( \pm {2 \over 3 V(\phi)} \right)$ (see Fig.6.).
Consider first the tunneling wave function $(-)$. The integrand
becomes very small as $\phi$ approaches $ \phi_{min} $ and it is
clear that by far the largest contribution to the integral in the
denominator comes from the region $\phi>\phi_{suf}$. One therefore
has $P \approx 1$, and sufficient inflation {\it is} a prediction
of the tunneling wave function.

Now consider the no-boundary wave function $(+)$. The integrand
diverges as $\phi$ approaches zero, but is cut off by
$\phi_{min}$. If, as we are assuming, $\phi_{min}$ is very small,
the main contribution to the integral in the denominator will come
from the region very close to $\phi_{min}$. One therefore has $P
<<1 $ for the no-boundary wave function, and sufficient inflation
{\it is not} a prediction.

The above conclusion about the no-boundary wave function is not
the one reached by Hawking and Page in their analysis (Hawking and
Page, 1986). They concluded that sufficient inflation has
probability unity. The difference with the analysis given here
(which is based on that of Vilenkin (1989)), is that Hawking and
Page did not restrict to $\phi < \phi_p $. For $\phi> \phi_p$, the
integrands in (9.4) level off to $1$; thus although in the range
$\phi_{min} < \phi < \infty $ the integrands in the denominator
are strongly peaked at $\phi_{min}$, the contribution to the
integral from this region is overwhelmingly outweighed by that
from very large values of $\phi$. (9.4) would therefore yield the
value $1$ for both wave functions.

A new aspect to the no-boundary/tunneling debate was recently
exposed by Grishchuk and Rozhansky (1989) (see also Grishchuk and
Rozhansky (1988)). They asked whether the above calculation of the
no-boundary wave function, which involves the approximate
solutions to the classical Euclidean field equations for the
scalar field model, is really valid down to $\phi$ close to zero.
The conclusion they came to is that they are not, and that the
above expression for the no-boundary wave function makes sense
only in the range $\phi> \phi_*$, for some critical value of
$\phi$, $\phi_*$ which they estimated. Although less than
$\phi_{suf}$, $\phi_*$ is much greater than $\phi_{min}$, the
lower bound we imposed to ensure that the universe expanded to a
``reasonable" size. The value of $\phi_*$ is model-dependent. For
the massive scalar field model $ \phi_* \approx 1 $.

Their conclusion was reached by giving a more careful treatment of
the motion of the scalar field, which was taken to be
approximately constant in the above analysis. For large $\phi$, $
|V'/V|<<1$, and the Euclidean solution for $a(\tau)$ is given
approximately by (8.19). The trajectories start at $a=0$ with some
value of $\phi$, expand, and then turn around and recollapse. In
particular, along the curve $ a^2 V(\phi)=1 $, neighbouring
Euclidean trajectories intersect -- they form a {\it caustic}.
Because the real Euclidean trajectories dominating the path cannot
reach immediately beyond the caustic, {\it i.e.} into the region
$a^2 V(\phi)>1 $, a complex solution is necessary in order to get
there. This means that the wave function becomes oscillatory in
this region. Suppose however, one follows the caustic to smaller
values of $\phi$. It departs from the curve $a^2 V(\phi)=1 $, and
in fact has a singularity at $\phi= \phi_*$, breaking into two
branches there. This seems to invalidate the form of the wave
function used above, and in fact, Grishchuk and Rozhansky claimed
that it implies that the wave function fails to predict the
emergence of any real Lorentzian trajectories for $\phi< \phi_*$.
Moreover, their analysis also applies, they claim, to the
tunneling wave function.

What this all means is that the conditions used above in the
calculation of the probability of sufficient inflation should be
replaced by the conditions $ \phi_* < \phi < \phi_p $. Most
importantly the region very close to $\phi=\phi_{min}$, in which
the no-boundary and tunneling wave function differ most severely,
is excised. This has the consequence that the predictions of these
two wave functions are not as different as previously believed.
Although the predictions of the tunneling wavefunction are little
affected by this result, for the no-boundary wave function it is
now not so obvious that $P<<1$. In particular, what one would hope
to find is that $\phi_*> \phi_{suf}$. This would have the
consequence that {\it all} the classical Lorentzian solutions the
wave function corresponds to have sufficient inflation; thus
sufficient inflation would be predicted with probability $1$,
irrespective of whether an upper cut-off is imposed. The value of
$\phi_*$ is, however, model dependent, and a model for which
$\phi_*> \phi_{suf}$ is yet to be found.\footnote{$^{\dag}$} {More
detailed calculations with the massive scalar field model, for
which $\phi_* \sim 1 $ and $\phi_{suf} \sim 4$ indicate that the
previous conclusions ({\it i.e.} pre-Grishchuk-Rozhansky)
concering the probability of inflation are in fact largely
unaffected (J.Fort, private communication).}

This is an interesting development which deserves further study.

\def\x{{\bf x}}
\def\a{{\alpha}}

\head{\bf 10. BEYOND MINISUPERSPACE}

For most of these lectures, we have largely concentrated on the
application of the formalism of quantum cosmology to
minisuperspace models. These models, with the appropriate boundary
conditions, have been reasonably successful -- in predicting
inflation, for example. However, the universe we see today is not
exactly described by the homogeneous metrics of the type
considered in minisuperspace models. There are local deviations
from homogeneity because matter is clumped into galaxies and other
large scale structures. In conventional galaxy formation
scenarios, this large scale structure can arise as a result of
small density perturbations $\delta \rho / \rho  \sim 10^{-4} $ in
an otherwise homogeneous universe at very early times. The hot big
bang model offered no explanation as to the origin of these
perturbations, but had to assume them as initial conditions. The
inflationary universe scenario shed considerable light on the
situation by showing that they could have arisen from
pre-inflationary quantum fluctuations in the scalar field hugely
amplified by inflation. To be more precise, the density
fluctuations in inflationary universe models are calculated from a
quantity of the form $ \langle 0 | \Phi^2 |0 \rangle $, using
standard methods of quantum field theory in a curved (usually de
Sitter-like) spacetime. However, a point that was not emphasized
in the early studies of this problem is that the form and
magnitude of the density fluctuations calculated in this way
depend rather crucially on the particular vacuum state $|0\rangle$
one uses, and in most curved spacetimes, there is no unique
natural choice. Since this is clearly a question of initial
conditions, one would expect to gain new insight into this issue
from the perspective of quantum cosmology. It is therefore of
considerable interest to go beyond minisuperspace to the full,
infinite dimensional superspace. It would of course be very
difficult to do this in full generality, but for the purposes of
describing density fluctuations and gravitational waves, it is
sufficient consider linearized fluctuations about a homogeneous
isotropic minisuperspace background. This is the subject of this
section.

We will find that there are two things that come out of this.
Firstly, we will see that in the semi-classical limit, quantum
cosmology reduces to the familiar formalism of quantum field
theory for the fluctuations on a classical minisuperspace
background. Secondly, the boundary conditions on the wave function
of the universe imply a particular choice of vacuum state for the
quantum fields.

\subhead {\bf Quantum Field Theory in Curved Spacetime}

Before going on to study perturbations about minisuperspace in
quantum cosmology, let us begin by reviewing some basic aspects of
quantum field theory in curved spacetime (see, for example,
Birrell and Davies (1982)). For definiteness, let us consider
scalar field theory described by the action
$$
S_m = - \half \int d^4 x \sqrt{-g} \left[ (\partial \Phi)^2 +m^2
\Phi^2) \right] \eqno(10.1)
$$
This theory is normally quantized in the Heisenberg picture in a
given background by introducing a set of mode functions
$u_k(\x,t)$ satisfying a wave equation of the form
$$
\left( \square - m^2 \right) u_k(\x,t) = 0 \eqno(10.2)
$$
The field operator $\hat \Phi$ is then expanded in terms of these
mode functions
$$
\hat \Phi(\x,t) = \sum_k \left( \hat a_k u_k(\x,t) + \hat
a_k^{\dag} u_k^*(\x,t) \right) \eqno(10.3)
$$
where $ \hat a_k^{\dag} $ and $ \hat a_k$ are the usual creation
and annihilation operators. The vacuum state is then defined to be
the state $ |0\rangle $ for which
$$
\hat a_k |0\rangle = 0 \eqno(10.4)
$$
The vacuum state is determined by the choice of mode functions
$u_k$.

In Minkowski space, there is a unique vacuum state which is
invariant under the Poincare group, and so is the agreed vacuum
state for all inertial observers. However, in an arbitrary curved
spacetime, there is no unique vacuum state. Any expectation value
will generally depend rather crucially on the particular choice of
state.

There is another perhaps less familiar way of doing quantum field
theory in curved spacetime which is closer to quantum cosmology
than the Heisenberg picture outlined above. This is the functional
Schrodinger picture (Brandenberger, 1984; Burges, 1984;,
Floreanini et al., 1987; Freese et al., 1985; Guth and Pi, 1985;
Ratra, 1985). This picture is based very much on the $(3+1)$
decomposition we also used for quantum cosmology earlier. The
$(3+1)$ form of the scalar field action (10.1) (in the gauge
$N^i=0$) is
$$
S_m = \half \int d^3x dt Nh^{\half} \left[ {\dot \Phi^2 \over N^2}
- h^{ij} \partial_i \Phi \partial_j \Phi - m^2 \Phi^2 \right]
\eqno(10.5)
$$
Defining canonical momenta $ \pi_{\Phi} $ in the usual way, one
readily derives the Hamiltonian
$$
H_m = \half \int d^3x Nh^{\half} \left[ h^{-1} \pi_{\Phi}^2 +
h^{ij} \partial_i \Phi \partial_j \Phi + m^2 \Phi^2 \right]
\eqno(10.6)
$$
In the functional Scr\"odinger quantization, the quantum state of
the scalar field is represented by a wave functional
$\Psi_m[\Phi(\x),t]$, a functional of the field configuration $
\Phi(\x) $ on the surface $t= constant$. The evolution of the
quantum state is governed by the functional Schr\"odinger equation
$$
i {\partial \Psi_m \over \partial t} = H_m \Psi_m \eqno(10.7)
$$
where the operator appearing on the right-hand side is the
Hamiltonian (10.6) with the momenta replaced by operators in the
usual way,
$$
\pi_{\Phi} (\x) \rightarrow -i {\delta \over \delta \Phi(\x)}
\eqno(10.8)
$$

There are two differences between the representation of states in
the two picture outlined above. Firstly, Heisenberg picture states
are time-independent, whereas Schr\"odinger picture states are not
(at least, in the flat space case -- in curved backgrounds
Heisenberg states may acquire time-dependence through the
gravitational field). They are related by
$$
| \Psi_S (t) \rangle = \exp\left( - i \int^t dt' H_m(t') \right) |
\Psi_H \rangle \eqno(10.9)
$$
Secondly, the Schr\"odinger picture states are represented at each
moment of time by wave functionals $\Psi[\Phi(\x)]$ rather than
abstract Hilbert space elements $|\Psi \rangle $. The relationship
between these two is found by introducing a complete set of {\it
field states} $ |\Phi(\x)\rangle $, defined to be the eigenstates
of the field operator $\hat \Phi $ at a moment of time
$$
\hat \Phi |\Phi(\x)\rangle  = \Phi(\x) |\Phi(\x)\rangle
\eqno(10.10)
$$
The wave functionals $\Psi[\Phi(\x)]$ are then defined to be the
coefficients in the expansion of the abstract Hilbert space
elements in terms of the complete set of field states:
$$
|\Psi_S\rangle = \int {\cal D} \Phi(\x) \ |\Phi(\x)\rangle \
\langle \Phi(\x)| \Psi_S\rangle
$$
$$
\equiv \int {\cal D} \Phi(\x) \ |\Phi(\x)\rangle \
\Psi_S[\Phi(\x)] \eqno(10.11)
$$

The question of choosing a vacuum state $|0\rangle $ in the
Heisenberg picture becomes the question of choosing a solution to
the functional Schr\"odinger equation (10.7) in the functional
Schr\"odinger picture.

With these preliminaries in mind, let us now turn to perturbations
about minisuperspace.

\subhead {\bf Inhomogeneous Perturbations about Minisuperspace}

Now we will study inhomogeneous perturbations about
minisuperspace. We primarily follow Halliwell (1987b), Halliwell
and Hawking (1985), and Hartle (1986), but many more references
are given in Section 13. To see how this works, it is simplest to
consider a particular example. Namely, we will consider
perturbations about the scalar field model considered earlier.
There, the minisuperspace ansatz involved writing
$$
h_{ij} = e^{2\alpha} \Omega _{ij} , \quad  \Phi (\x, t) = \phi(t)
$$
$$
N(\x,t) = N_0(t), \quad N^i(\x,t) = 0 \eqno(10.12)
$$
where $\Omega_{ij}$ is the metric on the unit three-sphere. To go
beyond this perturbatively we write
$$
h_{ij}=e^{2\alpha} \left(\Omega_{ij}+\epsilon_{ij}\right) ,\
\Phi(\x,t) =\phi(t)+\delta\phi(\x, t)
$$
$$
N(\x,t) = N_0(t) + \delta N(\x,t) \eqno(10.13)
$$
and in addition, we allow non-zero $N^i(\x,t)$, which is regarded
as a small perturbation. The easiest way to deal with the
inhomogeneous perturbations is to expand in harmonics on the
three-sphere. So, for example, one writes the scalar field
perturbation as
$$
\delta\phi(\x,t)=\sum_{nlm} f_{nlm}(t) Q_{lm}^n (\x) \eqno(10.14)
$$
where $ Q_{lm}^n $ are three-sphere harmonics. They satisfy
$$
^{(3)}\Delta Q_{lm}^n = - (n^2-1) Q_{lm}^n \eqno(10.15)
$$
where $^{(3)}\Delta$ is the Laplacian on the three-sphere. The sum
in (10.14) excludes the homogeneous mode, $ n=1 $. The details of
this expansion are not important in what follows, and may be found
in Halliwell and Hawking (1985).

Inserting the above ansatz into the Einstein-scalar action, and
expanding to quadratic order in the perturbations, one obtains a
result of the form
$$
S[g_{\mu\nu}, \Phi] = S_0[q^{\a},N_0] +
S_2[q^{\a},N_0,\epsilon_{ij}, \delta \phi, \delta N, N^i]
\eqno(10.16)
$$
where as before, we use $q^{\alpha}$ to denote the minisuperspace
coordinates. $S_0$ is the original minisuperspace action and $S_2$
is the action of the perturbations, and is quadratic in them. The
total Hamiltonian following from (10.16) is then found to be of
the form
$$
H_T = N_0 \left( H_0 + \int d^3 x \ {\cal H}_{(2)} \right)
$$
$$
+ \int d^3 x \ \delta N(\x) \ {\cal H}_{(1)} (\x) + \int d^3 x \
N^i(\x) \ {\cal H}_i (\x) \eqno(10.17)
$$
From this one may see that first of all, there is a non-trivial
momentum constraint at every point $\x$ in the three-surface
$$
{\cal H}_i(\x) = 0 \eqno(10.18)
$$
It is linear in the perturbations. Secondly, the Hamiltonian
constraint has split into two parts. There is a part linear in the
perturbations, at each point $\x$ of the three-surface,
$$
{\cal H}_{(1)} (\x) = 0 \eqno(10.19)
$$
and there is a part consisting of the original minisuperspace
Hamiltonian plus a term quadratic in the perturbations integrated
over the three-surface:
$$
H_0 + \int d^3 x \ {\cal H}_{(2)}  \equiv H_0 + H_2 = 0
\eqno(10.20)
$$

Quantization procedes by introducing a wave function $\Psi(q^{\a},
\epsilon_{ij}, \delta \Phi) $ and insisting that it be annihilated
by the operator versions of the constraints, (10.18)-(10.20). The
procedure is complicated by the fact that, in addition to the
invariance under reparametrizations present in the minisuperspace
case, the perturbations involve gauge degree of freedom. There are
numerous ways of dealing with this. For example, the gauge degrees
of freedom for the perturbations generated by (10.18) and (10.19)
may be fixed classically, and the constraints solved for the
physical degrees of freedom of the perturbations. This then leaves
only the constraint (10.20), which now depends only on the
minisuperspace variable and the unconstrained physical degrees of
freedom of the metric and matter perturbations. One way or
another, (10.20) ends up being the most important equation, and it
is this that we now concentrate on.

A useful example to bear in mind is the case of purely scalar
field perturbations about a purely gravitational background
consisting of a Robertson-Walker metric with scale factor
$e^{\a}$. Then the Hamiltonian $H_2$ is given by
$$
H_2=\sum_{nlm} {1\over 2} e^{-3\alpha} \left[-{\partial^2\over
\partial f^2_{nlm}} + (m^2 e^{6\alpha} + (n^2-1)e^{4\alpha} )
f_{nlm}^2\right] \eqno(10.21)
$$
after expansion in harmonics.

The Wheeler-DeWitt equation resulting from (10.20) is of the form
$$ \left[-{1\over 2m_p^2} \nabla^2 +m_p^2 U(q)+H_2\right]\Psi(q,\Phi)=0
\eqno(10.22)
$$
For convenience, we will consider only scalar field perturbations
$\Phi$, but what follows is equally applicable to the case of
gravitational wave perturbations. The operator $\nabla$ operates
only on $q^{\a}$, not on the perturbations. We are interested in
the solution to the Wheeler-DeWitt equation in the region of
superspace where the minisuperspace variables $q^{\a}$ are
approximately classical, but the perturbations may be quantum
mechanical. We therefore look for solutions of the form
$$
\Psi(q, \Phi)=\exp\left(im_p^2 S_0(q)+i S_1(q)\right)\
\psi(q,\Phi) + O(m_p^{-2}) \eqno(10.23)
$$
where $S_0(q)$ is real,\footnote{$^{\dag}$} {As in Section 7, we
could also allow a slowly varying exponential prefactor, but this
may be absorbed into the definition of $S_1$} but $S_1$ and $\psi$
may be complex. Inserting (10.23) into (10.22), and equating
powers of the Planck mass, one obtains the following. At lowest
order, once again one gets the Hamilton-Jacobi equation for $ S_0
$,
$$
{1\over 2} (\nabla S_0)^2+U(q)=0. \eqno(10.24)
$$
This shows that, to lowest order in $m_p^2$, the wave function
(10.23) is, as in the minisuperspace case, peaked about the
ensemble of solutions to the classical field equations with
Hamilton-Jacobi function $S_0$. It is convenient to introduce the
tangent vector to these classical solutions,
$$
{\partial \over \partial t} = \nabla S_0 \cdot \nabla \eqno(10.25)
$$

At the next order, one obtains the equation:
$$
\psi \left[ \nabla S_0 \cdot \nabla S_1 - {i \over 2} \nabla^2 S_0
\right] = i \nabla S_0 \cdot \nabla \psi - H_{2} \psi \eqno(10.26)
$$
This is one equation for the two unknowns $S_1$ and $\psi$, so
there is the freedom to impose some restrictions on them. We are
anticipating that the $\psi$ will be matter wave functionals for
the scalar field $\Phi$. Let us therefore introduce an inner
product between matter wave functionals, at each point of
minisuperspace, $q^{\a}$:
$$
(\psi_1, \psi_2) = \int {\cal D} \Phi(\x) \ \psi_1^*(q,\Phi(\x)) \
\psi_2(q,\Phi(\x)) \eqno(10.27)
$$
Note that this involves an integral only over $\Phi(\x)$ and {\it
not} over the minisuperspace variables $q^{\a}$. This is
acceptable because we expect the appropriate matter wave
functionals to be normalizable in the matter modes. We do not,
however, expect any part of the wave function to be normalizable
in the large, minisuperspace modes, so we do not attempt to
introduce an inner product involving an integral over $q^{\a}$.
Using the freedom available in $\psi$, let us demand that
$$
{d \over dt} (\psi,\psi) = 0 \eqno(10.28)
$$
That is, the norm of $\psi$ is preserved along the classical
minisuperspace trajectories. This seems like a reasonable
restriction if we are to recover quantum field theory for matter.
We may therefore take $(\psi,\psi)=1$. Differentiating out
(10.28), it is readily seen that
$$
\left( i {\partial \psi \over \partial t}, \psi \right) = \left(
\psi, i {\partial \psi \over \partial t} \right) \eqno(10.29)
$$
and hence that this quantity is real.

Armed with this information, let us now return to (10.26). Taking
the inner product of (10.26) with $\psi$, and making the
reasonable assumption that the perturbation Hamiltonian $H_2$ is
hermitian in the above inner product, we quickly discover that,
apart from the $\psi$, the left-hand side must be real. Its
imaginary part must therefore vanish,
$$
\nabla S_0 \cdot \nabla (ImS_1) - \half \nabla^2 S_0 = 0
\eqno(10.30)
$$
If we write $C=\exp(- ImS_1)$, then $C$ is the usual real
minisuperspace WKB prefactor, obeying (6.26), and is unaffected by
the perturbations.

Subtracting (10.30) from (10.26), and using the definition
(10.25), one obtains the following equation for $\psi$:
$$
i {\partial \psi \over \partial t} = \left[ H_2 + {\partial \over
\partial t} ( Re S_1 ) \right] \psi \eqno(10.31)
$$
Finally, by writing $\tilde \psi = e^{iReS_1} \psi $, we discover
that $\tilde \psi$ obeys the functional \break Schr\"odinger
equation along the classical trajectories in minisuperspace about
which the wave function is peaked:
$$
i{\partial\tilde \psi\over\partial t} =H_2\tilde\psi \eqno(10.32)
$$

This derivation may be concisely summarized as follows: the WKB
solution to the Wheeler-DeWitt equation (10.22) is of the form
$$
\Psi (q,\Phi) = C(q) e^{im_p^2 S_0(q)} \tilde \psi (q,\Phi)
\eqno(10.33)
$$
where $S_0(q)$ is a solution to the Hamilton-Jacobi equation,
indicating that the wave function to leading order is peaked about
a set of classical trajectories, $C(q)$ is the usual unperturbed
WKB prefactor, and $\tilde \psi$ satisfies the functional
Sch\"odinger equation (10.32) along the classical trajectories
about which the wave function is peaked.

What we have shown, therefore, is that the Wheeler-DeWitt equation
reduces, in the semi-classical limit, to the familiar formalism of
quantum field theory for the fluctuations $ \Phi $ in a fixed
classical background. This shows that quantum cosmology is
consistent with the standard approach, which involves quantum
field theory on a fixed background. See Section 13 for references
to the large number of papers on this issue.

\def\x{{\bf x}}

\head {\bf 11. VACUUM STATES FROM QUANTUM COSMOLOGY}

We have shown that quantum cosmology reduces, in the
semi-classical limit, to the formalism of quantum field theory for
the matter modes in a fixed curved spacetime background. So far we
have therefore done little new, except to demonstrate consistency
with that we already know. However, there is a bonus. Boundary
conditions on the wave function define a particular solution to
the Wheeler-DeWitt equation of the form (10.33), where $\tilde
\psi$ is a solution to the functional Schr\"odinger equation for
the perturbations. This means that boundary conditions on the wave
function of the universe will pick out a particular solution to
the functional Schr\"odinger equation; that is, {\it they define a
particular vaccum state for matter}, with which to do quantum
field theory.

The natural question to ask now, is what is the nature of the
vacuum state picked out by the no-boundary and tunneling boundary
conditions in a given background? The background of particular
interest as far as inflation is concerned is de Sitter space, or
spacetimes that are very nearly de Sitter. For that background it
may be shown that the vacuum state defined by both of these
proposals is a vacuum state known as the ``Euclidean" or
``Bunch-Davies" vacuum. This is the vacuum state that is often
assumed when calculating density fluctuations, and leads to a
reasonable spectrum for the emergence of large scale structure.

Before seeing exactly how the above proposals define this vacuum
state, let us first explain how it is defined.\footnote{$^{\dag}$}
{For a useful discussion of de Sitter-invariant vacua, see Allen
(1985), and references therein.}

\subhead {\bf De Sitter-Invariant Vacua}

Minkowski space has as its isometry group the 10 parameter
Poincare group. There is a vacuum which is invariant under this
group, and thus is the agreed vacuum for all inertial observers.
It is unique, up to trivial Bogoliubov transformations. The
isometry group of de Sitter space, which also has 10 parameters,
is the de Sitter group, $SO(4,1)$. In choosing vacuum states with
which to do quantum field theory in de Sitter space, it is
therefore natural to seek vacua invariant under the de Sitter
group.

A convenient way of characterizing vacua is through the symmetric
two-point function in a state $ |\lambda \rangle $:
$$
G_{\lambda} (x,y) = \langle \lambda | ( \Phi(x) \Phi(y) + \Phi(y)
\Phi(x) ) | \lambda \eqno(11.1)
$$
The state $|\lambda \rangle$ is then said to be de Sitter
invariant if the two-point function depends on $x$ and $y$ only
through $\mu(x,y)$, the geodesic distance between $x$ and $y$:
$$
G_{\lambda} (x,y) = f_{\lambda}(\mu) \eqno(11.2)
$$
Using the fact that $\Phi$ obeys the Klein-Gordon equation, a
second order ordinary differential equation for $ f_{\lambda}(\mu)
$ is readily derived. From it, it may be shown that there is not
just one de Sitter-invariant vacuum, but there is a one-parameter
family of inequivalent de Sitter-invariant vacua.

For this one-parameter family, the function $f_{\lambda}(\mu)$
generally has two poles: one when $y$ is on the light-cone of $x$,
the other when $y$ is on the light cone of $\bar x$, the point in
de Sitter space antipodal to $x$. However, amongst the
one-parameter family, there is one member for which
$f_{\lambda}(\mu)$ has just one pole, when $y$ is on the
light-cone of $x$. This member is called the ``Euclidean" or
``Bunch-Davies" vacuum, and has the nicest analytic properties. As
mentioned above, it is this one that is always used in
calculations of density fluctations in inflationary universe
models.

There is another equivalent way of characterizing the Euclidean
vacuum that will be most convenient for our purposes. This is a
definition in terms of a particular choice of mode functions.
Suppose we expand the scalar field operator in terms of a set of
modes functions $\{u_{nlm}(\x,t)\}$, say,
$$
\hat \Phi (\x,t) = \sum_{nlm} \left( u_{nlm}(\x,t) \hat a_{nlm} +
u_{nlm}^*(\x,t) \hat a_{nlm}^{\dag} \right) \eqno(11.3)
$$
The vacuum state $|0\rangle$ corresponding to this particular
choice of mode functions is defined by
$$
\hat a_{nlm} | 0 \rangle =0 \eqno(11.4)
$$
To define the Euclidean vacuum, one first chooses the mode
functions
$$
u_{nlm} (\x,t) = y_n(t) Q^n_{lm} (\x) \eqno(11.5)
$$
where the $ Q^n_{lm} (\x) $ are three-sphere harmonics, and the
$y_n(t)$ satisfy the equation
$$
\ddot y_n + 3 {\dot a \over a} \dot y_n + \left( {n^2-1 \over a^2}
+m^2 \right) y_n = 0 \eqno(11.6)
$$
Here, $a(t)=H^{-1} \cosh (Ht) $ is the scale factor for de Sitter
space. The normalization of the $y_n(t)$ is fixed through the
Wronskian condition
$$
y_n \dot y_n^* - y_n^* \dot y_n = {i \over a^3} \eqno(11.7)
$$
The Euclidean section of de Sitter space is the four-sphere, and
may be obtained by writing $ t = -i (\tau - {\pi \over 2H}) $,
which turns $a(t)=H^{-1} \cosh(Ht)$ into $a(\tau) = H^{-1}
\sin(H\tau)$. The Euclidean vacuum is then defined by the
requirement that the $y_n(t)$ are {\it regular} on the Euclidean
section. The $y_n(t)$ actually become real on the Euclidean
section, so one may equivalently demand that the $y_n^*(t)$ are
regular there.

There is a third possible way of dicussing de Sitter-invariant
vacua, which is conceptually the most transparent way. This is to
explicitly construct the de Sitter generators and demand that the
state be annihilated by them, but we will not consider this here
(Burges, 1984; Floreanini et al., 1986).

\subhead {\bf The No-Boundary Vacuum State}

Now let us explicitly calculate the matter state wave functional
for a massive minimally coupled scalar field in a de Sitter
background, using the no-boundary proposal. We follow Laflamme
(1987a). We regard all the modes of the scalar field, including
the homogeneous mode, as perturbations on a homogeneous isotropic
background with scale factor $a(t)$, driven by a cosmological
constant. The no-boundary wave function is given by a path
integral of the form
$$
\Psi_{NB}(\tilde a, \tilde \Phi) = \int {\cal D} g_{\mu\nu} {\cal
D} \Phi \exp\left( - I_g[g_{\mu\nu}] - I_m[g_{\mu\nu},\Phi]
\right) \eqno(11.8)
$$
In the saddle-point approximation to the integral over metrics,
this leads to an expression of the form
$$
\Psi_{NB} (\tilde a, \tilde \Phi) \approx \exp\left( -I_g[\bar
g_{\mu\nu}] \right) \int {\cal D} \Phi \exp \left( - I_m[ \bar
g_{\mu\nu}, \Phi] \right) \eqno(11.9)
$$
where $\bar g_{\mu\nu}$ is the saddle-point metric. When $a H<1$,
$\bar g_{\mu\nu}$ is real and is the metric on the section of
four-sphere closing off a three-sphere of radius $a$. When $ a H>1
$, $\bar g_{\mu\nu}$ is complex, and corresponds to a section of
de Sitter space with minimum radius $a$ matched onto half a
four-sphere at its maximum radius.

Comparing (11.9) with (10.23), one may see that the matter wave
functionals are given by the path integral
$$
\psi[\tilde a, \tilde \Phi] = \int {\cal D} \Phi \exp \left( -
I_m[ \bar g_{\mu\nu}, \Phi] \right) \eqno(11.10)
$$
The no-boundary proposal implies that the integral over matter
modes is over fields $\Phi(\x,\tau)$ that match $\tilde \Phi(\x)$
on the three-sphere boundary. As in Section 8, we shall demand
that the saddle-point of the functional integral over $\Phi$ in
(11.10) corresponds to a {\it regular} solution to the scalar
field equation on the given background geometry.

The scalar field is most easily handled by expanding in
three-sphere harmonics
$$
\Phi(\x,\tau) = \sum_{nlm} f_{nlm}(\tau) Q^n_{lm}(\x) \eqno(11.11)
$$
In terms of the coefficients $f_{nlm}(\tau)$, the Euclidean action
is
$$
I_m[a(\tau), \Phi] = \half \sum_{nlm} \int_0^1 d \tau Na^3 \left[
{1\over N^2} \left( {df_{nlm} \over d \tau} \right)^2 + \left(
{n^2-1 \over a^2} + m^2 \right) f^2_{nlm} \right]
$$
$$
\equiv \sum_{nlm} I_{nlm} [a(\tau), f_{nlm}] \eqno(11.12)
$$
The Euclidean field equations are
$$
{d^2f_{nlm} \over d\tau^2} + {3 \over a} {da \over d \tau} {d
f_{nlm} \over d \tau} - N^2 \left( {n^2-1 \over a^2} + m^2 \right)
f_{nlm} = 0 \eqno(11.13)
$$
Here, $a(\tau),N$ is the solution to the field equation and
constraint for the background satsifying $a(0)=0$, $a(1)=\tilde
a$. Explicitly,
$$
a(\tau) = {1 \over H} \sin (NH \tau) , \quad N = {1 \over H}\left(
{\pi \over 2} - \cos^{-1}(\tilde a H) \right) \eqno(11.14)
$$

The solutions to (11.13) may be written down explicitly in terms
of hypergeometric functions, although this is not necessary for
our purposes. They are regular everywhere, with the possible
exception of the region near $ \tau =0$. In this region, $a(\tau)
\sim N \tau$, and it is easily shown that the solutions to (11.13)
behave like $\tau^{-n-1}$, or $\tau^{n-1}$. Clearly only one of
these is regular. It may be picked out by imposing the initial
condition
$$
f_{nlm}(0) = 0, \quad {\rm for } \quad n=2,3,..., \quad {\rm and}
\quad {d f_{nlm} \over d \tau} (0) = 0, \quad {\rm for} \quad n=1.
\eqno(11.15)
$$
These are the initial conditions on the histories implied by the
no-boundary proposal. The histories also satisfy the final
condition
$$
f_{nlm}(1) = \tilde f_{nlm} \eqno(11.16)
$$

Because the modes decouple, we may write
$$
\psi [\tilde a, \tilde \Phi(\x)] = \prod_{nlm} \psi_{nlm} (\tilde
a, \tilde f_{nlm} ) \eqno(11.17)
$$
From (11.10) it then follows that
$$
\psi (\tilde a, \tilde f_{nlm}) = \int {\cal D} f_{nlm}
e^{-I_{nlm}} \eqno(11.18)
$$
Because $I_{nlm}$ is quadratic in the scalar field modes, the path
integral (11.18) may be evaluated exactly to yield an expression
of the form
$$
\psi (\tilde a, \tilde f_{nlm}) = A_{nlm}(\tilde a) \exp\left( -
\bar I_{nlm}(\tilde a, \tilde f_{nlm} ) \right) \eqno(11.19)
$$
Here, $ \bar I_{nlm}(\tilde a, \tilde f_{nlm} )$ is the action of
the solution to the Euclidean field equations satisfying the
boundary conditions (11.15), (11.16). Let us denote this solution
by $g_n(\tau)$. It is independent of $l,m$, because the field and
equations and boundary conditions are. Then it is readily shown
that
$$
I_{nlm}(\tilde a, \tilde f_{nlm}) = \half \left[ a^3(\tau)
g_n(\tau) { d g_n(\tau) \over d \tau} \right]_0^1 = \half \tilde
a^3 \tilde f^2_{nlm} \left[ {1 \over g_n } {d g_n \over d \tau}
\right]_{\tau=1} \eqno(11.20)
$$
The matter wave functional defined by the no-boundary proposal is
therefore given by (11.18), with
$$
\psi_{nlm} (\tilde a, \tilde f_{nlm}) = A_{nlm}(\tilde a)
\exp\left( - \half \tilde a^3 \tilde f^2_{nlm} \left[ {1 \over g_n
} {d g_n \over d \tau} \right]_{\tau=1} \right) \eqno(11.21)
$$
The key point to note is that it involves the expression $\dot
g_n/ g_n$, evaluated at the upper end-point, where the $g_n(\tau)$
are solutions to the field equations which are {\it regular} on
the Euclidean section.

We now need to show that this matter wave functional corresponds
to the Euclidean vacuum state defined above. This basically
involves determining what the vacuum state $|0\rangle$ defined by
(11.4) looks like in the functional Schr\"odinger picture. To this
end, first compare the expansions (11.3) and (11.11) of the scalar
field. Turning (11.11) into an operator, one may therefore write
$$
\hat f_{nlm} (t) = y_n(t) \hat a_{nlm} + y_n^*(t) \hat
a_{nlm}^{\dag} \eqno(11.22)
$$
The momentum operator conjugate to this is
$$
\hat \pi_{nlm} (t) = a^3 {\dot {\hat f}}_{nlm} = a^3 \dot y_n(t)
\hat a_{nlm} + a^3 \dot y_n^*(t) \hat a_{nlm}^{\dag} \eqno(11.23)
$$
(11.22) and (11.33) are readily inverted to yield
$$
\hat a_{nlm} = -i y_n^* \left( a^3 {\dot y_n^* \over y_n^*} \hat
f_{nlm} -\hat \pi_{nlm} \right) \eqno(11.24)
$$
By inserting a complete set of field states $\{|f_{nlm}\rangle\}$
in (11.4), we thus obtain the following equation for the vacuum
state $\psi_{nlm}(f_{nlm}) \equiv \langle f_{nlm}|0 \rangle $:
$$
\left( a^3 {\dot y_n^* \over y_n^*} f_{nlm} +i {\partial \over
\partial f_{nlm} } \right) \psi_{nlm} (f_{nlm}) = 0 \eqno(11.25)
$$
It is readily solved to yield
$$
\psi_{nlm} = \exp\left(  {i \over 2} a^3 {\dot y_n^* \over y_n^*}
f^2_{nlm} \right) \eqno(11.26)
$$
This, therefore, is the Euclidean vacuum in the functional
Schr\"odinger picture. Going to the Euclidean section, one thus
obtains
$$
\psi_{nlm} = \exp\left(- \half a^3 {1 \over y_n^*} {d y_n^* \over
d \tau} f^2_{nlm} \right) \eqno(11.27)
$$
The equivalence of (11.27) and (11.21) immediately follows from
the definition of the Euclidean vacuum, which is that the $y_n$,
and hence the $y_n^*$, are solutions to the field equations which
are regular on the Euclidean section. This completes the
demonstration that the vacuum state defined by the no-boundary
proposal is the de Sitter-invariant Euclidean vacuum.

A more heuristic argument for the de Sitter invariance of the
no-boundary matter wave functionals may also be given. This
argument shows that the de Sitter invariance is an inevitable
consequence of the very geometrical nature of the no-boundary
proposal, and is therefore true of most types of matter fields
(D'Eath and Halliwell, 1987).

Suppose one asks for the quantum state of the matter field on a
three-sphere of radius $a<H^{-1}$. The no-boundary state is
defined by a path integral of the form (11.10). One sums over all
matter fields regular on the section of four-sphere interior to
the three-sphere which match the prescribed data on the
three-sphere boundary. The resulting state will depend on the
geometry only through the radius of the three-sphere, and not on
its intrinsic location or orientation on the four-sphere. One thus
has the freedom to move the three-sphere around on the four-sphere
without changing the quantum state -- at each location one is
summing over exactly the same field configurations to define it.
These different locations are related to each other by the
isometry group of the four-sphere, $SO(5)$. It follows that the
state is $SO(5)$-invariant on the Euclidean section. On
continuation back to the Lorentzian section, one thus finds that
the state is invariant under $SO(4,1)$, the de Sitter group; that
is, the state is de Sitter invariant. This argument may be made
mathematically precise, although we will not go into that here.

It may be shown that the tunneling wave function picks out the
same vacuum state. This follows essentially from the imposition of
a regularity requirement on the matter wave functionals
(Vachaspati, 1989; Vachaspati and Vilenkin, 1988; Vilenkin, 1988).

\head {\bf 12. SUMMARY }

The purpose of these lectures has been to describe the route from
a quantum theory of cosmological boundary conditions to a
classical universe with the potential for evolving into one
similar to that in which we live.

We began in Section 2 with a brief introductory tour of quantum
cosmology by way of a simple example. This simple model
illustrated the need for a quantum theory of initial conditions.
The general formalism of quantum cosmology was briefly outlined in
Sections 3 and 4. The full theory is very difficult to handle in
practice, so in Section 5, we restricted to the case of
minisuperspace models. The canonical and path integral formalism
for minisuperspace models was described. In Section 6, we
discussed the most important prediction a quantum theory of
cosmology should make -- the emergence of classical spacetime. The
emergence of classical spacetime is very much contingent on
boundary conditions on the wave function, and occurs only in
particular regions of configuration space. These ideas were
further developed in Section 7, in which the WKB approximation was
described. Wave functions of oscillatory WKB form correspond to
classical spacetime in that they are peaked about a set of
classical solutions to the Einstein equations. Moreover, this set
of solutions is a subset of the general solution; thus boundary
conditions on the wave function of the universe effectively imply
initial conditions on the set of classical solutions. We discussed
the way in which the wave function may be used to construct a
measure on this set of classical solutions.

In Section 8, certain boundary condition proposals were described
-- the no-boundary proposal of Hartle and Hawking, and the
tunneling boundary condition of Linde and of Vilenkin. Each of
these proposals suffers from imprecision or lack of generality,
although with a certain amount of license, each may be
successfully used to calculate wave functions in simple models. We
calculated the no-boundary and tunneling wave functions for the
scalar field model introduced in Section 2. These wave functions
were compared in Section 9. The two wave functions are peaked
about the same set of classical solutions, but they give rather
different measures on this set of solutions. In particular, they
may give very different values for the likelihood of sufficient
inflation. The comparison of these two wave functions was
inconclusive, but this merely reflects the fact that no consensus
of opinion has yet emerged.

In Sections 10 and 11 we described how one goes beyond
minisuperspace by considering inhomogeneous perturbations. There
are two things that come out of this. First, one finds that in the
limit in which gravity becomes classical, one recovers quantum
field theory for the perturbations in a fixed classical
gravitational background. Secondly, boundary conditions on the
wave function of the universe are found to imply a particular
choice of vacuum state for the perturbations. In particular, in
the case of a de Sitter background, the no-boundary and tunneling
proposals pick out the de Sitter-invariant Euclidean vacuum. The
density perturbations arising from this particular choice are of
the correct form for the subsequent emergence of large scale
structure.

Finally, I would like to emphasize the rather open-ended nature of
many of the issues in quantum cosmology covered in these lectures.
One might get the impression from reading the literature on the
subject that certain aspects of the field are complete and neatly
tied up beyond criticism. In my opinion this is most certainly not
the case, and I have tried to indicate areas of difficulty at the
appropriate points throughout the text. There is, I believe,
considerable scope for development and improvement in many parts
of the field. For example, the methods used in quantum cosmology
to extract predictions from the wave function, as described in
Section 6, are rather crude, and it would be much more satisfying
to apply methods such as those described by Hartle in his lectures
(Hartle, 1990). Another example concerns the use of the path
integral in quantum cosmology. Although the role it plays is
supposedly very central, especially in the formulation of the
no-boundary proposal, it is I think reasonable to say that, with
but a few exceptions, its use in quantum cosmology has been for
the most part rather heuristic. A more careful approach using the
path integral in a serious way would very desirable. Furture
investigation of these and other issues is likely to be very
profitable.

\head{\bf ACKNOWLEDGEMENTS}

I am very grateful to numerous people for useful conversations and
for comments on early drafts of the manuscript, including Bruce
Allen, Dalia Goldwirth, Jim Hartle, Jorma Louko, Robert Myers, Don
Page, Tanmay Vachaspati and Alex Vilenkin. I would also like to
thank Sidney Coleman, Jim Hartle and especially, Tsvi Piran, for
organizing the school.

This work was supported in part by funds provided by the U.S.
Department of Energy (D.O.E.) under contract No.
DE-AC02-76ER03069.

\head{\bf 13. A GUIDE TO THE LITERATURE}

\subhead {\bf General}

Some of the earlier works in the field of quantum cosmology
include those of DeWitt (1967), Misner (1969a, 1969b, 1969c, 1970,
1972, 1973) and Wheeler (1963, 1968). Early reviews are those of
MacCallum (1975), Misner (1972) and Ryan (1972). More recent
introductory or review accounts are those of Fang and Ruffini
(1987), Fang and Wu (1986), Halliwell (1988b), Hartle (1985d,
1986), Hawking (1984b), Linde (1989a, 1989b), Narlikar and
Padmanabhan (1986) and Page (1986a).

\subhead {\bf Minisuperspace Models}

The literature contains a vast number of papers on minisuperspace.
Models with {\bf scalar fields} have been considerd by Blyth and
Isham (1975), del Campo and Vilenkin (1989b), Carow and Watamura
(1985), Christodoulakis and Zanelli (1984b), Esposito and Platania
(1988), Fakir (1989), Gibbons and Grishchuk (1988), Gonzalez-Diaz
(1985), Hartle and Hawking (1983), Hawking (1984a), Hawking and Wu
(1985), Moss and Wright (1984), Page (1989a), Poletti (1989),
Pollock (1988a), Yokoyama et al. (1988) and Zhuk (1988). The
scalar field model of Section 2 is described in, for example,
Hawking (1984a) and Page (1986a).

{\bf Anisotropic} minisuperspace models are considered in the
papers by Amsterdamski (1985), Ashtekar and Pullin (1990), Berger
(1975, 1982, 1984, 1985, 1988, 1989), Berger and Vogeli (1985),
Bergamini and Giampieri (1989), del Campo and Vilenkin (1989a),
Duncan and Jensen (1988), Fang and Mo (1987), Furusawa (1986),
Halliwell and Louko (1990), Hawking and Luttrell (1984), Hussain
(1987, 1988), Kodama (1988b), Laflamme (1987b), Laflamme and
Shellard (1987), Louko (1987a, 1987b, 1988a), Louko and Ruback
(1989), Louko and Vachaspati (1988), Matsuki and Berger (1989),
Misner (1969c, 1973), Moss and Wright (1985) and Schleich (1988).

The extension to {\bf Kaluza-Klein} theories has been considered
by Beciu (1985), Bleyer at al. (1989), Carow-Watamura et al.
(1987), Halliwell (1986, 1987a), Hu and Wu (1984, 1985, 1986),
Ivashchuk et al.(1989), Lonsdale (1986), Matzner and Mezzacappa
(1986), Okada and Yoshimura (1986), Pollock (1986), Shen (1989a),
Wu (1984, 1985a, 1985b, 1985c) and Wudka (1987a).

In these lectures we concentrated on Einstein gravity.
Minisuperspace models involving {\bf higher derivative} actions
have been studied by Coule and Miji\'c (1988), Hawking (1987a),
Hawking and Luttrell (1984b), Horowitz (1985), Hosoya (1989),
Miji\'c et al. (1989), Pollock (1986, 1988b, 1989b) and Vilenkin
(1985a).

Other minisuperspace models not obviously falling into any of the
above categories include those of Brown (1989), Li and Feng
(1987), Liu and Huang (1988), Mo and Fang (1988) and Wudka
(1987b).

The question of the {\bf validity} of minisuperspace, when
considered as an approximation to the full theory, has been
addressed by Kucha{$\rm \check r$} and Ryan (1986, 1989).

\subhead {\bf Inhomogeneous Peturbations about Minisuperspace}

Perturbative models of the type described in Section 10 have been
studied by Anini (1989a, 1989b), Banks et al.(1985), D'Eath and
Halliwell (1987), Fischler et al. (1985), Halliwell and Hawking
(1985), Morris (1988), Ratra (1989), Rubakov (1984), Shirai and
Wada (1988), Vachaspati and Vilenkin (1988), Vilenkin (1988) and
Wada (1986, 1986c, 1987).

An important feature of this type of model is the {\bf derivation
of the \break Schr\"odinger equation} from the Wheeler-DeWitt
equation and the emergence of {\bf quantum field theory in curved
spacetime} This sort of issue has been considered by Banks (1985),
Brout (1987), Brout et al. (1987), Brout and Venturi (1989),
DeWitt (1967), Halliwell (1987c), Halliwell and Hawking (1985),
Laflamme (1987a), Lapchinsky and Rubakov (1979), Vachaspati (1989)
and Wada (1987).

In Section 10 we only derived the dynamics of the perturbation
modes on a minisuperspace background. However, one can go one step
further than that and ask how the perturbation modes react back on
the minisuperspace background. In principle, one may thus attempt
to derive the {\bf semi-classical Einstein equations}. This area
seems to be somewhat confused, and no completely clear derivation
has yet been given. The relevant papers are those of Brout (1987),
Brout et al. (1987), Brout and Venturi (1989), Castagnino et al.
(1988), Halliwell (1987b), Hartle (1986), Padmanabhan (1989a),
Padmanabhan (1989c), Padmanabhan and Singh (1988) and Singh and
Padmanabhan (1989).

\subhead {\bf Black Holes and Spherically Symmetric Systems }

One is normally interested in cosmological models, but spherically
symmetric systems, including black holes have been studied by
Allen (1987), Fang and Li (1986), Laflamme (1987b), Nagai (1989),
Nambu and Sasaki (1988) and Rodrigues et al. (1989). The
connection between the path integral for the no-boundary wave
function and that for the partition function for a black hole in a
box is discussed by Halliwell and Louko (1990).

\subhead {\bf Quantum Cosmology and String Theory}

String-inspired models have been studied by Enqvist et al. (1987,
1989), Gonzalez-Diaz (1988), Lonsdale and Moss (1987) and Pollock
(1989a, 1989b). The formal resemblances between quantum cosmology
and string theory have been explored by Birmingham and Torre
(1987), Luckock et al. (1988) and Matsuki and Berger (1989).

\subhead {\bf Fermionic Matter and Supersymmety}

Most papers involve bosonic matter sources, but the inclusion of
fermions and supersymmetric aspects have been studied by
Christodoulakis and Papadopoulos (1988), Christodoulakis and
Zanelli (1984b), D'Eath and Halliwell (1987), D'Eath and Hughes
(1988), Elitzur et al. (1986), Furlong and Pagels (1987), Isham
and Nelson (1974), Macias et al. (1987), Shen (1989b) and Shen and
Tan (1989).

\subhead {\bf Interpretation}

The rather basic interpretation mentioned in Section 4 (that we
regard a strong peak in the wave function as a prediction) comes
from Hartle (1986), Geroch (1984) and Wada (1988a). Other relevant
papers include those of Barbour and Smolin (1989), Barrow and
Tipler (1986), DeWitt and Graham (1973), Drees (1987), Ellis et
al. (1989), Everett (1957), Gell-Mann and Hartle (1989), Halliwell
(1987b, 1989b), Hartle (1988a, 1988b, 1988c, 1990), Kazama and
Nakayama (1985), Markov and Mukhanov (1988), Tipler (1986, 1987),
Wald and Unruh (1988), Vilenkin (1989) and Wada (1986a, 1988b).

The {\bf decoherence} requirement discussed in Section 6, for
quantum cosmology, has been considered by Calzetta (1989),
Fukuyama and Morikawa (1989), Gell-Mann and Hartle (1989),
Halliwell (1989b), Joos (1986), Kiefer (1987, 1988, 1989a, 1989c),
Mellor (1989), Padmanabhan (1989b), Morikawa (1989) and Zeh (1986,
1988, 1989a, 1989b). Further discussions of this and related
issues are those of Hu (1989) (which also includes extensive
references on statistical effects) and Kandrup (1988).

Decoherence as considered in the above references involves the
notion of diagonalization of a reduced density matrix. {\bf
Density matrices} in quantum cosmology have been considered in a
somewhat different context by Hawking (1987b), Page (1986b).

For more general discussions of decoherence in quantum mechanics,
see Gell-Mann and Hartle (1990), Joos and Zeh (1985), Unruh and
Zurek (1989) and Zurek (1981, 1982).

In an attempt to see how classical behaviour emerges, some authors
have constructed {\bf wavepacket} solutions to the Wheeler-DeWitt
equation, including Kiefer (1988, 1989d), Kazama and Nakayama
(1985) and Wada (1985).

The first requirement for classical behaviour discussed in Section
6 (peaking about classical configurations) was discussed using the
{\bf Wigner} function by Halliwell (1987b), Kodama (1988a) and
Singh and Padmanabhan (1989). Use of the Wigner function in this
way has been criticised by Anderson (1990). A somewhat different
approach using the Wigner function is that of Calzetta and Hu
(1989).

\subhead {\bf The Issue of Time}

Various authors have addressed the issue of time in quantum
cosmology and quantum gravity more generally. The sorts of
question one is interested in are along the following lines: Does
the theory possess an intrinsic time? If it does not, can one
quantize it? Does time emerge from a theory that has no time in it
to start with ? Many of these questions are discussed by Banks
(1985), Brout (1987), Brout et al. (1987), Brout and Venturi
(1989), Brown and York (1989), Castagnino (1989), Englert (1989),
Fukuyama and Kamimura (1988), Fukuyama and Morikawa (1989),
Greensite (1989a, 1989b), Halliwell (1989a), Hartle (1988a, 1988b,
1988c, 1990), Jacobson (1989), Kucha{$\rm \check r$} (1989),
Sorkin (1987, 1989) and Unruh and Wald (1988).

A related issue is the connection of the cosmological {\bf arrow
of time} with the thermodynamic arrow in quantum cosmology. This
has been studied by Fukuyama and Morikawa (1989), Hawking (1985),
Page (1984, 1985), Qadir (1987), Wada (1989) and Zeh (1986, 1988,
1989a, 1989b).

\subhead {\bf Path Integrals and the Wheeler-DeWitt Equation}

The explicit construction of the path integral for the wave
function of the universe and the derivation of the associated
Wheeler-DeWitt equation have been considered by Barvinsky (1986),
Barvinsky and Ponomariov (1986), Barvinsky (1987), Halliwell
(1988), Halliwell and Hartle (1990), Teitelboim (1980, 1982,
1983a, 1983b, 1983c) and Woodard (1989). The detailed construction
of the path integral described in Section 4 (Eq. (4.7)) is
described by Teitelboim (1982, 1983a). The discussion of the
minisuperspace path integral in Section 5 is based on Halliwell
(1988).

The issue of finding {\bf complex contours} to make the Euclidean
path integral converge has been studied by Gibbons, Hawking and
Perry (1978), Halliwell and Hartle (1989), Halliwell and Louko
(1989a, 1989b, 1990), Halliwell and Myers (1989), Hartle (1984,
1989), Hartle and Schleich (1987), Mazur and Mottola (1989) and
Schleich (1985, 1987, 1989).

Other papers involving path integrals are those of Arisue et al.
(1987), Berger (1985), Berger and Vogeli (1985), Duncan and Jensen
(1988), Farhi (1989), Giddings (1990), Hajicek (1986a, 1986b),
Hartle (1984, 1988a, 1988b, 1988c), Louko (1988a, 1988b, 1988c,
1988d), Narlikar and Padmanabhan (1983) and Suen and Young (1989).

\subhead {\bf Quantization Methods and Superspace}

One most commonly uses the Dirac quantization procedure in quantum
cosmology, in which one takes the wave function to be annihilated
by the operator versions of the constraints. However, one could in
principle use the ADM (or reduction) method, in which one solves
the constraints classical before quantizing. The connections
between these methods for systems like gravity has been considered
by Ashtekar and Horowitz (1982), Gotay (1986), Gotay and Demaret
(1983), Gotay and Isenberg (1980), Hajicek (1989), Isenberg and
Gotay (1981) and Kaup and Vitello (1974).

The properties of {\bf superspace} and quantization methods in it
have been discussed by DeWitt (1970), Fisher (1970), Giulini
(1989), Isham (1976), and Kucha{$\rm \check r$} (1981). The
article by Kucha{$\rm \check r$} also contains a useful guide to
the literature on canonical quantization.

\subhead {\bf Topological Aspects}

Goncharov and Bytsenko (1985, 1987), Gurzadyan and Kocharyan
(1989), Li Miao (1986), Mkrtchyan (1986), and Starobinsky and
Zel'dovich (1984), considered the possibilities of non-trivial
topologies in quantum creation of the universe. Other interesting
toplogical aspects of the no-boundary proposal have been
considered by Hartle and Witt (1988) (see also Louko and Ruback
(1989)).

\subhead {\bf Singularities}

Numerous authors have been interested in singularities in quantum
cosmology and their possible avoidance, including Laflamme and
Shellard (1987), Lemos (1987), Louko (1987a), Narlikar (1983,
1984) and Smith and Bergman (1988).

\subhead {\bf Boundary Condition Proposals}

We concentrated exclusively on the boundary condition proposals of
Hartle and Hawking (Hartle and Hawking, 1983; Hawking 1982,
1984a), Linde (1984a, 1984b, 1984c) and Vilenkin (1982, 1983,
1984, 1985b, 1986, 1988), but there are others (see for example,
Suen and Young (1989)).

\subhead {\bf Quantum Creation of the Universe}

Some of the older papers on quantum creation of the universe are
those of Atkatz and Pagels (1982), Brout, Englert and Gunzig
(1978, 1979), Brout, Englert and Spindel (1979), Casher and
Englert (1981), Gott (1982) and Tryon (1973). Various aspects of
the quantum creation of the universe as a tunneling event have
been explored by Goncharov et al. (1987), Grishchuk (1987),
Grishchuk and Sidorov (1988, 1989), Grishchuk and Zel'dovich
(1982), Lavrelashvili, Rubakov, Serebryakov and Tinyakov (1989),
Lavrelashvili, Rubakov, and Tinyakov (1985), Rubakov (1984) and
Rubakov and Tinyakov (1988).

\subhead {\bf Measures}

The measure coming from quantum cosmology on sets of inflationary
solutions, and also classical measures, have been studied Gibbons
et al. (1987) and Hawking and Page (1986, 1988). Gibbons and
Grishchuk (1988) introduced a measure on the set of solutions to
the Wheeler-DeWitt equation.

\subhead {\bf Operator Ordering }

The issue of operator ordering in the Wheeler-DeWitt equation has
been studied in minisuperspace by Halliwell (1988), Misner (1972)
and Moss (1988). More generally, see Christodoulakis and Zanelli
(1986a, 1986b, 1987), Friedman and Jack (1988), Hawking and Page
(1986) and Tsamis and Woodard (1987).

\subhead {\bf Creating a Universe in the Laboratory}

The possibility of quantum creation of an inflationary universe in
the laboratory has bee studied by Farhi et al. (1989) and Fischler
et al. (1989). See also Hiscock (1987) and Sato et al. (1982).

\subhead {\bf Miscellaneous}

{\bf Regge calculus} minisuperspace models have been studied by
Hartle (1985a, 1985b, 1985c, 1989). In {\bf $(2+1)$} dimensions,
gravity becomes essentially quantum mechanical. This has been
studied from a quantum cosmology viewpoint by Hosoya and Nakao
(1989) and Martinec (1984). Considerable simplifications appear to
occur in general relativity using the {\bf Ashtekar variables}
(Ashtekar, 1987). Their application to cosmologies has been
considerd by Ashtekar and Pullin (1990), Hussain and Smolin (1989)
and Kodama (1988b). The relationship between the wave function of
the universe and the {\bf stochastic approach} to inflation have
been studied by Goncharov et al. (1987), Goncharov and Linde
(1986) and Miji\'c (1988a, 1988b, 1989). Many classical
cosmologies exhibit {\bf chaos}. Quantization of such cosmologies
has been studied by Berger (1989) and Furusawa (1986). Finally,
mention should be made of the extensive contributions of Narlikar,
Padmanabhan and collaborators, much of which concentrates on
quantization of the conformal part of the metric, including
Narlikar (1981, 1983, 1984), Padmanabhan (1981, 1982a, 1982b,
1983a, 1983b, 1983c, 1983d, 1983e, 1983f, 1984a, 1984b, 1985a,
1985b, 1986, 1987, 1988), Padmanabhan and Narlikar (1981, 1982),
Padmanabhan et al. (1989), Singh and Padmanabhan (1987).


\def\here#1\par{\parshape=2.0in 15.2 truecm 1.0 truecm 14.2 truecm {#1}\par}

\gdef\ref#1, #2, #3, #4, 1#5#6#7.{\rm\here#1 (1#5#6#7), #2 \rm{\bf
#3}, #4.\it}

\gdef\refp#1, #2, 1#3#4#5.{\rm\here#1 (1#3#4#5), #2.\it}

\gdef\refc#1{\rm\here#1}

\def\pr{\sl Phys.Rev.}
\def\prl{\sl Phys.Rev.Lett.}
\def\prep{\sl Phys.Rep.}
\def\jmp{\sl J.Math.Phys.}
\def\rmp{\sl Rev.Mod.Phys.}

\def\np{\sl Nucl.Phys.}
\def\pl{\sl Phys.Lett.}
\def\annp{\sl Ann.Phys.(N.Y.)}
\def\cqg{\sl Class.Quantum Grav.}
\def\grg{\sl Gen.Rel.Grav.}
\def\nc{\sl Nuovo Cimento}
\def\ijmp{\sl Intern.J.Mod.Phys.}
\def\fp{\sl Found.Phys.}
\def\ijtp{\sl Int.J.Theor.Phys.}
\def\ptp{\sl Prog.Theor.Phys.}

\parskip=0pt
\parindent=0pt

\head{\bf REFERENCES}

The following is a list of almost 400 papers in the field of
quantum cosmology, plus a small number of other references
relevant to these lectures (the latter are generally distinguised
by the fact that their title is not included). A substantially
similar list of references may be found in ``A Bibliography of
Papers on Quantum Cosmology" ({\sl Int.J.Mod.Phys.A}, 1990, to
appear).

\ref {Allen, B.}, \pr, D32, 3136, 1985.

\ref {Allen, M.}, \cqg, 4, 149, 1987. Canonical quantization of a
spherically symmetric, massless scalar field interacting with
gravity in (2+1) dimensions.

\ref {Amsterdamski, P.}, \pr, D31, 3073, 1985. Wave function of an
anisotropic universe.

\refp {Anderson, A.}, Utah preprint, 1990. On predicting
correlations from Wigner functions.

\refp {Anini, Y.}, ICTP preprint IC/89/219, 1989a. Quantum
cosmological origin of large scale structure.

\refp {Anini, Y.}, ICTP preprint IC/89/307, 1989b. The initial
quantum state of matter perturbations about a de Sitter
background.

\ref {Arisue, H., Fujiwara, T., Kato, M. and Ogawa, K.}, \pr, D35,
2309, 1987. Path integral and operator formalism in quantum
gravity.

\ref {Ashtekar, A.}, \pr, D36, 1587, 1987.

\ref {Ashtekar, A. and Horowitz, G.T.}, \pr, D26, 3342, 1982. On
the canonical approach to quantum gravity.

\refc {Ashtekar, A. and Pullin, J. (1990)}, in {\it Nathen Rosen
Festschrift} (Israel Physical Society). {\it Bianchi cosmologies:
A new description}

\ref {Atkatz, D. and Pagels, H.}, \pr, D25, 2065, 1982. Origin of
the universe as a quantum tunneling event.

\ref {Banks, T.}, \np, B249, 332, 1985. TCP, quantum gravity, the
cosmological constant and all that...

\ref {Banks, T., Fischler, W. and Susskind, L.}, \np, B262, 159,
1985. Quantum cosmology in 2+1 and 3+1 dimensions.

\refc {Barbour, J.B. and Smolin, L. (1989)}, in {\it Proceedings
of the Osgood Hill Meeting on Conceptual Problems in Quantum
Gravity}, eds. A.Ashtekar and J.Stachel (Birkhauser, Boston). {\it
Can quantum mechanics be sensibly applied to the universe as a
whole?}

\refc {Barrow, J.D. and Tipler, F. (1986)}, {\it The Anthropic
Cosmological Principle} (Oxford University Press, Oxford).

\ref {Barvinsky, A.O.}, \pl, B175, 401, 1986. Quantum
geometrodynamics: The Wheeler-DeWitt equation for the wave
function of the universe.

\ref {Barvinsky, A.O. and Ponomariov, V.N.}, \pl, B167, 289, 1986.
Quantum geometrodynamics: The path integral and the initial value
problem for the wave function of the universe.

\ref {Barvinsky, A.O.}, \pl, B195, 344, 1987. The wave function
and the effective action in quantum cosmology: Covariant loop
expansion.

\ref {Beciu, M.I.}, \nc, B90, 223, 1985. Canonical quantum
Kaluza-Klein theory.

\ref {Belinsky, V.A., Grishchuk, L.P., Khalatnikov, I.M. and
Zel'dovich, Ya.B.}, \pl, B155, 232, 1985.

\ref {Bergamini, R. and Giampieri, G.}, \pr, D40, 3960, 1989.
Influence of rotation on the initial state of the universe.

\ref {Berger, B.}, \annp, 83, 458, 1974. Quantum graviton creation
in a model universe.

\ref {Berger, B.}, \pr, D11, 2770, 1975. Quantum cosmology: Exact
solution for the Gowdy $T^3$ model.

\ref {Berger, B.}, \pl, B108, 394, 1982. Singularity avoidance in
the semi-classical Gowdy $T^3$ model.

\ref {Berger, B.}, \annp, 156, 155, 1984. Quantum effects in the
Gowdy $T^3$ cosmology.

\ref {Berger, B.}, \pr, D32, 2485, 1985. Path integral quantum
cosmology II: Bianchi type I with volume-dependent source

\ref {Berger, B. and Vogeli, C.}, \pr, D32, 2477, 1985. Path
integral quantum cosmology I: Vacuum Bianchi type I.

\ref {Berger, B.}, \grg, 20, 755, 1988. Monte Carlo simulation of
a quantized universe.

\ref {Berger, B.}, \pr, D39, 2426, 1989. Quantum chaos in the
mixmaster universe.

\refc {Birrell, N.D.and Davies, P.C.W. (1982)}, {\it Quantum
Fields in Curved Space} (Cambridge University Press, Cambridge).

\ref {Birmingham, D. and Torre, C.}, \pl, B194, 49, 1987. An
application of the Hartle-Hawking proposal to string theory.

\refp {Bleyer, U., Liebscher, D.E., Schmidt, H.J. and Zhuk, A.L.},
Potsdam preprint ZIAP 89-11, 1989. On the Wheeler-DeWitt equation
in multi-dimensional cosmology with phenomenological matter.

\ref {Blyth, W.F. and Isham, C.J.}, \pr, D11, 768, 1975.
Quantization of a Friedman universe filled with a scalar field.

\ref {Brandenberger, R.}, \np, B245, 328, 1984.

\ref {Brandenberger, R.}, \ijmp, A2, 77, 1987.

\ref {Brandenberger, R.}, {\sl J.Phys.G}, 15, 1, 1989.

\ref {Brout, R.}, {\sl Found.Phys.}, 17, 603, 1987. On the concept
of time and the origin of cosmological temperature.

\ref {Brout, R., Englert, F. and Gunzig, E.}, \annp, 115, 78,
1978. The creation of the universe as a quantum phenomenon.

\ref {Brout, R., Englert, F. and Gunzig, E.}, \grg, 10, 1, 1979.
The causal universe.

\ref {Brout, R., Englert, F. and Spindel, P.}, \prl, 43, 417,
1979. Cosmological origin of the grand-unified mass scale.

\ref {Brout, R., Horwitz, G. and Weil, D.}, \pl, B192, 318, 1987.
On the onset of time and temperature in cosmology.

\ref {Brout, R. and Venturi, G.}, \pr, D39, 2436, 1989. Time in
semi-classical gravity.

\ref {Brown, J.D.}, \pr, D41, 1125, 1990. Tunneling in perfect
fluid (minisuperspace) quantum cosmology.

\ref {Brown, J.D. and York, J.W.}, \pr, D40, 3312, 1989. Jacobi's
action and the recovery of time in general relativity.

\ref {Burges, C.J.C.}, \np, B244, 533, 1984.


\ref {del Campo, S. and Vilenkin, A.}, \pl, B224, 45, 1989a.
Tunneling wave function for anisotropic universes.

\ref {del Campo, S. and Vilenkin, A.}, \pr, D40, 688, 1989b.
Initial conditions for extended inflation.

\ref {Calzetta, E.}, \cqg, 11, L227, 1989. Memory loss and
asymptotic behavior in minisuperspace cosmological models.

\ref {Calzetta, E. and Hu, B.L.}, \pr, D40, 380, 1989. Wigner
distribution and phase space formulation of quantum cosmology.

\ref {Carow, U. and Watamura, S.}, \pr, D32, 1290, 1985. A quantum
cosmological model of the inflationary universe.

\ref {Carow-Watamura, U., Inami, T. and Watamura, S.}, \cqg, 4,
23, 1987. A quantum cosmological approach to Kaluza-Klein theory
and the boundary condition of no boundary.

\ref {Caves, C.M.}, \pr, D33, 1643, 1986.

\ref {Caves, C.M.}, \pr, D35, 1815, 1987.

\ref {Casher, A. and Englert, F.}, \pl, B104, 117, 1981. The
quantum era.

\ref {Castagnino, M.}, \pr, D39, 2216, 1989. Probabilistic time in
quantum gravity.

\ref {Castagnino, M., Mazzitelli, D. and Yastremiz, C.}, \pl,
B203, 118, 1988. On the graviton contribution to the back-reaction
Einstein equations.

\ref {Christodoulakis, T. and Papadopoulos, C.G.}, \pr, D38, 1063,
1988. Quantization of Robertson-Walker geometry coupled to a
spin-3/2 field.

\ref {Christodoulakis, T. and Zanelli, J.}, \pl, A102, 227, 1984a.
Quantum mechanics of the Robertson-Walker geometry.

\ref {Christodoulakis, T. and Zanelli, J.}, \pr, D29, 2738, 1984b.
Quantization of Robertson-Walker geometry coupled to fermionic
matter.

\ref {Christodoulakis, T. and Zanelli, J.}, \nc, B93, 1, 1986a.
Operator ordering in quantum mechanics and quantum gravity.

\ref {Christodoulakis, T. and Zanelli, J.}, \nc, B93, 22, 1986b.
Consistent algebra for the constraints of quantum gravity.

\ref {Christodoulakis, T. and Zanelli, J.}, \cqg, 4, 851, 1987.
Canonical approach to quantum gravity.

\ref {Coule, D. and Miji\'c, M.B.}, \ijmp, A3, 617, 1988. Quantum
fluctuations and eternal inflation in the $R^2$ model.

\ref {D'Eath, P.D. and Halliwell, J.J.}, \pr, D35, 1100, 1987.
Fermions in quantum cosmology.

\ref {D'Eath, P.D. and Hughes, D.}, \pl, B214, 498, 1988.
Supersymmetric minisuperspace.

\ref {DeWitt, B.S.}, \pr, 160, 1113, 1967. Quantum theory of
gravity I. The canonical theory.

\refc {DeWitt, B.S. (1970)}, in {\it Relativity}, eds. M.Carmeli,
S.Fickler and L.Witten  \break (Plenum,  New York). {\it Spacetime
as a sheaf of geodesics in superspace.}

\refc {DeWitt, B.S. and Graham, N. (eds.) (1973)}, {\it The Many
Worlds Interpretation of Quantum Mechanics} (Princeton University
Press, Princeton).

\ref {Drees, W.B.}, \ijtp, 26, 939, 1987. Interpretation of the
wave function of the universe.

\ref {Duncan, M.J. and Jensen, L.G.}, \np, B312, 662, 1988. The
quantum cosmology of an anisotropic universe.

\ref {Duncan, M.J. and Jensen, L.G.}, \np, B328, 171, 1989. Is the
universe Euclidean?

\ref {Elitzur, E., Forge, A. and Rabinovici, E.}, \np, B274, 60,
1986. The wave functional of a super-clock.

\ref {Ellis, J., Mohanty, S. and Nanopoulos, D.V.}, \pl, B221,
113, 1989. Quantum gravity and the collapse of the wave function.

\ref {Englert, F.}, \pl, B228, 111, 1989. Quantum physics without
time.

\ref {Enqvist, K., Mohanty, S. and Nanopoulos, D.V.}, \pl, B192,
327, 1987. Quantum cosmology of superstings.

\ref {Enqvist, K., Mohanty, S. and Nanopoulos, D.V.}, \ijmp, A4,
873, 1989. Aspects of superstring quantum cosmology.

\ref {Esposito, G. and Platania, G.}, \cqg, 5, 937, 1988.
Inflationary solutions in quantum cosmology.

\ref {Everett, H.}, \rmp, 29, 454, 1957. Relative state
formulation of quantum mechanics.

\refp {Fakir, R.}, UBC preprint, 1989. Quantum creation of
universes with non-minimal coupling.

\ref {Fang, L.Z. and Li, M.}, \pl, B169, 28, 1986. Formation of
black holes in quantum cosmology.

\ref {Fang, L.Z. and Mo, H.J.}, \pl, B186, 297, 1987. Wave
function of a rotating universe.

\refc {Fang, L.Z. and Ruffini, R.(eds.) (1987)}, {\it Quantum
Cosmology}, Advanced Series in Astrophysics and Cosmology No.3
(World Scientific, Singapore).

\ref {Fang, L.Z. and Wu, Z.C.}, \ijmp, A1, 887, 1986. An overview
of quantum cosmology.

\ref {Farhi, E.}, \pl, B219, 403, 1989. The wave function of the
universe and the square root of minus one.

\refp {Farhi, E., Guth, A.H. and Guven, J.}, CTP preprint
CTP-1690, 1989. Is it possible to create a universe in the
laboratory by quantum tunneling?

\refc {Fisher, A.E. (1970)}, in {\it Relativity}, eds. M.Carmeli,
S.Fickler and L.Witten (Plenum, New York). {\it The theory of
superspace.}

\refp {Fischler, W., Morgan, D. and Polchinksi, J.}, Texas
preprint UTTG-27-89, 1989. Quantum nucleation of false vacuum
bubbles.

\ref {Fischler, W., Ratra, B. and Susskind, L.}, \np, B259, 730,
1985. \rm ({\it errata} \np, {\bf B268}, 747 (1986)). {\it Quantum
mechanics of inflation.}

\ref {Floreanini, R., Hill, C.T. and Jackiw, R.}, \annp, 175, 345,
1987.

\ref {Freese, K., Hill, C.T. and Mueller, M.}, \np, B255, 639,
1985.

\ref {Friedman, J.L. and Jack, I.}, \pr, D37, 3495, 1988. Formal
commutators of the gravitational constraints are not well defined:
A translation of Ashtekar's ordering to the Schr\"odinger
representation.

\ref {Fukuyama, T. and Kamimura, K.}, \ijmp, A3, 333, 1988.
Dynamical time variable in cosmology.

\ref {Fukuyama, T. and Morikawa, M.}, \pr, D39, 462, 1989.
Two-dimensional quantum cosmology: Directions of dynamical and
thermodynamic arrows of time.

\refp {Furlong, R.C. and Pagels, R.H.}, Rockefeller University
preprint \break RU86/B1/185, 1987. A super minisuperspace model
for quantum cosmology.

\ref {Furusawa, T.}, \ptp, 75, 59, 1986. Quantum chaos of
mixmaster universe.

\refc {Gell-Mann, M. and Hartle, J.B. (1990)} in, {\it Complexity,
Entropy and the Physics of Information}, Santa Fe Institute
Studies in the Sciences of Complexity, vol IX, edited by W.H.Zurek
(Addison Wesley); also in, {\it Proceedings of the Third
International Symposium on Foundations of Quantum Mechanics in the
Light of New Technology}, edited by S.Kobayashi (Japan Physical
Society). {\it Quantum cosmology and quantum mechanics}.

\ref {Gerlach, U.H.}, \pr, 177, 1929, 1969. Derivation of the ten
Einstein field equations from the semi-classical approximation to
quantum geometrodynamics.

\ref {Geroch, R.}, {\sl No{$\sl \hat u$}s}, 18, 617, 1984. The
Everett interpretation.

\ref {Gibbons, G.W. and Grishchuk, L.P.}, \np, B313, 736, 1988.
What is a typical wave function for the universe?

\refp {Gibbons, G.W. and Hartle, J.B.}, UCSB preprint, 1989. Real
tunneling geometries and the large-scale topology of the universe.

\ref {Gibbons, G.W., Hawking, S.W. and Perry, M.J.}, \np, B138,
141, 1978.

\ref {Gibbons, G.W., Hawking, S.W. and Stewart, J.M.}, \np, B281,
736, 1987. A natural measure on the set of all universes.

\refp {Giddings, S.}, Harvard preprint HUTP-89/A056, 1989. The
conformal factor and the cosmological constant.

\refc {Giulini, D. (1989)}, PhD Thesis, University of Cambridge.

\ref {Gleiser, M., Holman. R. and Neto, N.}, \np, 294B, 1164,
1987. First order formalism for quantum gravity.

\ref {Goncharov, Y.P. and Bytsenko, A.A.}, \pl, B160, 385, 1985.
The supersymmetric Casimir effect and quantum creation of the
universe with non-trivial topology.

\ref {Goncharov, Y.P. and Bytsenko, A.A.}, \pl, B182, 20, 1986.
Inflation, oscillation and quantum creation of the universe in
gauge extended supergravities.

\ref {Goncharov, Y.P. and Bytsenko, A.A.}, \cqg, 4, 555, 1987.
Casimir effect in supergravity theories and quantum birth of the
universe with non-trivial topology.

\ref {Goncharov, A.S. and Linde, A.D.}, {\sl Fiz.Elem.Chastits
At.Yadra}, 17, 837, 1986. \rm ({\sl Sov.J.Part.Nucl.}, {\bf 17},
369 (1986)). {\it Tunneling in an expanding universe: Euclidean
and Hamiltonian approaches.}

\ref {Goncharov, A.S., Linde, A.D. and Mukhanov, V.F.}, \ijmp, A2,
561, 1987. Global structure of the inflationary universe.

\ref {Gonzalez-Diaz, P.F.}, \pl, B159, 19, 1985. On the wave
function of the universe.

\ref {Gonzalez-Diaz, P.F.}, {\sl Hadronic J.}, 9, 199, 1986.
Lie-admissable structure of small distance quantum cosmology.

\ref {Gonzalez-Diaz, P.F.}, \pr, D38, 2951, 1988. Initial
conditions of a stringy universe.

\ref {Gotay, M.J.}, \cqg, 3, 487, 1986. Negative energy states in
quantum gravity?

\ref {Gotay, M.J. and Demaret, J.}, \pr, D28, 2402, 1983. Quantum
cosmological singularities.

\ref {Gotay, M.J. and Isenberg, J.A.}, \pr, D22, 235, 1980.
Geometric quantization and gravitational collapse.

\ref {Gott, J.R.}, {\sl Nature}, 295, 304, 1982. Creation of open
universes from de Sitter space.

\refp {Greensite, J.}, San Francisco preprint SFSU-TH-89/1, 1989a.
Conservation of probability in quantum cosmology.

\refp {Greensite, J.}, San Francisco preprint SFSU-TH-89/3, 1989b.
Time and probability in quantum cosmology.

\ref {Grishchuk, L.P.}, {\sl Mod.Phys.Lett.}, A2, 631, 1987.
Quantum creation of the universe can be observationally verified.

\ref {Grishchuk, L.P. and Rozhansky, L.V.}, \pl, B208, 369, 1988.
On the beginning and the end of classical evolution in quantum
cosmology.

\refp {Grishchuk, L.P. and Rozhansky, L.V.}, Caltech preprint
GRP-207, 1989. Does the Hartle-Hawking wave function predict the
universe we live in?

\ref {Grishchuk, L.P. and Sidorov, Yu.V.}, {\sl
Zh.Eksp.Teor.Fiz.}, 94, 29, 1988. \rm ({\sl Sov.Phys. JETP}, {\bf
67}, 1533 (1988)). \it Boundary conditions for the wave function
of the universe.

\ref {Grishchuk, L.P. and Sidorov, Yu.V.}, \cqg, 6, L155, 1989.
Relic gravitons and the birth of the universe.

\refc {Grishchuk, L.P. and Zeldovich, Ya.B.(1982)}, in {\it
Quantum Structure of Space and Time}, eds. M.J.Duff and C.J.Isham
(Cambridge University Press, Cambridge). {\it Complete
cosmological theories}.

\ref {Gurzadyan, V.G and Kocharyan, A.A.}, {\sl
Zh.Eksp.Teor.Fiz.}, 95, 3, 1989. \rm ({\sl Sov.Phys. JETP}, {\bf
68}, 1 (1989)). \it With what topology could the universe be
created?

\ref {Guth, A.H.}, \pr, D28, 347, 1981.

\ref {Guth, A.H and Pi, S-Y.}, \pr,  D31, 1899, 1985.

\ref {Hajicek, P.}, \pr, D34, 1040, 1986a. Origin of non-unitarity
in quantum gravity.

\ref {Hajicek, P.}, \jmp, 27, 1800, 1986b. Elementary properties
of a new kind of path integral.

\refp {Hajicek, P.}, Bern preprint BUTP-89/23, 1989. Dirac
quantization of systems with quadratic constraints.

\ref {Halliwell, J.J.}, \np, B266, 228, 1986. The quantum
cosmology of Einstein-Maxwell theory in six dimensions.

\ref {Halliwell, J.J.}, \np, B286, 729, 1987a. Classical and
quantum cosmology of the Salam-Sezgin model.

\ref {Halliwell, J.J.}, \pr, D36, 3626, 1987b. Correlations in the
wave function of the universe.

\refp {Halliwell, J.J.}, DAMTP preprint, 1987c. Quantum field
theory in curved space-time as the semi-classical limit of quantum
cosmology.

\ref {Halliwell, J.J.}, \pr, D38, 2468, 1988a. Derivation of the
Wheeler-DeWitt equation from a path integral for minisuperspace
models.

\refp {Halliwell, J.J.}, ITP preprint NSF-ITP-88-131, 1988b.
Quantum cosmology: An introductory review.

\refc {Halliwell, J.J. (1989a)}, in {\it Proceedings of the Osgood
Hill Meeting on Conceptual Problems in Quantum Gravity}, eds.
A.Ashtekar and J.Stachel (Birkhauser, Boston). {\it Time in
quantum cosmology.}

\ref {Halliwell, J.J.}, \pr, D39, 2912, 1989b. Decoherence in
quantum cosmology.

\ref {Halliwell, J.J. and Hartle, J.B.}, \pr, D41, 1815, 1990.
Integration contours for the no-boundary wave function of the
universe.

\refp {Halliwell, J.J. and Hartle, J.B.}, ITP preprint, 1990. Wave
functions constructed from an invariant sum-over-histories satisfy
constraints.

\ref {Halliwell, J.J. and Hawking, S.W.}, \pr, D31, 1777, 1985.
Origin of structure in the universe.

\ref {Halliwell, J.J. and Louko, J.}, \pr, D39, 2206, 1989a.
Steepest-descent contours in the path-integral approach to quantum
cosmology I: The de Sitter minisuperspace model.

\ref {Halliwell, J.J. and Louko, J.}, \pr, D40, 1868, 1989b.
Steepest-descent contours in the path-integral approach to quantum
cosmology II: Microsuperspace.

\refp {Halliwell, J.J. and Louko, J.}, CTP preprint, 1990.
Steepest-descent contours in the path-integral approach to quantum
cosmology III: A general method with applications to some
anisotropic models.

\ref {Halliwell, J.J. and Myers, R.}, \pr, D40, 4011, 1989.
Multiple-sphere configurations in the path-integral representation
of the wave function of the universe.

\refc {Hanson, A., Regge, T. and Teitelboim, T. (1976)}, {\it
Constrained Hamiltonian Systems}, Contributi del Centro Linceo
Interdisciplinare di Scienze Matematiche e loro Applicazioni N.22
(Accademia Nazionale dei Lincei, Rome).

\ref {Hartle, J.B.}, \pr, D29, 2730, 1984. Ground state wave
function of linearized gravity.

\ref {Hartle, J.B.}, \jmp, 26, 804, 1985a. Simplicial
minisuperspace I: General discussion.

\ref {Hartle, J.B.}, \jmp, 27, 287, 1985b. Simplicial
Minisuperspace II: Some classical solutions on simple
triangulations.

\ref {Hartle, J.B.}, \cqg, 2, 707, 1985c. Unruly topologies in
two-dimensional quantum gravity.

\refc {Hartle, J.B. (1985d)}, in {\it High Energy Physics 1985:
Proceedings of the Yale Summer School,} eds. M.J.Bowick and
F.Gursey (World Scientific, Singapore). {\it Quantum Cosmology.}

\refc {Hartle, J.B. (1986)}, in {\it Gravitation in Astrophysics,
Cargese, 1986}, eds. B.Carter and J.Hartle (Plenum, New York).
{\it Prediction and observation in quantum cosmology}.

\ref {Hartle, J.B.}, \pr, D37, 2818, 1988a. Quantum kinematics of
spactime I: Non-relativistic theory.

\ref {Hartle, J.B.}, \pr, D38, 2985, 1988b. Quantum kinematics of
spacetime II: A model quantum cosmology with real clocks.

\refp {Hartle, J.B.}, Santa Barbara preprint, 1988c. Quantum
kinematics of spacetime III: General relativity.

\ref {Hartle, J.B.}, \jmp, 30, 452, 1989. Simplicial
minisuperspace III: Integration contours in a five-simplex model.

\refc {Hartle, J.B. (1990)}, this volume.

\ref {Hartle, J.B. and Hawking, S.W.}, \pr, D28, 2960, 1983. Wave
function of the universe.

\refc {Hartle, J.B. and Schleich, K. (1987)}, in {\it Quantum
Field Theory and Quantum Statistics: Essays in Honour of the
Sixtieth Birthday of E.S.Fradkin}, eds. I.A.Batalin,
G.A.Vilkovisky and C.J.Isham (Hilger, Bristol). {\it The conformal
rotation in linearized gravity}.

\ref {Hartle, J.B. and Witt, D.M.}, \pr, D37, 2833, 1988.
Gravitational $\theta$-states and the wave function of the
universe.

\refc {Hawking, S.W. (1982)}, in {\it Astrophysical Cosmology},
eds. H.A.Br\"uck, G.V.Coyne and M.S.Longair (Pontifica Academia
Scientarium, Vatican City).
{\it The boundary conditions of the universe.}

\ref {Hawking, S.W.}, \np, B239, 257, 1984a. The quantum state of
the universe.

\refc {Hawking, S.W. (1984b)}, in {\it Relativity, Groups and
Topology II, Les Houches Session XL}, eds. B.DeWitt and R.Stora
(North Holland, Amsterdam). {\it Lectures on quantum cosmology.}

\ref {Hawking, S.W.}, \pr, D32, 2489, 1985. The arrow of time in
cosmology.

\refc {Hawking, S.W. (1987a)}, in {\it Quantum Field Theory and
Quantum Statistics: Essays in Honour of the Sixtieth Birthday of
E.S.Fradkin}, eds. I.A.Batalin, G.A.Vilkovisky and C.J.Isham
(Hilger, Bristol). {\it Who's afraid of (higher derivative)
ghosts?}

\ref {Hawking, S.W.}, {\sl Physica Scripta}, T15, 151, 1987b. The
density matrix of the universe.

\ref {Hawking, S.W. and Luttrell, J.C.}, \np, B247, 250, 1984a.
Higher derivatives in quantum cosmology I: The isotropic case.

\ref {Hawking, S.W. and Luttrell, J.C.}, \pl, B143, 83, 1984b. The
isotropy of the universe.

\ref {Hawking, S.W. and Page, D.N.}, \np, B264, 185, 1986.
Operator ordering and the flatness of the universe.

\ref {Hawking, S.W. and Page, D.N.}, \np, B298, 789, 1988. How
probable is inflation?

\ref {Hawking, S.W. and Wu, Z.C.}, \pl, B151, 15, 1985. Numerical
calculations of minisuperspace cosmological models.

\ref {Higgs, P.W.}, \prl, 1, 373, 1958. Integration of secondary
constraints in quantized general relativity.

\ref {Hiscock, W.A.}, \pr, D35, 1161, 1987. Can black holes
nucleate vacuum phase transitions?

\ref {Horowitz, G.T.}, \pr, D31, 1169, 1985. Quantum cosmology
with a positive definite action.

\refp {Hosoya, A.}, Osaka University preprint OU-HET-84, 1984.
Synchronous time gauge in quantum cosmology.

\refc {Hosoya, A. (1989)}, in {\it Wandering in the Fields}, eds.
K.Kawarabayashi and A. Ukawa. {\it Quantum Cosmology of $R^2$
theory of gravity}.

\refp {Hosoya, A. and Nakao, K.}, Hiroshima preprint RRK 89-16,
1989. \break (2+1)-dimensional quantum gravity.

\refp {Hu, B.L.}, Cornell preprint, CLNS 89/941, 1989. Quantum and
statistical effects in superspace cosmology.

\ref {Hu, X.}, \pl, 135A, 245, 1989. An equivalent form of
conformal factor quantum gravity in cosmology.

\ref {Hu, X.M. and Wu, Z.C.}, \pl, B149, 87, 1984. Quantum
Kaluza-Klein cosmology II.

\ref {Hu, X.M. and Wu, Z.C.}, \pl, B155, 237, 1985. Quantum
Kaluza-Klein cosmology III.

\ref {Hu, X.M. and Wu, Z.C.}, \pl, B182, 305, 1986. Quantum
Kaluza-Klein cosmology IV.

\ref {Hussain, V.}, \cqg, 4, 1587, 1987. Quantum effects on the
singularity of the Gowdy cosmology.

\ref {Hussain, V.}, \pr, D38, 3314, 1988. The Weyl tensor and
gravitational entropy.

\ref {Hussain, V. and Smolin, L.}, \np, B327, 205, 1989. Exact
quantum cosmologies from two Killing field reductions of general
relativity.

\ref {Isenberg, J.A. and M.J. Gotay}, \grg, 13, 301, 1981. Quantum
cosmology and geometric quantization.

\ref {Isham, C.J.}, {\sl Proc.R.Soc.Lond.}, A351, 209, 1976. Some
quantum field theory aspects of the superspace quantization of
general relativity.

\ref {Isham, C.J. and Nelson, J.E.}, \pr, D10, 3226, 1974.
Quantization of a coupled fermi field and Robertson-Walker metric.

\ref {Isham, C.J. and Kucha{$\rm \check r$}, K.V.}, \annp, 164,
288, 1985a.

\ref {Isham, C.J. and Kucha{$\rm \check r$}, K.V.}, \annp, 164,
316, 1985b.

\refp {Ivashchuk, V.D., Melnikov, V.N., and Zhuk, A.I.}, Potsdam
preprint PRE-EL-89-04, 1989. On the Wheeler-DeWitt equation in
multidimensional cosmology.

\refc {Jacobson, T. (1989)}, in {\it Proceedings of the Osgood
Hill Meeting on Conceptual Problems in Quantum Gravity}, eds.
A.Ashtekar and J.Stachel (Birkhauser, Boston). {\it Unitarity,
Causality and Quantum Gravity}.

\ref {Joos, E.}, \pl, A116, 6, 1986. Why do we observe a classical
spacetime?

\ref {Joos, E. and Zeh, H.D.}, {\sl Zeit.Phys.}, B59, 223, 1985.

\ref {Jordan, R.D}, \pr, D36, 3604, 1987. Expectation values in
quantum cosmology.

\ref {Kandrup, H.E.}, \cqg, 5, 903, 1988. Conditional
probabilities and entropy in (minisuperspace) quantum cosmology.

\ref {Kaup, D.J. and Vitello, A.P.}, \pr, D9, 1648, 1974. Solvable
quantum cosmologival models and the importance of quantizing in a
special canonical frame.

\ref {Kazama, Y. and Nakayama, R.}, \pr, D32, 2500, 1985. Wave
packet in quantum cosmology.

\ref {Kiefer, C.}, \cqg, 4, 1369, 1987. Continuous measurement of
minisuperspace variables by higher multipoles.

\ref {Kiefer, C.}, \pr, D38, 1761, 1988. Wave packets in
minisuperspace.

\ref {Kiefer, C.}, \cqg, 6, 561, 1989a. Continuous measurement of
intrinsic time by fermions.

\ref {Kiefer, C.}, \pl, B225, 227, 1989b. Non-minimally coupled
scalar fields and the initial value problem in quantum gravity.

\ref {Kiefer, C.}, \pl, A139, 201, 1989c. Quantum gravity and
Brownian motion.

\refp {Kiefer, C.}, Heidelberg preprint, 1989d. Wave packets in
quantum cosmology and the cosmological constant.

\ref {I.Klebanov, L.Susskind and T.Banks}, \np, B317, 663, 1989.
Wormholes and the cosmological constant.

\refp {Kodama, H.}, Kyoto University preprint KUCP-0014, 1988a.
Quantum cosmology in terms of the Wigner function.

\ref {Kodama, H.}, {\sl Prog.Theor.Phys}, 80, 1024, 1988b.
Specialization of Ashtekar's formalism to Bianchi Cosmology.

\refc {Kucha{$\rm \check r$}, K.V. (1981)}, in {\it Quantum
Gravity 2: A Second Oxford Symposium}, eds. C.J.Isham, R.Penrose
and D.W.Sciama (Clarendon Press, Oxford). {\it Canonical methods
of quantization.}

\ref {Kucha{$\rm \check r$}, K.V.}, \jmp, 22, 2640, 1981. General
relativity: Dynamics without symmetry.

\ref {Kucha{$\rm \check r$}, K.V.}, {\sl Found.Phys.}, 16, 193,
1986.

\refc {Kucha{$\rm \check r$}, K.V. (1989)}, in {\it Proceedings of
the Osgood Hill Meeting on Conceptual Problems in Quantum
Gravity}, eds. A.Ashtekar and J.Stachel (Birkhauser, Boston). {\it
The problem of time in canonical quantization of relativistic
systems.}

\refc {Kucha{$\rm \check r$}, K.V. and Ryan, M.P.} (1986), in
Yamada Conference XIV, eds. H.Sato and T.Nakamura (World
Scientific). {\it Can minisuperspace quantization be justified?}

\ref {Kucha{$\rm \check r$}, K.V. and Ryan, M.P.}, \pr, D40, 3982,
1989. Is minisuperspace quantization valid? Taub in Mixmaster.

\ref {Laflamme, R.}, \pl, B198, 156, 1987a. Euclidean vacuum:
Justification from quantum cosmology.

\refc {Laflamme, R. (1987b)}, in {\it Origin and Early History of
the Universe: Proceedings of the 26th Liege International
Astrophysical Colloquium}, ed. J.Demaret. {\it Wave function of a
black hole interior.}

\ref {Laflamme, R.}, \np, B324, 233, 1989. Geometry and
thermofields.

\ref {Laflamme, R. and Shellard, E.P.S.}, \pr, D35, 2315, 1987.
Quantum cosmology and recollapse.

\ref {Lapchinsky, V. and Rubakov, V.A.}, {\sl Acta.Phys.Polonica},
B10, 1041, 1979. Canonical quantization of gravity and quantum
field theory in curved space-time.

\refp {Lavrelashvili, G.A., Rubakov, V.A., Serebryakov, M.S. and
Tinyakov, P.G.}, Institute for Nuclear Research preprint P-0637,
1989. Negative Euclidean action. Instantons and pair creation in
strong background fields.

\ref {Lavrelashvili, G.A., Rubakov, V.A. and Tinyakov, P.G.}, \pl,
B161, 280, 1985. Tunneling transitions with gravitation: Breaking
of the quasi-classical approximation.

\ref {Lemos, N.A.}, \pr, D36, 2364, 1987. Conservation of
probability and quantum cosmological singularities.

\ref {Li, A. and Feng, L-L.}, {\sl Commun.Theor.Phys.}, 7, 175,
1987. Wave function of the universe with quantum corrections of
matter fields.

\ref {Linde, A.D.}, {\sl Zh.Eksp.Teor.Fiz.}, 87, 369, 1984a. \rm
({\sl Sov.Phys.JETP} {\bf 60}, 211, 1984). \break \it Quantum
creation of an inflationary universe.

\ref {Linde, A.D.}, \nc, 39, 401, 1984b. Quantum creation of
inflationary universe.

\ref {Linde, A.D.}, {\sl Rep.Prog.Phys}, 47, 925, 1984c. The
inflationary universe.

\refc {Linde, A.D. (1989a)}, {\it Inflation and Quantum Cosmology}
(Academic Press, Boston).

\refc {Linde, A.D. (1989b)}, {\it Particle Physics and
Inflationary Cosmology} (Gordon and Breach, New York).

\ref {Liu, L. and Huang, C-G.}, \grg, 583, 20, 1988. The quantum
cosmology of Brans-Dicke theory.

\ref {Lonsdale, S.R.}, \pl, B175, 312, 1986. Wave function of the
universe for N=2 D=6 supergravity.

\ref {Lonsdale, S.R. and Moss, I.G.}, \pl, B189, 12, 1987. A
superstring cosmological model.

\ref {Louko, J.}, \pr, D35, 3760, 1987a. Fate of singularities in
Bianchi type-III quantum cosmology.

\ref {Louko, J.}, \cqg, 4, 581, 1987b. Propagation amplitude in
homogeneous quantum cosmology.

\ref {Louko, J.}, \annp, 181, 318, 1988a. Semi-classical path
measure and factor ordering in quantum cosmology.

\ref {Louko, J.}, \pl, B202, 201, 1988b. Canonizing the
Hartle-Hawking proposal.

\ref {Louko, J.}, \pr, D38, 478, 1988c. Quantum cosmology and
electromagnetism.

\ref {Louko, J.}, \cqg, 5, L181, 1988d. A Feynman prescription for
the Hartle-Hawking proposal.

\refp {Louko, J. and Ruback, P.}, CERN preprint TH-5501/89, 1989.
Spatially flat quantum cosmology.

\ref {Louko, J. and Vachaspati, T.}, \pl, B223, 21, 1989. On the
Vilenkin boundary condition proposal in anisotropic universes.

\ref {Luckock, H., Moss, I.G. and Toms, D.}, \np, B297, 748, 1988.
Polyakov strings in background fields: Wave functions and quantum
cosmology.

\refc {M.A.H.MacCallum (1975)}, in {\it Quantum Gravity}, eds.
C.J.Isham, R.Penrose and D.W.Sciama (Clarendon Press, Oxford).
{\it Quantum cosmological models.}

\ref {A.Macias, O.Obregon and M.P.Ryan Jr.}, \cqg, 4, 1477, 1987.
Quantum cosmology - the supersymmetric square root.

\ref {M.A.Markov and V.F.Mukhanov}, \pl, 127A, 251, 1988.
Classical preferable basis in quantum mechanics.

\ref {E.Martinec}, \pr, 30, 1198, 1984. Soluble systems in quantum
gravity.

\ref {T.Matsuki and B.K.Berger}, \pr, D39, 2875, 1989. Consistency
of quantum cosmology for models with plane symmetry.

\ref {R.A.Matzner and A.Mezzacappa}, {\sl Found.Phys}, 16, 227,
1986. Professor Wheeler and the crack of doom: Closed cosmologies
in the five-dimensional Kaluza-Klein theory.

\refp {P.Mazur and E.Mottola}, Florida preprint UFIFT-AST-89-3,
1989. The gravitational measure, conformal factor problem and
stability of the ground state of quantum gravity.

\refp {F.Mellor}, Newcastle preprint NCL-89-TP/19, 1989.
Decoherence in quantum \break Kaluza-Klein theories.

\refp {Li Miao}, ICTP preprint IC/85/103, 1985. Proposal on
entropy in quantum cosmology.

\ref {Li Miao}, \pl, B173, 229, 1986. On topological aspects of
the birth of the universe.

\refp {Miji\'c, M.B.}, UBC preprint 88-0892, 1988a. On the
probability for a small cosmological constant in a
post-inflationary Friedmann universe.

\refc {Miji\'c, M.B. (1988b)}, in Yamada Conference XX, eds.
S.Hayakawa and K.Sato (Universal Academy Press). {\it Existence of
eternal inflation and quantum cosmology.}

\refp {Miji\'c, M.B.}, SISSA preprint, 1989. Quantum and
inflationary cosmology.

\ref {Miji\'c, M.B., M.S.Morris and W-M.Suen}, \pr, D39, 1496,
1989. Initial conditions for the $R+\epsilon R^2$ cosmology.

\ref {Misner, C.W.}, \pr, 186, 1319, 1969a. Quantum cosmology I.

\ref {Misner, C.W.}, \pr, 186, 1328, 1969b. Quantum cosmology II.

\ref {Misner, C.W.}, \prl, 22, 1071, 1969c. Mixmaster universe.

\refc {Misner, C.W. (1970)}, in {\it Relativity}, eds. M.Carmeli,
S.Fickler and L.Witten \break (Plenum, New York). {\it Classical
and quantum dynamics of a closed universe.}

\refc {Misner, C.W. (1972)}, in {\it Magic Without Magic: John
Archibald Wheeler, a Collection of Essays in Honor of his 60th
Birthday}, ed. J.Klauder (Freeman, San Francisco). {\it
Minisuperspace.}

\ref {Misner, C.W.}, \pr, D8, 3271, 1973. A minisuperspace example
- the Gowdy $T^3$ universe.

\refc {Misner, C.W., Thorne, K. and Wheeler, J.A. (1970)}, {\it
Gravitation} (W.Freeman, San Francisco).

\ref {Mkrtchyan, R.L.}, \pl, B172, 313, 1986. Topological aspects
of the birth of the universe.

\ref {Mo, H.J. and Fang, L.Z.}, \pl, B201, 321, 1988. Cosmic wave
function for induced gravity.

\ref {Morikawa, M.}, \pr, D40, 4023, 1989. Evolution of the cosmic
density matrix.

\ref {Morris, M.S.}, \pr, D39, 1511, 1988. Initial conditions for
perturbations in the $R+\epsilon R^2$ cosmology.

\ref {Moss, I.G.}, {\sl Ann.Inst. Henri Poincar\'e}, 49, 341,
1988. Quantum cosmology and the self-observing universe.

\refp {Moss, I.G. and Poletti, S.}, Newcastle preprint
NCL-89-TP25, 1989. Boundary conditions for quantum cosmology.

\ref {Moss, I.G. and Wright, W.A.}, \pr, D29, 1067, 1984. Wave
function of the inflationary universe.

\ref {Moss, I.G. and Wright, W.A.}, \pl, B154, 115, 1985. The
anisostropy of the universe.

\ref {Nagai, H.}, \ptp, 82, 322, 1989. Wave function of the de
Sitter-Schwarzschild universe.

\ref {Nambu, Y. and Sasaki, M.}, {\sl Prog.Th.Phys.}, 79, 96,
1988. The wave function of a collapsing dust sphere inside the
black hole horizon.

\ref {Narlikar, J.V.}, \fp, 11, 473, 1981. Quantum conformal
fluctuations near the classical spacetime singularity.

\ref {Narlikar, J.V.}, \pl, 96A, 107, 1983. Elimination of the
standard big bang singularity and particle horizons through
quantum conformal fluctuations.

\ref {Narlikar, J.V.}, \fp, 14, 443, 1984. The vanishing
likelihood of spacetime singularity in quantum conformal
cosmology.

\ref {Narlikar, J.V. and  Padmanabhan, T.}, \prep, 100, 151, 1983.
Quantum cosmology via path integrals.

\refc {Narlikar, J.V. and Padmanabhan, T. (1986)}, {\it
Gravitation, Gauge Theories and Quantum Cosmology} (D.Reidel,
Dordrecht).

\ref {Y.Okada and M.Yoshimura}, \pr, D33, 2164, 1986. Inflation in
quantum cosmology in higher dimensions.

\ref {Padmanabhan, T.}, \grg, 13, 451, 1981. Quantum fluctuations
and non-avoidance of the singularity in Bianchi Type I cosmology.

\ref {Padmanabhan, T.}, \grg, 14, 549, 1982a. Quantum stationary
states in Bianchi universes.

\ref {Padmanabhan, T.}, \pl, 87A, 226, 1982b. Friedmann universe
in a quantum gravity model.

\ref {Padmanabhan, T.}, \ijtp, 22, 1023, 1983a. Quantum conformal
fluctuations and stationary states.

\ref {Padmanabhan, T.}, \grg, 15, 435, 1983b. Quantum cosmology
and stationary states.

\ref {Padmanabhan, T.}, \pl, 93A, 116, 1983c. Instability of flat
space and origin of conformal fluctuations.

\ref {Padmanabhan, T.}, \pl, 96A, 110, 1983d. Quantum gravity and
the flatness problem of standard big bang cosmology.

\ref {Padmanabhan, T.}, \pr, D28, 745, 1983e. An approach to
quantum gravity.

\ref {Padmanabhan, T.}, \pr, D28, 756, 1983f. Universe before
Planck time: A quantum gravity model.

\ref {Padmanabhan, T.}, \cqg, 1, 149, 1984a. Quantum stationary
states and avoidance of singularities.

\ref {Padmanabhan, T.}, \pl, 104A, 196, 1984b. Inflation from
quantum gravity.

\ref {Padmanabhan, T.}, \grg, 17, 215, 1985a. Planck length as a
lower bound to all physical length scales.

\ref {Padmanabhan, T.}, \annp, 165, 38, 1985b. Physical
significance of the Planck length.

\ref {Padmanabhan, T.}, \cqg, 3, 911, 1986. Role of general
relativity in uncertainty principle.

\ref {Padmanabhan, T.}, \cqg, 4, L107, 1987. Limitations on the
operational definitions of spacetime events and quantum gravity.

\ref {Padmanabhan, T.}, \prl, 60, 2229, 1988. Acceptable density
perturbations from inflation due to quantum gravitational damping.

\ref {Padmanabhan, T.}, \cqg, 6, 533, 1989a. Semi-classical
approximations for gravity and the issue of back-reaction.

\ref {Padmanabhan, T.}, \pr, D39, 2924, 1989b. Decoherence in the
density matrix describing quantum three-geometries and the
emergence of classical spacetime.

\ref {Padmanabhan, T.}, \ijmp, A4, 4735, 1989c. Some fundamental
aspects of semiclassical and quantum gravity.

\ref {Padmanabhan, T. and Narlikar, J.V.}, \pl, 84A, 361, 1981.
Stationary states in a quantum gravity model

\ref {Padmanabhan, T. and Narlikar, J.V.}, {\sl Nature}, 295, 677,
1982. Quantum conformal fluctuations in a singular spacetime.

\ref {T.Padamanabhan, Seshadri, T.R. and Singh, T.P.}, \pr, D39,
2100, 1989. Making inflation work: Damping of density
perturbations due to Planck-energy cutoff.

\refp {Padmanabhan, T. and Singh, T.P.}, Tata preprint TIFR-TAP-4,
1988. On the semi-classical limit of the Wheeler-DeWitt equation.

\ref {Page, D.N.}, {\sl Int.J.Theor.Phys.}, 23, 725, 1984. Can
inflation explain the second law of thermodynamics?

\ref {Page, D.N.}, \pr, D32, 2496, 1985. Will entropy decrease if
the universe recollapses?

\refc {Page, D.N. (1986a)}, in {\it Quantum Concepts in Space and
Time}, eds. R.Penrose and C.J.Isham (Clarendon Press, Oxford).
{\it Hawking's wave function for the universe.}

\ref {Page, D.N.}, \pr, D34, 2267, 1986b. Density matrix of the
universe.

\refp {Page, D.N.}, ITP preprint NSF-ITP-89-207, 1989a.
Minisuperspaces with conformally and minimally coupled scalar
fields.

\refp {Page, D.N.}, ITP preprint NSF-ITP-89-18, 1989b. Time as an
inaccessible observable.

\ref {Pilati, M.}, \pr, D26, 2645, 1982. Strong coupling quantum
gravity I: Solution in a particular gauge.

\ref {Pilati, M.}, \pr, D28, 729, 1983. Strong coupling quantum
gravity II: Solution without gauge fixing.

\ref {Piran, T. and Williams, R.}, \pl, B163, 331, 1985.

\ref {Poletti, S.}, \cqg, 6, 1943, 1989. Double inflation and
quantum cosmology.

\ref {Pollock, M.D.}, \pl, B167, 301, 1986. On the creation of an
inflationary universe from nothing in a higher-dimensional theory
of gravity with higher derivative terms.

\ref {Pollock, M.D.}, \pl, B215, 635, 1988a. On the initial
condtions for super-exponential inflation.

\ref {Pollock, M.D.}, \np, B306, 931, 1988b. On the semi-classical
approximation to the wave function of the universe and its
probabilistic interpretation.

\ref {Pollock, M.D.}, \np, B135, 528, 1989a. On the relationship
between quantum cosmology and the dimensions of the superstring.

\ref {Pollock, M.D.}, \np, B324, 187, 1989b. On the quantum
cosmology of the superstring theory including the effects of
higher derivative terms.


\ref {Qadir, A.}, \pl, 121A, 113, 1987. The arrow of time and the
expansion of the universe.


\ref {Ratra, B.}, \pr, D31, 1931, 1985.

\ref {Ratra, B.}, \pr, D40, 3939, 1989. Quantum mechanics of
exponential potential inflation.

\ref {Rodrigues, L.M.C.S., Soares I.D. and Zanelli, J.}, \prl, 62,
989, 1989. Black hole decay and topological stability in quantum
gravity.

\ref {Rubakov, V.A.}, \pl, B148, 280, 1984. Quantum mechanics of
the tunneling universe.

\ref {Rubakov, V.A. and Tinyakov, P.G.}, \pl, B214, 334, 1988.
Gravitational instantons and creation of expanding universes.

\refc {Ryan, M. (1972)}, {\it Hamiltonian Cosmology}, Lecture
Notes in Physics No.13 \break (Springer, New York).


\ref {Sakharov, A.D.}, {\sl Sov.Phys. JETP}, 60, 214, 1984.
Cosmological transitions with a change in metric signature.

\ref {Sato, K., Kodama, H., Sasaki, M. and Maeda, K.}, \pl, B108,
103, 1982. Multiproduction of universes by first-order phase
transition of a vacuum.

\ref {Schleich, K.}, \pr, D32, 1889, 1985. Semi-classical wave
function of the universe at small 3-geometries.

\ref {Schleich, K.}, \pr, D36, 2342, 1987. The conformal rotation
in perturbative gravity.

\ref {Schleich, K.}, \pr, D39, 2192, 1989. Conformal rotation in
Bianchi type-I quantum cosmology.

\ref {Shen, Y.G.}, {\sl Chin.Phys.}, L6, 43, 1989a. Cosmic wave
function for the coupling scalars field in the theory of
Kaluza-Klein.

\ref {Shen, Y.G.}, {\sl Sci.China}, A32, 847, 1989b. Quantum
cosmology with scalar-spinor interaction field.

\ref {Shen, Y.G. and Tan, Z.Q.}, {\sl Chin.Phys.}, L6, 289, 1989.
Wave function of the universe for the spinor field in the induced
theory of gravity.

\ref {Shirai, I. and Wada, S.}, \np, B303, 728, 1988. Cosmological
perturbations and quantum fields in curved space-time.

\ref {Singh, T.P. and Padmanabhan, T.}, \pr, D37, 2993, 1987.
Semiclassical cosmology with a scalar field.

\ref {Singh, T.P. and Padmanabhan, T.}, \annp, 196, 296, 1989.
Notes on semiclassical gravity.

\ref {Smith, G.J. and Bergman, P.G.}, \pr, D33, 3570, 1988.
Quantum blurring of cosmological singularities.

\refc {Sorkin, R.D. (1987)}, in {\it History of Modern Gauge
Theories}, eds. M.Dresden and A.Rosenblum (Plenum). {\it On the
role of time in the sum-over-histories framework for gravity}

\refc {Sorkin, R.D. (1989)} in {\it Proceedings of the Osgood Hill
Meeting on Conceptual Problems in Quantum Gravity},  eds.
A.Ashtekar and J.Stachel  (Birkh\"auser,  \break Boston). {\it
Problems with causality in the sum-over-histories framework for
quantum mechanics}.

\ref {Starobinsky, A.A. and Zel'dovich, Ya.B.}, {\sl
Sov.Astrn.L.}, 10, 135, 1984. Quantum creation of a universe with
non-trivial topology.

\ref {Suen, W-M. and Young, K.}, \pr, D39, 2201, 1989. Wave
function of the universe as a leaking system.

\ref {Teitelboim, T.}, \pl, B96, 77, 1980. Proper time approach to
the quantization of the gravitational field.

\ref {Teitelboim, T.}, \pr, D25, 3159, 1982. Quantum mechanics of
the gravitational field.

\ref {Teitelboim, T.}, \pr, D28, 297, 1983a. Proper-time gauge in
the quantum theory of gravitation.

\ref {Teitelboim, T.}, \pr, D28, 310, 1983b. Quantum mechanics of
the gravitational field in asymptotically flat space.

\ref {Teitelboim, T.}, \prl, 50, 705, 1983c. Causality versus
gauge invariance in quantum gravity and supergravity.

\refc {Teitelboim, T. (1990)}, this volume.

\ref {Tipler, F.}, \prep, 137, 231, 1986. Interpreting the wave
function of the universe.

\ref {Tipler, F.}, \cqg, 4, L189, 1987. Non-Schrodinger forces and
pilot waves in quantum cosmology.

\ref {Tryon, E.P.}, {\sl Nature}, 246, 396, 1973. Is the universe
a vacuum fluctuation?

\ref {Tsamis, N.C. and Woodard, R.}, \pr, D36, 3641, 1987. The
factor-ordering problem must be regulated.

\ref {Unruh, W.G. and Wald, R.}, \pr, D40, 2598, 1989. Time and
the interpretation of quantum gravity.

\ref {Unruh, W.G. and Zurek, W.H.}, \pr, D40, 1071, 1989.

\ref {Vachaspati, T.}, \pl, B217, 228, 1989. De Sitter-invariant
states from quantum cosmology.

\ref {Vachaspati, T.and Vilenkin, A.}, \pr, D37, 898, 1988.
Uniqueness of the tunneling wave function of the unuverse.

\ref {Vilenkin, A.}, \pl, B117, 25, 1982. Creation of universes
from nothing.

\ref {Vilenkin, A.}, \pr, D27, 2848, 1983. Birth of inflationary
universes.

\ref {Vilenkin, A.}, \pr, D30, 509, 1984. Quantum creation of
universes.

\ref {Vilenkin, A.}, \pr, D32, 2511, 1985a. Classical and quantum
cosmology of the Starobinsky model.

\ref {Vilenkin, A.}, \np, B252, 141, 1985b. Quantum origin of the
universe.

\ref {Vilenkin, A.}, \pr, D33, 3560, 1986. Boundary conditions in
quantum cosmology.

\ref {Vilenkin, A.}, \pr, D37, 888, 1988. Quantum cosmology and
the initial state of the universe.

\ref {Vilenkin, A.}, \pr, D39, 1116, 1989. The interpretation of
the wave function of the universe.

\ref {Wada, S.}, \cqg, 2, L57, 1984. Quantum cosmology and
classical solutions in the two-dimensional higher derivative
theory.

\refp {Wada, S.}, Tokyo preprint 85-0185, 1985. Wave packet of a
universe.

\ref {Wada, S.}, {\sl Prog.Theor.Phys.}, 75, 1365, 1986a.
Quantum-classical correspondence in wave functions of the
universe.

\ref {Wada, S.}, \pr, D34, 2272, 1986b. Consistency of the
canonical quantization of gravity and boundary conditions for the
wave function of the universe.

\ref {Wada, S.}, \np, B276, 729, 1986c. Quantum cosmological
perturbations in pure gravity.

\ref {Wada, S.}, \prl, 59, 2375, 1987. Natural quantum state of
matter fields in quantum cosmology.

\ref {Wada, S.}, {\sl Mod.Phys.Lett}, A3, 645, 1988a.
Interpretation and predictability of quantum mechanics and quantum
cosmology.

\ref {Wada, S.}, {\sl Mod.Phys.Lett}, A3, 929, 1988b.
Macroscopicity and Classicality of Quantum Fluctuations in de
Sitter space.

\refp {Wada, S.}, Tokyo preprint, 1989. The arrow of time in
quantum cosmology and van Hove's theorem.

\refc {Wheeler, J.A. (1963)}, in {\it Relativity, Groups and
Topology}, eds. C.DeWitt and B.DeWitt (Gordon and Breach, New
York).

\refc {Wheeler, J.A.(1968)}, in {\it Batelles Rencontres}, eds.
C.DeWitt and J.A.Wheeler (Benjamin, New York). {\it Superspace and
the nature of quantum geometrodynamics.}

\ref {Woo, C.H.}, \pr, D39, 3174, 1989. Comment on ``Quantum
cosmology and the initial state of the universe".

\refp {Woodard, R.}, Brown preprint HET-723, 1989. Enforcing the
Wheeler-DeWitt constraint the easy way.

\ref {Wu, Z.C.}, \pl, B146, 307, 1984. Quantum Kaluza-Klein
cosmology I.

\ref {Wu, Z.C.}, \pr, D31, 3079, 1985a. Dimension of the universe.

\ref {Wu, Z.C.}, \grg, 17, 1217, 1985b. Space-time is
four-dimensional.

\refp {Wu, Z.C.}, preprint, 1985c. Primordial black hole.

\ref {Wudka, J.},  \pr, D35, 3255, 1987a. Quantum effects in a
model of cosmological compactification.

\ref {Wudka, J.}, \pr, D36, 1036, 1987b. Boundary conditions and
the cosmological constant.

\refp {Yokoyama, J., Maeda, K. and Futamase, T.}, Tokyo preprint
UTAP-78/88, 1988. Quantum cosmology with a non-minimally coupled
scalar field.

\ref {Zeh, H.D.}, \pl, A116, 9, 1986. Emergence of time from a
universal wave function.

\ref {Zeh, H.D.}, \pl, A126, 311, 1988. Time in quantum gravity.

\refc {Zeh, H.D. (1989a)}, {\it The Physical Basis of the
Direction of Time} (Springer, Heidelberg).

\refc {Zeh, H.D. (1989b)}, in {\it Complexity, Entropy and the
Physics of Information}, Santa Fe Institute Studies in the
Sciences of Complexity, edited by W.H.Zurek, vol.IX (Reading, MA,
Addison-Wesley). {\it Quantum measurements and entropy.}

\ref {Zhuk, A.}, \cqg, 5, 1357, 1988. Problem of boundary
conditions in quantum cosmology: A simple example.

\ref {Zurek, W.H.}, \pr, D24, 1516, 1981.

\ref {Zurek, W.H.}, \pr, D26, 1862, 1982.

\vfil \eject

\nopagenumbers

\epsfxsize=4.5in\epsfbox{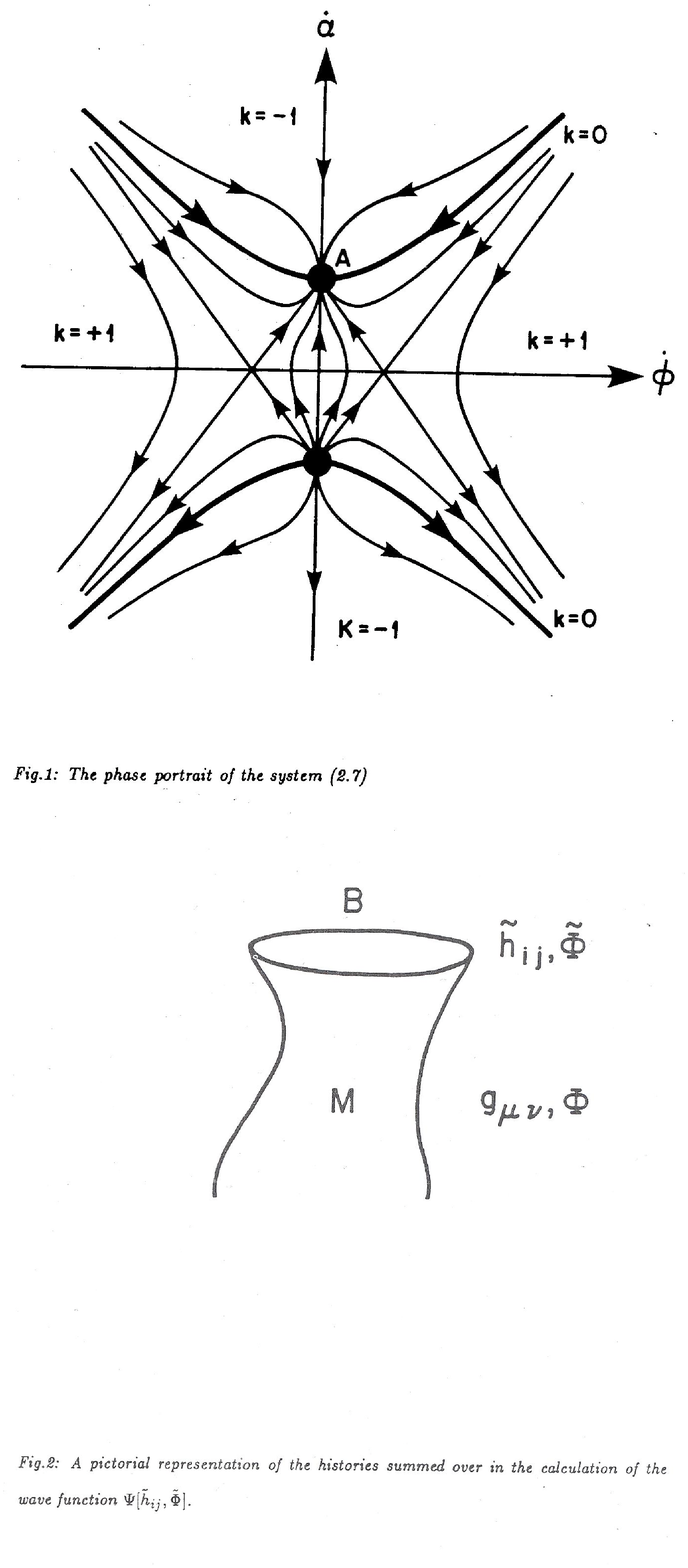}

\epsfxsize=4.5in\epsfbox{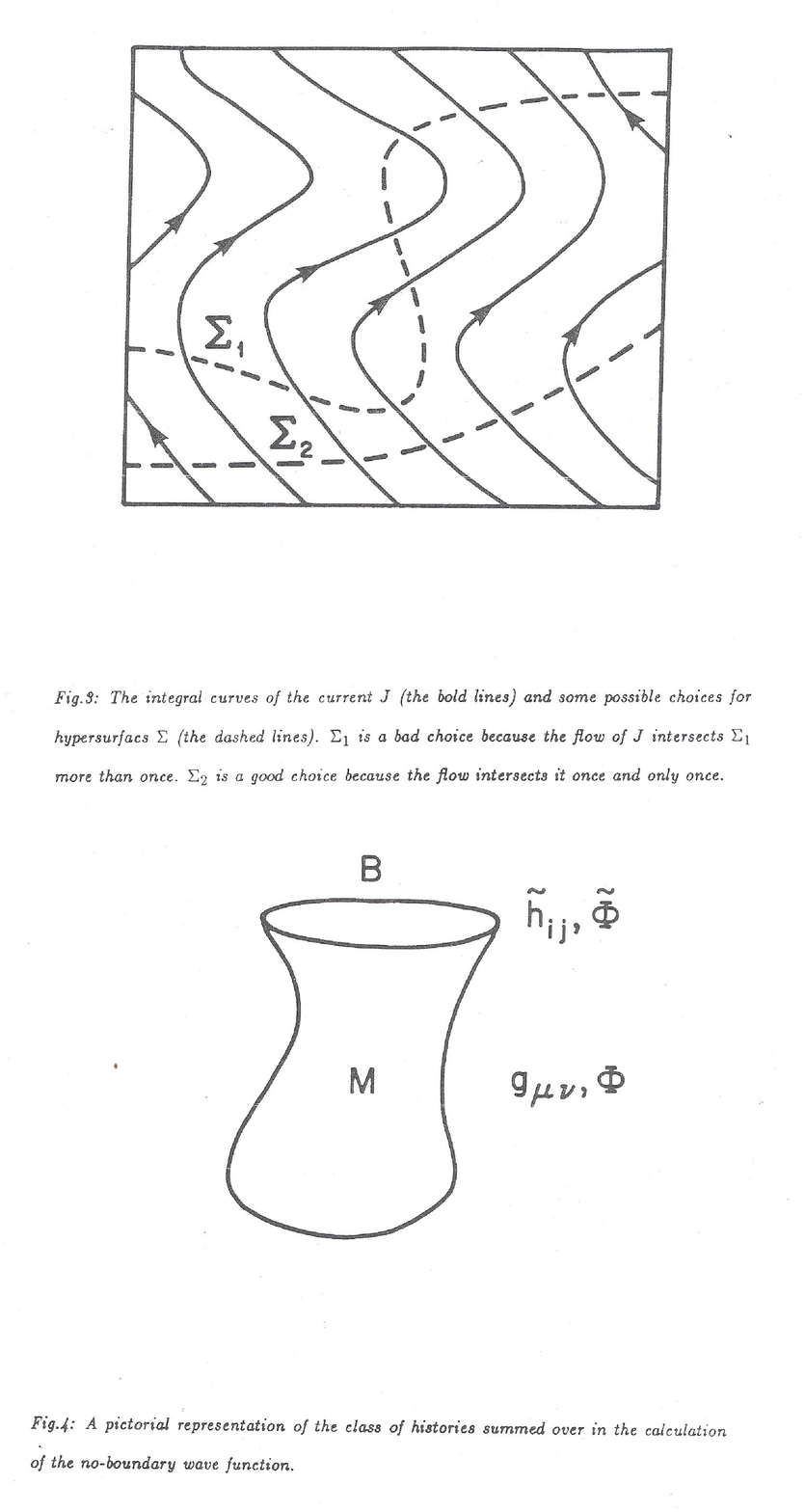}

\epsfxsize=4.5in\epsfbox{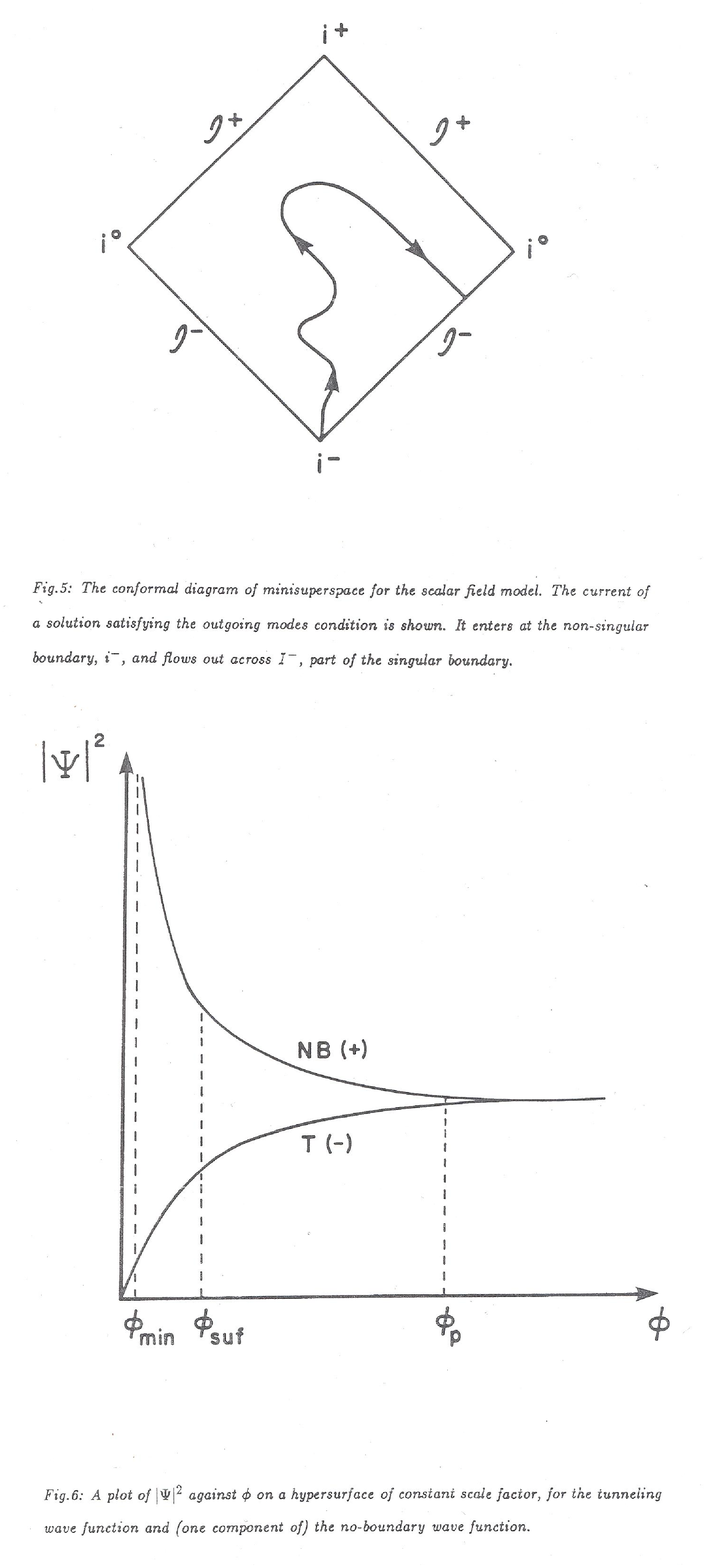}

\end

\end